\documentclass[11pt]{article}
\pdfoutput=1 
\usepackage{./aug/jheppub}

\usepackage{graphicx,subcaption}
\usepackage{amsfonts,amsmath,amssymb}
\usepackage{tikz}
\usepackage{longtable}
\usepackage{verbatim}
\usepackage{array,setspace,mathrsfs,yfonts,dsfont,bbm,colonequals,amscd}
\usepackage{relsize,suffix,mathtools,cancel,bbm}

\newcommand{\be}{\begin{eqnarray}}
\newcommand{\ee}{\end{eqnarray}}
\newcommand{\bea}{\begin{eqnarray}}
\newcommand{\eea}{\end{eqnarray}}

\newcommand{\nn}{\nonumber}
\newcommand{\bn}{\begin{enumerate}}
\newcommand{\en}{\end{enumerate}}

\def\IT{\mathbb{T}}

\def\CI{{\cal I}}

\def\CN{{\cal N}}
\def\CO{{\cal O}}

\def\CS{{\cal S}}

\def\CW{{\cal W}}

\def\CZ{{\cal Z}}

\def\Tr{\mathop{\mathrm{Tr}}\nolimits}

\def\SU{\mathrm{SU}}
\def\SO{\mathrm{SO}}

\def\U{\mathrm{U}}

\def\det{\mathrm{det}}

\def\det{{\rm det}}

\def\rk{ {\rm rk } }

\def \nn{\nonumber}

\def \Tr{{\rm Tr}}

\definecolor{purple}{rgb}{0.6,0.4,0.8}

\newcommand{\NN}{\mathcal{N}}

\renewcommand{\Re}{\text{Re}}
\renewcommand{\Im}{\text{Im}}

\newcommand{\qPoc}[2]{\left(#1;#2\right)_\infty}
\newcommand{\GF}[3]{\Gamma\left(#1;#2,#3\right)}
\newcommand{\tGF}[3]{\tilde{\Gamma}\left(#1;#2,#3\right)}
\newcommand{\thf}[2]{\theta_0\left(#1;#2\right)}

\newcommand{\ou}{\overline{u}}
\newcommand{\Dbr}[2]{\left[\Delta_{#1}\right]_{#2}}
\newcommand{\br}[2]{\left[#1\right]_{#2}}
\newcommand{\tQ}{\tilde{Q} }
\newcommand{\tJ}{\tilde{J} }
\newcommand{\tL}{\tilde{\Lambda} }

\title{Black hole entropy function for toric theories\\ via Bethe Ansatz}

\preprint{}

\author{Assaf Lanir,\!}
\author{Anton Nedelin,\!}
\author{and Orr Sela}

\affiliation{Department of Physics, Technion, Haifa, 32000, Israel}

\emailAdd{slanira@campus.technion.ac.il}
\emailAdd{anton.nedelin@physics.uu.se}
\emailAdd{sorrsela@campus.technion.ac.il}

\abstract
{
We evaluate the large-$N$ behavior of the superconformal indices of toric quiver gauge theories, and use it to find the entropy functions of the dual electrically charged rotating $\mathrm{AdS}_5$ black holes. To this end, we employ the recently proposed Bethe Ansatz method, and find a certain set of solutions to the Bethe Ansatz Equations of toric theories. This, in turn, allows us to compute the large-$N$ behavior of the index for these theories, including the infinite families $Y^{pq}$, $X^{pq}$ and $L^{pqr}$ of quiver gauge theories. Our results are in perfect agreement with the predictions made recently using the Cardy-like limit of the superconformal index. We also explore the index structure in the space of chemical potentials and describe the pattern of Stokes lines arising in the conifold theory case.
}

\begin{document} 

\maketitle
\flushbottom


\section{Introduction}
\label{Int}

The pursuit of a theory of quantum gravity is one of the greatest challenges of modern 
theoretical physics. The description of the microscopic origin of black hole entropy 
constitutes a highly important step towards such a theory. 
One of the tremendous successes in this direction was achieved in string theory which 
provides a microscopic explanation for the entropy of a class of asymptotically 
flat black holes \cite{Strominger:1996sh}. 

Another source that might shed light on this question could be the study of the 
entropy of asymptotically $\mathrm{AdS}$ black holes. This is by virtue of the $\mathrm{AdS/CFT}$ 
correspondence which establishes a connection between a quantum theory of gravity and 
a conformal field theory (CFT) living on the boundary of the $\mathrm{AdS}$-space. This connection provides us with 
the natural playground for the study of black hole microstates. In this setting the problem 
of microstate counting of asymptotically $\mathrm{AdS}$ black holes translates into the counting of 
BPS states in the dual CFT. For a long time, many approaches employed in the attempt to match these dual descriptions had fallen short. However, recent years have seen substantial progress in this direction. 

First, the Bekenstein-Hawking entropy of static dyonic BPS black holes in $\mathrm{AdS}_4$ was matched with the twisted index \cite{Benini:2015noa,Benini:2016hjo,Closset:2016arn} of the dual ABJM theory on $S^2\times S^1$ with a topological twist on $S^2$ \cite{Benini:2016rke,Benini:2015eyy}. Subsequently, similar calculations were generalized to numerous cases of dualities between magnetically charged $\mathrm{AdS}$ black holes and twisted indices of CFTs in various dimensions 
\cite{Hosseini:2016tor,Hosseini:2017fjo,Hosseini:2016cyf,Hosseini:2018uzp,Crichigno:2018adf,Benini:2017oxt,Azzurli:2017kxo,Hosseini:2018usu,Fluder:2019szh,Hosseini:2016ume}. Some of the calculations even included first quantum corrections to the black hole entropy \cite{Liu:2018bac,Liu:2017vbl,Liu:2017vll}. An extensive review of this subject can be found in \cite{Zaffaroni:2019dhb}.

While considerable progress was made in the context of static magnetically charged black holes (in the $\mathrm{AdS/CFT}$ setting), the situation as far as rotating electrically charged black holes are concerned had remained unclear. The simplest and canonical example of this kind is that of BPS black holes arising in type IIB supergravity on $\mathrm{AdS}_5\times S^5$ \cite{Gutowski:2004yv,Gutowski:2004ez,Chong:2005hr,Chong:2005da}. On the dual side, microstate counting of these black holes should correspond to the counting of $\frac{1}{16}$-BPS states in $4d$ $\CN=4$ $\SU(N)$ Super Yang-Mills (SYM) theory on $S^3\times \mathbb{R}$.
The latter are counted by the superconformal index \cite{Romelsberger:2005eg,Kinney:2005ej,Rastelli:2016tbz}.
Yet, an attempt to match these two quantities has failed already at the leading order in $N$, since the number
of BPS states grew as $N^0$, which is much slower than the growth of $N^2$ corresponding to the black hole entropy.
This slow growth was explained by the huge cancellation between fermionic and bosonic states counted by the index. 
Later, indices of other theories including the conifold theory \cite{Nakayama:2005mf}, the $Y^{pq}$ family \cite{Gadde:2010en} and finally 
toric theories in general \cite{Eager:2012hx}, were also evaluated at the large-$N$ limit.
In each case the situation repeated the one of $\CN=4$ SYM case: the indices appeared to 
be of order $\CO(1)$ at large $N$ and even the matching with a supergravity index not involving
any black hole microstates took place. 

This puzzle has been recently resolved in several steps. The first inspiration came from the \textit{extremization principle} used to resolve the problem in the case of magnetically charged black holes. In particular, in \cite{Hosseini:2017mds} the \textit{entropy function} of the rotating $\mathrm{AdS}_5$ 
black hole with two angular momenta $J_{1,2}$ and three charges $Q_{1,2,3}$ was proposed. The entropy function is the Legendre transform of the black hole entropy. It is a function of the chemical potentials for the electric charges $\Delta_{1,2,3}$ and angular momenta $\omega_{1,2}$, correspondingly. The chemical potentials, in turn, appear to be constrained by the relation $\Delta_1+\Delta_2+\Delta_3=\omega_1+\omega_2-1$. Later on, similar constructions were proposed for electrically-charged black holes in other dimensions \cite{Hosseini:2018dob,Choi:2018fdc}. Furthermore, it was also realized that the black hole entropy function may be reproduced by the evaluation of the complexified supergravity on-shell action in the bulk \cite{Cabo-Bizet:2018ehj}. 

On the dual CFT side, the entropy function has a clear formulation in terms of the corresponding field theory
parameters, but a precise relation to the counting of BPS states had remained obscure. An answer to this 
question appeared in \cite{Choi:2018hmj} where the authors considered the superconformal index with complex
fugacities. The latter allowed them to avoid large cancellations between different states and to obtain an
$N^2$ scaling for the growth in the number of BPS states. More importantly, considering the \textit{Cardy-like limit}\cite{DiPietro:2014bca},
also referred to as the high-temperature limit, the authors were able to identify the superconformal index of $\CN=4$ SYM given by 
\be
\CS=-\pi i N^2\frac{\Delta_1 \Delta_2 \Delta_3}{\omega_1\omega_2}\,,
\label{cardy:kim:result}
\ee
with the corresponding entropy function of the electrically charged $\mathrm{AdS}$ black hole. Here $\Delta_a$ and $\omega_a$ are chemical potentials conjugated to three $R$-charges of $\SO(6)$ $R$-symmetry and two angular momenta $J_{1,2}$ on $S^3$, correspondingly. Just as in the entropy function, these chemical potentials satisfy the constraint mentioned above.  

This important result was subsequently extended to various $4d$ $\CN=1$ SCFTs \cite{ArabiArdehali:2019tdm,Honda:2019cio,Kim:2019yrz,Cabo-Bizet:2019osg,Amariti:2019mgp}. For the purposes of the present paper, the most relevant is the observation made in \cite{Amariti:2019mgp}, where the authors considered a Cardy-like limit of the superconformal index for the class of  toric quiver gauge theories. The following expression was obtained at the large $N$ limit
\be
\CS=-i \pi N^2 \frac{C_{IJK}\Delta_I \Delta_J \Delta_K}{6\omega_1\omega_2}\,,
\label{antonio:result}
\ee
where $\Delta_I$ are chemical potentials for $(d-1)$ $\U(1)_I$ global symmetries satisfying the constraint $\sum_{I=1}^{d-1} \Delta_I-\omega_1-\omega_2=1$. 
The coefficients $C_{IJK}$ are associated with the triangle anomalies of the corresponding $\U(1)$ global symmetries that can be obtained directly from the toric data of the gauge theory quiver. Interestingly, expression (\ref{antonio:result}) was already proposed for the entropy function of $\mathrm{AdS}_5$ black holes with $d$ electric charges in Appendix A of \cite{Hosseini:2018dob}. 

Despite the great success in computing the superconformal index at the Cardy-like limit, the importance of that particular limit and the reason why it precisely reporduced the black hole entropy, even at finite temperatures, were unclear. The proper way to treat the problem is to calculate the large $N$ limit of the index with complex fugacities, relying upon no further restrictions or limits. This kind of computation was performed in \cite{Benini:2018ywd} for the canonical case of $\CN=4$ SYM, for which the authors exploited the \textit{Bethe Ansatz} technique \cite{Benini:2018mlo,Closset:2017bse}. Its main idea is to reformulate the integral representation of the superconformal index as a sum of residues of the integrand. The positions of the integrand's poles, in turn, are defined by the 
solutions to the \textit{Bethe Ansatz Equations} (BAEs) system. Generally, it is very hard to solve this system of equations. However, in \cite{Benini:2018ywd} it was shown that using a particular class of relatively simple solutions to the BAEs, it is possible to reproduce the entropy function (\ref{cardy:kim:result}) at large $N$, but finite temperature limit, within a certain range of values for the chemical potentials $\Delta_I$. 

Another intriguing observation made in \cite{Benini:2018ywd} is the highly complicated structure of the large $N$ index in the space of complex chemical potentials. In particular, there are many competing exponential contributions to the index. The full space of chemical potentials divides into separate regions with one of the exponential contributions dominating in each of them. These regions are separated by codimension one surfaces, \textit{Stokes lines}, on top of which different exponents contribute the same and hence there is no dominant contribution. This also explains why the calculation in \cite{Kinney:2005ej} resulted in an $\CO(N)$ index since it has been performed precisely on such Stokes lines where $\CO(N^2)$ saddles cancel each other.

The BAE approach is considerably more involved than Cardy limit calculations, but it provides us with a large $N$ answer that is precise in temperature. It further grants access to the Stokes phenomenon and the full large $N$ index structure. It is therefore natural to apply this approach to theories beyond $\CN=4$ SYM. Here, similarly to \cite{Amariti:2019mgp} we consider the superconformal index of various toric theories. In particular, we show that the basic solutions to the BAEs for $\CN=4$ SYM used in \cite{Benini:2018ywd} also solve the BAEs for all toric quiver theories. This allows us to readily evaluate the leading order contribution of the basic solution to the superconformal index of these theories. Note that in principle there might be other solutions contributing at the same order of $N$ which are difficult to find due to the complicated structure of the BAEs. However, just as in the $\CN=4$ SYM case, we find that the Cardy limit prediction of \cite{Amariti:2019mgp} presented in (\ref{antonio:result}) is exactly reproduced by the basic solution contribution, and is precise in temperature for a certain range of values for the chamical potentials. We thus provide a more definite support in favor of the conjectured form of the multi-charge black hole entropy function proposed in \cite{Hosseini:2018dob}.

Unfortunately, entropy functions (\ref{antonio:result}) appear to be quite complicated for performing extremization. In particular, in \cite{Amariti:2019mgp} the only cases considered in detail were those of $Y^{pp}$ and $L^{aba}$. However, the BAE approach also grants us access to the Stokes phenomenon. Since reproducing the full structure of the large-$N$ index in the space of chemical potentials seems too complicated a task, we consider certain limits. In particular, as was done in \cite{Benini:2018ywd} we consider the special case of equal chemical potentials $\Delta_I$ and 
also take $\omega_1=\omega_2=\tau$ as required by the BAE technique. Consequently, the only parameter on which the large-$N$ index depends is $\tau$, and we can define the positions of Stokes lines and regions of dominance of certain exponents. Interestingly, for the conifold theory we obtain the structure of the index in the complex plane of $\tau$ exactly reproducing the one presented in \cite{Benini:2018ywd} for the case of $\CN=4$ SYM. 

The paper is organized as follows. In Section \ref{Gen} we briefly review the BAE method for the calculation of the index and present the basic solutions to the BAEs for toric theories. We also present a general formula that can be used to evaluate the contribution of these solutions to the index at the large-$N$ limit. Next, in Section \ref{Coni} we study in detail the simplest possible case of all toric theories, namely the \textit{conifold theory}. We exploit the relative simplicity that characterizes this case, to clearly illustrate how most of the arguments presented in Section \ref{Gen} work. In Section \ref{Oth}, omitting details we directly apply the results of Section \ref{Gen} to many different cases of toric quiver theories, including the infinite families $Y^{pq},\, X^{pq},\, L^{pqr}$. Finally, in the last Section \ref{Extrem} we consider the large-$N$ index of the conifold theory derived in Section \ref{Coni} and study its structure in the case where all the chemical potentials $\Delta_I$ are equal. In particular, we find the positions of Stokes lines and regions of dominance for various contributions to the index.

{\bf Note added:} When we were finalizing our paper the  preprint \cite{Lezcano:2019pae}, which has a significant overlap with our work, has been posted on arXiv.
The authors' conclusions disagree with ours at various points.

\

\section{Large $N$ index for toric theories}
\label{Gen}

In this section we will consider some generalities of index computations that we will later use in order to obtain large-$N$ limits for the indices of toric theories. Let us consider general $\NN=1$ quiver gauge theories with gauge group $G$, flavour symmetry $G_F$ and non-anomalous $U(1)_R$-symmetry. The matter consists of $n_f$ chiral multiplets in the representation $R_a$ of the gauge group $G$. Since we are interested in toric theories, we will concentrate only on the cases of adjoint and bifundamental matter leaving the discussion of matter in the fundamental and other representations for the future. The integral representation of the superconformal index is then given by\cite{Dolan:2008qi,Rastelli:2016tbz} 
\be
\CI(p,q;v)=\frac{\qPoc{p}{p}^{\rk(G_a)}\qPoc{q}{q}^{\rk(G)}}{|\CW_G|}\oint\limits_{\IT^{\rk(G)}}\prod\limits_{i=1}^{\rk(G)}\frac{dz_i}{2\pi i z_i}
\prod\limits_{\alpha} \frac{1}{\GF{z^\alpha}{p}{q}}\times\nn\\
\prod\limits_{a=1}^{n_f}\prod\limits_{\rho_a\in R_a}\GF{(pq)^{r_a/2}z^{\rho_a}v^{\omega_a}}{p}{q}
\label{index:general}
\ee

Here $p$ and $q$ are complex fugacities for the angular momentum, $v_a$ are flavor fugacities with $a=1,\dots,\rk(G_F)$. The integration variables $z_i$ parametrize the maximal torus of $G$ with the integration contour given by $\rk(G)$ unit circles. The roots of $G$  are parametrized by vectors $\alpha$ while $\rho_a$ are the weights of the representations $R_a$ in which the chiral multiplets transform. Also, $|\CW_G|$ is the order of the Weyl group. Finally, $\GF{z}{p}{q}$ and 
$\qPoc{a}{q}$ are standard elliptic Gamma function \cite{Felder} 
\be
\GF{z}{p}{q}\equiv\prod\limits_{m,n=0}^\infty\frac{1-p^{m+1}q^{n+1}z^{-1}}{1-p^m q^n z}\,,
\label{gamma:def}
\ee
and $q$-Pochhammer symbol:
\be
\qPoc{z}{q}\equiv\prod\limits_{n=0}^\infty\left( 1-zq^n \right)\,.
\label{qPoc:def}
\ee

Throughout the paper, we will also be using the \textit{chemical potentials} $\tau, \sigma, \xi$ and \textit{complexified holonomies}  $u_i$ as follows
\be
p=e^{2\pi i \tau}\,,\quad q=e^{2\pi i \sigma}\,,\quad v_a=e^{2\pi i \xi_a}\,, \quad z_i=e^{2\pi i u_i}\,.
\label{chem:exp}
\ee

All the functions in the index (\ref{index:general}), and hence index itself, are well-defined for $|p|<1$ and $|q|<1$. 
We will be interested in  the large $N$ behavior of the index. In order to estimate it we will employ the  technique
of \textit{Bethe Ansatz Equations} (BAE) \cite{Closset:2017bse,Benini:2018mlo}, which has been recently used to find the large $N$ behavior of the superconformal index of ${\cal N}=4$ SYM \cite{Benini:2018ywd}. The obtained results in this case perfectly matched the entropy of the corresponding $AdS_5$ black hole. Below, we briefly describe this method closely following \cite{Benini:2018mlo,Benini:2018ywd}. 

The main observation behind the method is that for $\tau$ and $\sigma$ satisfying 
\be
\tau/\sigma\in \mathbb{Q}_+\,,
\label{tau:sigma:cond}
\ee
we can recast the integral representation (\ref{index:general}) of the index as the sum over poles located at the solutions to certain transcendental equations called BAEs. Notice that as was shown in \cite{Benini:2018mlo}, the set of $\tau$ and $\sigma$ satisfying (\ref{tau:sigma:cond}) is dense in the domain $\{|p|<1,|q|<1\}$ so the method is, in principle, applicable for any fugacities $p$ and $q$. 
However, even when $\tau$ and $\sigma$ are generic parameters satisfying (\ref{tau:sigma:cond}), it is hard to perform BAE calculations. Therefore, for simplicity and without loss of generality we will consider the simplest possible case of $\tau=\sigma$. 
Under this condition, as was shown in \cite{Closset:2017bse,Benini:2018mlo}, the poles of the integral (\ref{index:general}) are located at the solutions to the following equations written in terms of \textit{BA operators} $Q_i$:
\be
Q_i\left(u;\xi,\tau \right)&=&1\,,\qquad \forall i=1,\dots,\rk(G)\,,\nn\\
Q_i\left(u;\xi,\tau \right)&\equiv& \prod\limits_{a=1}^{n_f}\prod\limits_{\rho_a\in R_a} P\left( \rho_a(u)+\omega_a(\xi)+r_a\tau;\tau \right)^{\rho_a}\,,
\label{BAE:general}
\ee
where the function $P(u)$ is expressed through the usual $\thf{u}{\tau}\equiv\qPoc{e^{2\pi i u} }{e^{2\pi i \tau}}\qPoc{e^{2\pi i (\tau-u)} }{e^{2\pi i \tau}}$
as follows:
\be
P(u;\tau)\equiv\frac{e^{B(u;\tau) }}{\thf{u}{\tau}}\,,\quad B(u;\tau)=-\pi i \frac{u^2}{\tau}+\pi i u\,;
\label{P:function}
\ee

After finding all the solution to the equations (\ref{BAE:general}), we can evaluate the residues of the integrand (\ref{index:general}) and represent the supreconformal index in the following form:
\be
\CI(p,q;v)=\kappa_G\sum\limits_{\hat{u}\in\mathrm{BAE}} \CZ\left(\hat{u}-\tau;\xi,\tau \right)H\left( \hat{u};\xi,\tau \right)^{-1}\,,
\label{index:BAE:gen}
\ee
where we introduce the following notations. $\CZ\left(\hat{u};\xi,\tau\right)$ is the integrand of the index (\ref{index:general}) and can be generally written in the following form  
\be
\CZ\left(\hat{u};\xi,\tau\right)\equiv\prod\limits_{\alpha}\tGF{\alpha(u)}{\tau}{\tau}^{-1}  \prod\limits_{a=1}^{n_f}\prod\limits_{\rho_a\in R_a}
\tGF{\rho_a(u)+\omega(\xi)+r_a\tau}{\tau}{\tau}\,,
\label{integrand:general}
\ee
where we introduce a new $\tilde{\Gamma}$-function, which is just the usual $\Gamma$-function defined as the function of chemical potentials,
\be
\tGF{u}{\tau}{\sigma}\equiv\GF{e^{2\pi i u}}{e^{2\pi i \tau}}{e^{2\pi i \sigma}}\,.
\label{integrand:general}
\label{tGamma:def}
\ee
The $\kappa_G$ in (\ref{index:BAE:gen}) denotes the constant prefactor of the index integral representation in (\ref{index:general}) and is defined as 
\be
\kappa_G\equiv \frac{\qPoc{p}{p}^{\rk(G)}\qPoc{q}{q}^{\rk(G)}}{|\CW_G|}	\,.
\label{kappa:expression}
\ee
The final ingredient to be explained in (\ref{index:BAE:gen}) is  the \textit{Jacobian} $H\left(u;\,\xi,\,\tau \right)$ for the change of variables $u_i\mapsto Q_i(u)$:
\be
H\left( u;\,\xi,\,\tau \right)\equiv \det\left[ \frac{1}{2\pi i}\frac{\partial Q_i\left(u;\xi,\tau \right)}{\partial u_j} \right]\,.
\label{jacobian:general}
\ee

Notice that as promised we have concentrated on the particular case of $p=q$ for the index. However, all the equations above may be written for the more general case of $\tau/\sigma\in \mathbb{Q}_+$. All the corresponding expressions can be found in \cite{Benini:2018mlo}.

\subsection{Solutions to the BAE}

Let us start by very briefly reviewing the solution to the BAE for $\CN=4$ $\SU(N)$ SYM that was used in \cite{Benini:2018ywd} in order to match the black hole entropy with the large $N$ behavior of the index. The BAE (\ref{BAE:general}) in this case take the following form:
\be
1=Q_i\left( u;\Delta,\tau\right)=e^{2\pi i(\lambda+3\sum\limits_{j}u_{ij}  )}\prod\limits_{j=1}^N\frac{\thf{u_{ji}+\Delta_1}{\tau}\thf{u_{ji}+\Delta_2}{\tau}}
{\thf{u_{ij}+\Delta_1}{\tau}\thf{u_{ij}+\Delta_2}{\tau}}\times\nonumber\\
\frac{\thf{u_{ji}-\Delta_1-\Delta_2}{\tau}}{\thf{u_{ij}-\Delta_1-\Delta_2}{\tau}}\,,\qquad i=1,\dots,N\,,
\label{BAE:sym}
\ee
where $u_{ij}\equiv u_i-u_j$. Notice that due to the double periodicity of $Q_i$, if $u_i$ is the solution to the BAE then so are $u_i+1$ and $u_i+\tau$. Therefore, there is in fact an infinite number of solutions to the BAE, but we can group them into equivalence classes. In other words, our holonomies $u_i$ live on a torus with modular parameter $\tau$. Following \cite{Benini:2018mlo} we also introduce the convenient parametrization $\Delta$ for the flavor fugacities
defined as follows:
\be
\Delta_a=\omega(\xi_a)+r_a\frac{\tau+\sigma}{2}\,,
\label{delta:def}
\ee
where $r_a$ are the $R$-charges of the corresponding multiplets. The for the case of $\CN=4$ SYM, this parametrization just takes the form $\Delta_a=\xi_a+\frac{\tau+\sigma}{3}$. Finally, $\lambda$ in (\ref{BAE:sym}) is the \textit{Langrange multiplier} required here since we are considering 
$\SU(N)$ gauge group instead of $\U(N)$. For the same reason, the holonomies $u_i$ satisfy the following constraint:
\be
\sum\limits_{i=1}^N u_i=0\,~ ~\mathrm{mod}~ ~ \mathbb{Z}+\tau\mathbb{Z}\,.
\label{sun:constraint:holon}
\ee

The exact same equations appeared in \cite{Hosseini:2016cyf} in the context of the twisted index of $\CN=4$ $\SU(N)$ SYM on $T^2\times S^2$ with the topological twist on $S^2$. The authors also found the following solutions to these equation  at the high-temperature limit $\tau\to 0$: \footnote{In the context of the twisted index, the high-temperature limit corresponds to shrinking one of the cycles of $T^2$. In the context of the superconformal index, this 
	corresponds to shrinking radius of $S^1$ and hence taking the limit $\tau\to 0$.}  
\be
u_{ij}=\frac{\tau}{N}(j-i)\,,\qquad u_i=\frac{\tau}{N}\left(\frac{N+1}{2}-i\right)\,,\qquad \lambda=\frac{N-1}{2}\,.
\label{BAE:sol:sym}
\ee
In \cite{Benini:2018ywd} it was shown that the solution above is actually not limited to the high-temperature limit 
and, in fact, solves the BAE (\ref{BAE:sym}) exactly for both finite $N$ and $\tau$.

It was also shown in \cite{Hosseini:2016cyf} that a similar solution, namely
\be
u_i^{(a)}=-\frac{\tau}{N}i+\ou\,,\quad \ou\equiv \tau\frac{N+1}{2N}\,,\qquad i=1,\dots,\rk(G^{(a)})\,, 
\label{BAE:sol:u:gen}
\ee
is valid at the high temperature limit of the BAE for Klebanov-Witten (KW) (or conifold) theory \cite{Klebanov:1998hh}
and in general for any toric quiver. The index $(a)$ in the expression above refers to the node of the quiver. 
We will demonstrate here that, just as in the case of $\CN=4$ SYM, this high temperature solution is actually exact in $N$ and $\tau$. 
Finally, just as in the case of $\CN=4$ SYM we added the constant $\ou$ to satisfy the $\SU(N)$ constraint (\ref{sun:constraint:holon}) at each of the nodes. We will refer to the solution (\ref{BAE:sol:u:gen}) as \textit{basic solution}.

To understand how this works, notice that in toric theories each node has an equal number of bifundamentals and anti-bifundamentals. The contribution of each such pair to the BAE is given by
\be
Q_l^{(a)}\sim \prod\limits_{j=1}^N\frac{\thf{u^{(ba)}_{jl}+\Delta_{ba}}{\tau}}{\thf{u^{(ac)}_{lj}+\Delta_{ac}}{\tau}}e^{B\left( u^{(ac)}_{lj}+\Delta_{ac} \right)-
	B\left( u^{(ba)}_{jl}+\Delta_{ba};\right)}\,,
\ee
where $\Delta_{1,2}$ are fugacities corresponding to each of the chiral multiplet and $u^{(ab)}_{ij}\equiv u_i^{(a)}-u_j^{(b)}$. 
Again, all upper indices denote quiver nodes. In particular, the node $(a)$ is the one that we write the equation for, while $(b)$ and $(c)$ are the ones that connect to $(a)$ through bifundamental chirals. The same expression describes the contribution of one adjoint multiplet provided $b=c=a$. In this case, $\Delta_{aa}$ is just the chemical potential associated with this multiplet.

Substituting the basic solution (\ref{BAE:sol:u:gen}) in the expression above we can use the following chain of manipulations to drastically simplify it
\be
\frac{\prod_{j=1}^N\thf{u^{(ba)}_{jl}+\Delta_{ba}}{\tau}}{\prod_{j=1}^N\thf{u^{(ac)}_{lj}+\Delta_{ac}}{\tau}}=
\frac{\prod_{j=1}^N\thf{\frac{\tau}{N}(l-j)+\Delta_{ba}}{\tau}}{\prod_{j=1}^N\thf{\frac{\tau}{N}(j-l)+\Delta_{ac}}{\tau}}=
\nn\\
\frac{\prod_{k=0}^{l-1}\thf{\frac{\tau}{N}k+\Delta_{ba}}{\tau}\prod_{k=-1}^{l-N}\thf{\frac{\tau}{N}k+\Delta_{ba}}{\tau}}
{\prod_{k=0}^{N-l}\thf{\frac{\tau}{N}k+\Delta_{ac}}{\tau}\prod_{k=-1}^{1-l}\thf{\frac{\tau}{N}k+\Delta_{ac}}{\tau}}=
\nn\\
\frac{\prod_{k=0}^{N-1}\thf{\frac{\tau}{N}k+\Delta_{ba}}{\tau}\prod_{k=-1}^{l-N}\left( -q^{k/N}y_{ba} \right)}
{\prod_{k=0}^{N-1}\thf{\frac{\tau}{N}k+\Delta_{ac}}{\tau}\prod_{k=-1}^{1-l}\left( -q^{k/N}y_{ac} \right)}=
\nn\\
\frac{\prod_{k=0}^{N-1}\thf{\frac{\tau}{N}k+\Delta_{ba}}{\tau}}{\prod_{k=0}^{N-1}\thf{\frac{\tau}{N}k+\Delta_{ac}}{\tau}}\times
e^{2\pi i l\left( \tau-1-\Delta_{ac}-\Delta_{ba}\right)}e^{\pi i\left(\left(1-\tau  \right)\left( 1+N \right)+2\Delta_{ac}+2 N \Delta_{ba}  \right)}\,.
\label{theta:simplification}
\ee
Here, between the first line and the second we introduce new summation indices $k=i-j$ in the numerator and $k=j-i$ in the denominator. 
Between the second line and the third we performed a shift of the index $k\to k+N$ in the product $\prod_{k=-1}^{i-N}$ in the numerator and in the product $\prod_{k=-1}^{1-i}$ in the denominator. 

Now all the terms which do not depend on $l$, including the ratio of $\theta$-functions product, can be omitted since we can always absorb them into the Lagrange multiplier $\Lambda^{(a)}=e^{2\pi i \lambda^{(a)}}$.
Combining the $l$-dependent part with the one coming from the $B(u;\tau)$ parts we get:
\be
Q_l^{(a)}\sim e^{2\pi i l}=1\,,
\ee
{it i.e.} every dependence on $l$ goes away and we are left with terms independent of $l$ that can be absorbed into the Lagrange multiplier. This proves the validity of the basic solution for the class of non-chiral quivers with only bifundamental or adjoint matter. 

For the  case of $\CN=4$ SYM it was shown in \cite{Hong:2018viz} that the solution (\ref{BAE:sol:sym}) is a particular case of a wider class of solutions 
\be
u^{(a)}_{\hat{j}\hat{k}}=\frac{\hat{j}}{m}+\frac{\hat{k}}{n}\left( \tau+\frac{r}{m} \right)+\ou\,,
\label{BAE:sol:more:general}
\ee
where $\hat{j}=0,\dots,m-1$, $\hat{k}=0,\dots,n-1$ is the parametrization of the index $j=0,\dots,N-1$ in case $N$ can be represented as the product $N=nm$, and $0\leq r<n$. Since the BAE we get are similar to the simplest SYM case \footnote{The most important property that is required to generate the solutions (\ref{BAE:sol:more:general}) is modularity of the BAE, which is definitely present in our case as well since the equations are composed of $\theta$ functions only.} and the basic solution is just identical, it is straightforward that the same general class (\ref{BAE:sol:more:general}) is valid in our case. 

The family (\ref{BAE:sol:more:general}) is organized into $\mathrm{PSL}(2,\mathbb{Z})$ orbits with the following actions:
\be
T: \{m,n,r\}\to \{m,n,r+m\}\,,\qquad S:\{m,n,r\}\to \left\{\mathrm{gcd}(n,r),\frac{mn}{\mathrm{gcd}(n,r)},\frac{m(n-r)}{\mathrm{gcd}(n,r)}\right\}\,,
\nn\\
\label{PSL:action}
\ee
where the last entry of the triplet $\{m,n,r\}$ is always understood modulo $n$.

We will always be considering only the contribution of the basic solution (\ref{BAE:sol:u:gen}) and of the more general family (\ref{BAE:sol:more:general}). In principle, it is not obvious that there are no other solutions to the BAE contributing to the index at the leading $N$ order. This is already true even for the simplest case of $\CN=4$ SYM, let alone for complicated quiver theories, where the situation becomes more involved. Nevertheless, from \cite{Benini:2018ywd} we know that the class of solutions (\ref{BAE:sol:more:general}) gives the leading contribution to the index which matches the supergravity result. 
Motivated by this result, we will also proceed to check the contribution of (\ref{BAE:sol:more:general}) against $\mathrm{AdS}$ black hole entropy and previous high-temperature index computations \cite{Amariti:2019mgp}. 

\subsection{Large-$N$ superconformal index}

In this section we will describe the contributions to the index (\ref{index:general}) at large-$N$ of the basic solution (\ref{BAE:sol:u:gen}) and of the more general solutions (\ref{BAE:sol:more:general}) to the BAE. For this, we should just employ Eq. (\ref{index:BAE:gen}) which is straightforward though quite a tedious task. However, as we have shown in the previous section, the solutions we exploit are basically the same solutions that have been previously used in \cite{Benini:2018ywd} in order to find the large-$N$ behavior of the $\CN=4$ SYM index. Hence, we can exploit the derivations used in the latter case. Below, we briefly summarize the crucial expressions and derivations that are required for the calculation and explain, where relevant, the generalization to quiver theories. For all the details of the derivations we refer the reader to the original paper \cite{Benini:2018ywd}.

{\bf \textit{Contribution of the integrand.} }
The first and most important thing to estimate is the contribution of each $\tilde{\Gamma}$-function (\ref{tGamma:def}) in the integrand $\CZ\left(\hat{u};\xi,\tau\right)$ given in (\ref{integrand:general}) in the most general form. In \cite{Benini:2018ywd} it has been shown that upon substitution of the basic solution (\ref{BAE:sol:u:gen}), the leading term of the corresponding $\tilde{\Gamma}$-function in the large-$N$ limit is given by:
\be
& \hspace{-48mm}\sum\limits_{i,j=1}^N\log \left.\tGF{u_{ij}^{(ab)}+\Delta_{ab}}{\tau}{\tau}\right|_{u_{ij}^{(ab)}=\frac{\tau}{N}(j-i)}=
\nn\\
&-\pi i N^2\frac{\left(\Dbr{ab}{\tau}-\tau\right)\left(\Dbr{ab}{\tau}-\tau+\frac{1}{2}\right)\left(\Dbr{ab}{\tau}-\tau+1\right)}{3\tau^2}+\CO(N)\,,
\label{tGamma:large:N}
\ee
where we have introduced the periodic and discontinuous function
\be
\hspace{-9mm}{\displaystyle \Dbr{ }{\tau}\equiv \left( \Delta+n~ \left|~ n\in \mathbb{Z}\,,~ \Im\left( -\frac{1}{\tau} \right)>\Im\left( \frac{\Delta+n}{\tau}\right)>0\right.\right)~ \mathrm{for} ~
	\Im\left( \frac{\Delta}{\tau} \right)\not\in\mathbb{Z}\times \Im\left( \frac{1}{\tau}\right)}\,.
\label{square:brackets}
\ee
There is a number of useful properties of this function:
\be
\br{\Delta+1}{\tau}=\Dbr{}{\tau}\,,\quad \br{\Delta+\tau}{\tau}=\Dbr{ }{\tau}+\tau\,,\quad \br{-\Delta}{\tau}=-\Dbr{ }{\tau}-1\,,
\label{square:brackets:property}
\ee
i.e. the function $\Dbr{ }{\tau}$ shifts $\Delta$ along the real line and brings it to the \textit{fundamental domain}
\be
\Im\left( -\frac{1}{\tau}\right)>\Im\left( \frac{\Delta}{\tau} \right)>0\,.
\label{delta:stripe}
\ee

\begin{figure}[b]
	\centering
	\includegraphics[scale=0.5]{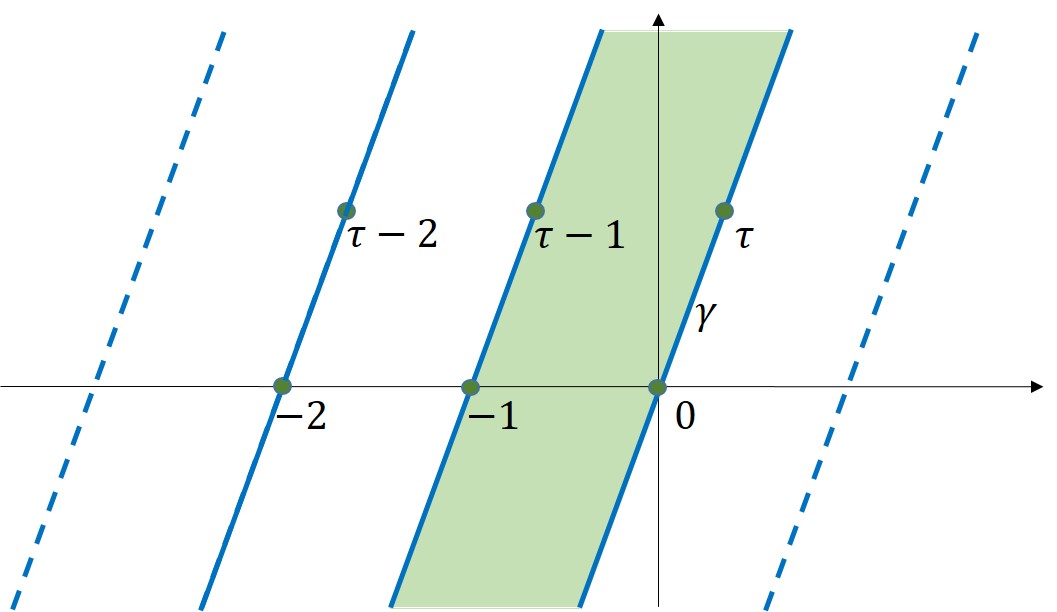}
	\caption{The definition of $\left[\Delta\right]_{\tau}$ divides the complex $\Delta$-plane into domains with boundaries which
		are the elements of the set $\gamma+\mathbb{Z}$, where $\gamma$ is the line passing through $0$ and $\tau$. 
		The green filled area is the fundamental domain, corresponding to $\textrm{Im}\left(-\frac{1}{\tau}\right)>\textrm{Im}\left(\frac{\Delta}{\tau}\right)>0$,
		where the function $\left[\Delta\right]_{\tau}$ takes its values.}
	\label{stripe:pic}
\end{figure}

For convenience we show this strip in Fig. \ref{stripe:pic}. We also introduce the notation $\gamma$ for the right boundary of this domain, {\it i.e.} the line going between the points $0$ and $\tau$. Notice that the function $\Dbr{ }{\tau}$ is discontinuous along the lines $\gamma+\mathbb{Z}$. The latter ones are \textit{Stokes lines} that divide the complex $\Delta$-plane into regions corresponding to different dominating large-$N$ contributions. Along these lines the different contributions compete with each other and a more accurate calculation including an evaluation of the subleading terms is required. 

As a result, expression (\ref{tGamma:large:N}) is not defined in the case $\Delta=0$ which sits exactly on the Stokes line. At the same time, this value of $\Delta$ is relevant for the contribution of the vector multiplets in (\ref{integrand:general}). In this case, as has been shown in \cite{Benini:2018ywd}, the leading-order contribution of the $\tilde{\Gamma}$ function is given by 
\be
\hspace{-4mm}\sum\limits_{i\neq j=1}^N\log \left.\tGF{u_{ij}^{(ab)}}{\tau}{\tau}\right|_{u_{ij}^{(ab)}=\frac{\tau}{N}(j-i)}=
\pi i N^2\frac{\tau\left( \tau-\frac{1}{2}\right)\left( \tau-1 \right)}{3\tau^2}+\CO(N\log N)\,.
\label{tGamma:large:N:vector}
\ee
These two expressions constitute all the building blocks we need for the evaluation of the integrand contribution to the superconformal index.

{\bf \textit{Contribution of the Jacobian.} }
The second possible contribution to the index comes from the Jacobian $H$ of the transformation between the $u_i$ and $Q_i$ variables given in (\ref{jacobian:general}). However, since we consider a quiver theory with $\SU(N)$ groups it is convenient to write the definition of the Jacobian in a clearer form as 
\be
H=\mathrm{det}\left[ \frac{1}{2\pi i}\frac{\partial\left( Q^{(1)}_1,\dots\,Q^{(1)}_N\,,Q^{(2)}_1,\dots\,,Q^{(p)}_N \right)}
{\partial\left(u_1^{(1)},\dots,\,u_N^{(1)},\,\lambda^{(1)}\,,u_1^{(2)},\dots,\,u_N^{(p)}\,,\lambda^{(p)}  \right)} \right]\,.
\label{jacobian:general:quiver}
\ee
In the expression above, we have first of all precisely written out the quiver indices $1,\dots,p$, where $p$ is the number of nodes. The resulting expression can be seen as a division of the full $Np\times Np$ matrix into $p\times p$ blocks of the size $N\times N$. Second, notice that due to the $\SU(N)$ gauge groups in the quiver, each set of holonomies $u_i^{(a)}$ should satisfy the constraint (\ref{sun:constraint:holon}) at each node. As a result, we have only $N-1$ independent holonomies and instead of the last holonomy $u_N^{(a)}$ we use the Lagrange multipliers $\lambda^{(a)}$

To evaluate expression (\ref{jacobian:general:quiver}), let us write the general form of the BAE operators $Q_i^{(a)}$ first. As we said, since our quiver is balanced each node $(a)$ has an equal number of outgoing (bifundamental chiral) arrows $(ab)$ and incoming (antibifundamental chiral) arrows $(ca)$. Then, we can write the corresponding operator $Q_i^{(a)}$ using the general expression (\ref{BAE:general}) in the following form:
\be
Q_l^{(a)}\left( \{u\},\{\Delta\},\tau \right)=e^{2\pi i \lambda^{(a)}} \frac{\prod\limits_{b\in (a,b) }
	\prod\limits_{j=1}^N \thf{u_{jl}^{(ba)}+\Delta_{ab}}{\tau} e^{-B\left( u_{jl}^{(ba)}+\Delta_{ab} \right)}}
{\prod\limits_{c\in (c,a) }\prod\limits_{j=1}^N \thf{u_{lj}^{(ac)}+\Delta_{ca}}{\tau} e^{-B\left( u_{lj}^{(ac)}+\Delta_{ca} \right)}}\,,
\label{BAE:operator:toric}
\ee
where we denote by $(a,b)$ the set of all arrows going from the node $(a)$ to the node $(b)$ of the quiver. Therefore, when we take the product over all $b\in (a,b)$ for example, the product is over all the nodes that have incoming arrows originating from the node $(a)$. As discussed before, the number of terms of the products in the numerator and the denominator of (\ref{BAE:operator:toric}) is the same since each node of the quiver has an equal number of incoming and outgoing arrows. Moreover, as before:
$u_{jl}^{(ab)}\equiv u_{j}^{(a)}-u_l^{(b)}$. 

Now, we can easily find the logarithmic derivative of the BAE operator (\ref{BAE:operator:toric}):
\be
\frac{\partial \log Q_l^{(a)}}{\partial u_m^{(b)}}=\sum\limits_{j=1}^N\left(\sum\limits_{c\in (a,c)}\frac{\partial u_{jl}^{(ca)}}{\partial u_m^{(b)}}
F\left(u_{jl}^{(ca)}+\Delta_{ac} \right)-\sum\limits_{d\in (d,a)}\frac{\partial u_{lj}^{(ad)}}{\partial u_m^{(b)}}
F\left(u_{lj}^{(ad)}+\Delta_{da} \right)\right)\,,
\label{BAE:op:log:der}
\ee
where for convenience we introduced the following function: 
\be
F(u)\equiv \frac{2\pi i}{\tau}u-\pi i+\frac{\theta_0'\left( u;\tau \right)}{\thf{u}{\tau}}\,.
\label{det:der:func}
\ee
Now we can estimate the contributions of each of the logarithmic derivatives (\ref{BAE:op:log:der}) to the Jacobian (\ref{jacobian:general:quiver}) upon the substitution of the basic solutions (\ref{BAE:sol:u:gen}) to the BAE. In particular, we should use 
\be
\frac{\partial u_{jl}^{(ca)}}{\partial u_m^{(b)}}=\delta_{jm}^{(cb)}-\delta_{lm}^{(ab)}-\delta_{jN}^{(cb)}+\delta_{kN}^{(ab)}\,.
\ee
Notice that the last two terms appear because we consider only $N-1$ holonomies $u_i^{(a)}$ as independent variables, while due to the $\SU(N)$ constraint at each node $u_N^{(a)}=-\sum_{i=1}^{N-1}u_i^{(a)}$. Then, we can rewrite the matrix elements of the Jacobian matrix $H$ as follows:
\be
\left.\frac{\partial Q_l^{(a)}}{\partial u_m^{(b)}}\right|_{\mathrm{BAE}}&=&
\left.\frac{\partial \log Q_l^{(a)}}{\partial u_m^{(b)}}\right|_{\mathrm{BAE}}=\left(\delta_{lm}^{(ab)}-\delta_{lN}^{(ab)}\right)A_l^{(a)}+A_{ml}^{(ab)}\,,
\nn\\
A_l^{(a)}&=&-\sum\limits_{j=1}^N\sum\limits_{c\in (a,c)}F\left( \frac{\tau}{N}(l-j)+\Delta_{ac} \right)
-\sum\limits_{j=1}^N\sum\limits_{d\in (d,a)}F\left( \frac{\tau}{N}(j-l)+\Delta_{da} \right)\,,
\nn\\
A_{lm}^{(ab)}&=&\delta_{b\in (a,b)} \left[F\left(\frac{\tau}{N}(l-m)+\Delta_{ab} \right)-F\left(\frac{\tau}{N}(l-N)+\Delta_{ab} \right)\right]+
\nn\\
&&\hspace{2cm}\delta_{b\in (b,a)}\left[F\left( \frac{\tau}{N}(m-l)+\Delta_{ba} \right)F\left(\frac{\tau}{N}(N-l)+\Delta_{ba} \right)\right]\,,
\label{jacobian:matrix:elements}
\ee
where in the first equality we have used the definition of the BAEs $Q_i^{(a)}=1$. In addition, we have introduced $\delta_{b\in(a,b)}$, which is one if $b$ is in the set of arrows going from node $(a)$ to node $(b)$ and zero otherwise.

Finally, we also need the elements of the Jacobian (\ref{jacobian:general:quiver}) containing  the derivatives with respect to the Lagrange multipliers $\lambda^{(a)}$:
\be
\left.\frac{\partial Q_l^{(a)}}{\partial \lambda^{(b)}}\right|_{\mathrm{BAE}}=2\pi i.
\label{jacobian:lambda:der}
\ee

Now, notice that $\left.F\left(u_{ij}^{(ab)}+\Delta_{ab} \right)\right|_{\mathrm{BAE}}\sim \CO(1)$. Indeed, upon 
substituting the basic solution (\ref{BAE:sol:u:gen}) in the function $F(u)$, we see that its argument is 
$\tau\frac{1}{N}+\Delta_{ab}<u_{ij}^{(ab)}+\Delta_{ab}<\tau\frac{N-1}{N}+\Delta_{ab}$. At the same time, potential divergences can arise only at zeros of the $\theta_0(u)$ function, which take place at $u=m+n\tau$, $m,n\in \mathbb{Z}$. As we see from the above, the latter condition can only be satisfied if the chemical potential $\Delta_{ab} \in m+n\tau$. In this case, on the boundaries of the Bethe roots the distribution $F\left( u_{ij}^{(ab)}+\Delta_{ab} \right)$ can grow faster then $N^0$. Nonetheless, if we consider generic $\Delta_{ab}$, all the functions $F(u)$ in (\ref{BAE:op:log:der}) are of order one\footnote{On top of this, as we will see for the values $\Delta=n\tau$, $n\in\mathbb{Z}$ we hit Stokes lines during the computation of the index and hence the whole calculation does not make much sense since the basic solution is not the only one contributing at the leading order. As a result, we will be doing our computations assuming the $\Delta$s do not take problematic values.}. 
Using this estimate for the functions $F(u)$ on the BA solutions, we can obtain $A_l^{(a)}\sim\CO(N)$ and $A_{lm}^{(ab)}\sim\CO(1)$. Therefore, only the diagonal elements and the elements of each $N$-th line are of order $N$. All the other elements are only of order one. For the determinant it gives at most an $N^{pN}$ contribution so that 
\be
\log H\sim pN\log N\,.
\label{jacobian:estimate}
\ee
As a result, the contribution of the Jacobian $H$ to the logarithm of the index $\CI$ given in (\ref{index:BAE:gen}) is always subleading with respect to the order $N^2$ integrand contribution $\CZ$ on the basic solution (\ref{BAE:sol:u:gen}). The remaining term $\kappa_G$ also contributes at order $p N\log N$, and so the only contribution of the basic solution that is relevant for us is the one coming from the integrand. 

\textit{ {\bf Contribution from $SL\left(2,\mathbb{Z}\right)$ solutions.} }
We should next understand what is the contribution of all the other solutions from the family (\ref{BAE:sol:more:general}).
In the case of $\CN=4$ SYM, it was argued that the $T$-transformed solutions 
\be
u_{ij}=\frac{\tau+r}{N}(j-i)\,,\qquad r=0,1,\dots,N-1\,,
\label{BAE:t:transform:sol}
\ee
would contribute at the leading order in $N$ \cite{Benini:2018ywd}. In particular, one can notice that both the integrand $\CZ$ and the Jacobian $H$ are invariant under $\tau\to \tau+r$. Hence, once again only the integrand $\CZ$ contributes to the index at the leading order and the value can be computed from (\ref{tGamma:large:N}) and (\ref{tGamma:large:N:vector}) by simply substituting $\tau\to \tau+r$.

At the same time, in \cite{Benini:2018ywd} that $S$-transformed basic solutions 
\be
u_{ij}=\frac{j-i}{N}
\label{BAE:sol:S:transformed}
\ee
contribute to the index only at order $N$ and are thus subleading. Based on this, the authors argued that among all the solutions (\ref{BAE:sol:more:general}) labeled by $\{m,n,r\}$, only those obtained from the basic one (\ref{BAE:sol:sym}) contribute at leading order. 

Since our basic solutions and all the terms in the index (\ref{index:BAE:gen}) have exactly the same form as in the case of $\CN=4$ SYM, the arguments above are also valid in our case and we will assume that only the $T$-transformed solutions of (\ref{BAE:sol:u:gen}) contribute at leading order to the index. 

 {\bf \textit{ Summary.} } Gathering the above derivations, we can write down the leading $N^2$ term in the superconformal index (\ref{index:general}). Since we are now convinced that for all toric theories only the basic solution (\ref{BAE:sol:u:gen}) and its 
$T$-transform contribute at the leading order, we may now directly substitute these solutions into the index (\ref{index:BAE:gen}). 
Hence each time we have an adjoint or a chiral bifundamental multiplet with corresponding chemical potential $\Delta$, we should add a contribution (\ref{tGamma:large:N}) to $\log \CI$. For each vector multiplet, {it i.e.} for each node of the quiver, we should add a contribution of (\ref{tGamma:large:N:vector}) to the $\log\CI$. Since, as we saw, the Jacobian $H$ always contributes at the subleading order, these are all the relevant expressions we need to calculate the total contribution of the basic solution to the index, namely
\be
\left.\lim_{N\to\infty}\log\CI\right|_{\mathrm{basic~sol.}}&=&-\pi i N^2 \Theta\left(\{\Delta\},\tau\right)\,,
\nn\\
\Theta\left( \{\Delta\},\tau \right)&=&p\frac{\tau\left( \tau-\frac{1}{2}\right)\left( \tau-1 \right)}{3\tau^2}+
\nn\\
&&\sum\limits_{(a,b)}
\frac{\left(\Dbr{ab}{\tau}-\tau\right)\left(\Dbr{ab}{\tau}-\tau+\frac{1}{2}\right)\left(\Dbr{ab}{\tau}-\tau+1\right)}{3\tau^2}\,,
\label{index:general:BAE:basic}
\ee
where $p$ is the total number of the gauge group nodes and in the last term of the $\Theta(\{\Delta\};\tau)$ function, we summed over all oriented arrows in the quiver graph. This is done to avoid counting twice the contribution of each bifundamental chiral multiplet. In addition, if $a=b$ in the $(a,b)$ pair, then the adjoint multiplet is pictured as one arrow on the graph and should be counted only once. 

As discussed above, we should finally include the contributions of the $T$-transformed solutions (\ref{BAE:t:transform:sol}). 
This is done simply by shifting $\tau\to \tau+r$ in $\Theta(\{\Delta\};\tau)$. This corresponds to the contribution of 
one of the $T$-transformed solutions. In the final step we should find a $T$-transformation (or equivalently such $r\in \mathbb{Z}$)  
that maximaizes the real part of $-\pi i\Theta$:
\be
\log\CI_\infty\equiv\lim_{N\to\infty}\log\CI=\max\limits_{r\in\mathbb{Z}}\left( -\pi i N^2\Theta\left( \{\Delta\};\tau+r \right) \right)\,.
\label{index:BAE:gen:final}
\ee
This would give us the final expression for the dominant large-$N$ contribution to the index. Still, this expression is not highly informative on its own. In the following sections of the paper, we will thus be considering various toric theories case by case, and applying (\ref{index:BAE:gen:final}) we will be deriving a compact form for the final answer such that the relation between the $U(1)$ triangular anomalies and large-$N$ superconformal index would be clear. 

The function $\CI_\infty$ has a complicated structure. The space of complex parameters $(\{\Delta\},\tau)$ is divided into regions separated by \textit{Stokes lines}, which are codimension-one surfaces. In each of the regions one solution determined by the value of $r$ that maximizes $\Im\Theta\left(\{\Delta\},\tau+r  \right)$ dominates over the other solutions. On the Stokes lines themselves, however, two or more values of $r_1,r_2,\dots\in\mathbb{Z}$ yield the same contribution, {it i.e.} 
\be
\Im\Theta\left( \{\Delta\},\tau+r_1 \right)=\Im\Theta\left( \{\Delta\},\tau+r_2 \right)
\label{stokes:lines:condition}
\ee
In this case we have two or more equivalent exponents contributing to $\CI_\infty$. We should sum these exponents, but in this case their relative sign should be known in order to compute the final answer for $\CI_\infty$. That is because the contributions could potentially cancel each other, forcing us to consider the subleading orders.
The particular structure of the parameter space and Stokes lines depends on the particular theory considered. We will consider some details pertaining to the Stokes phenomenon for various toric theories in Section \ref{Extrem}.

\

\section{The conifold theory}
\label{Coni}

\begin{figure}[h]
        \centering
	\begin{minipage}{0.43\textwidth}
	\centering
	\includegraphics[width=\linewidth]{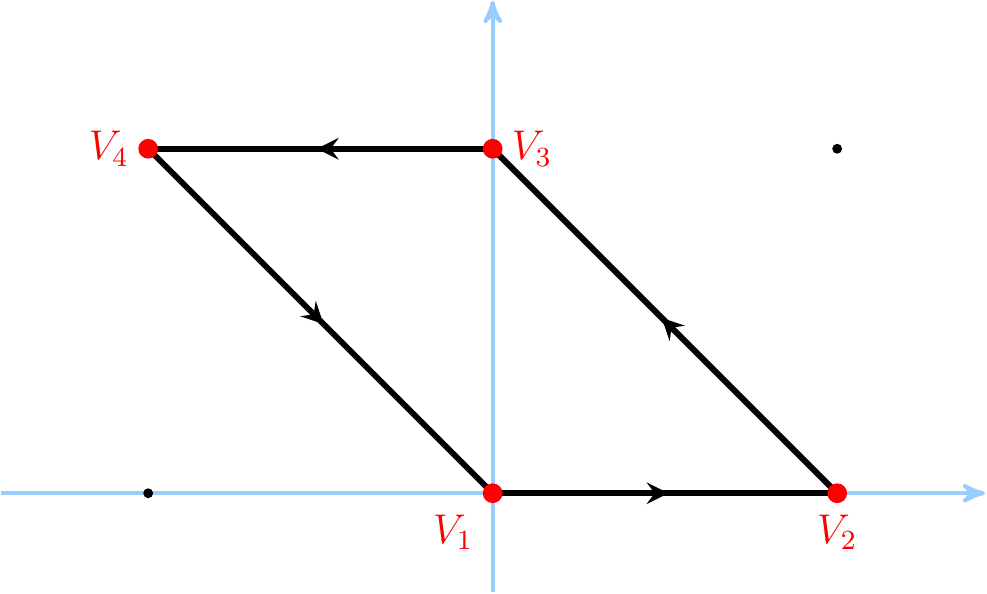}
	\caption{Toric digram of $T^{1,1}$ conifold.}
	\label{T11:toric:pic}
	\end{minipage}\hspace{20mm}
	\begin{minipage}{0.43\textwidth}
	\centering
	\includegraphics[width=0.8\linewidth]{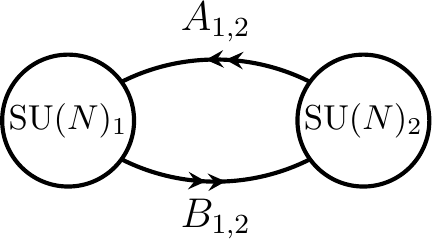}
	\caption{Quiver diagram of the conifold theory.}
	\label{conifold:quiver}
	\end{minipage}
\end{figure}

We are now ready to apply the derivations of the previous section to particular examples of toric theories. In principle, as discussed above we can directly apply the final expression (\ref{index:BAE:gen:final}) to any toric theory. However, in this section we will study in detail the BAEs, solutions to it and the large-$N$ index of the simplest possible toric theory beyond $\CN=4$ SYM, namely the \textit{conifold theory}.  
This theory was originally proposed in \cite{klebanov1998superconformal} as the worldvolume theory of  a stack of $N$ $D3$-branes
probing the tip of the conical singularity $xy-zt=0$. In the near-horizon limit this gives the $AdS_{5}\times T^{1,1}$ dual of the 
theory. Here $T^{1,1}$ is the \textit{conifold} which can be seen as the coset $\SU(2)\times \SU(2)/\U(1)$. The conifold is in fact a toric manifold with a toric diagram identified with the vectors:
\be
V_1=(0,0,1)\,,\qquad V_2=(1,0,1)\,,\qquad V_3=(0,1,1)\,,\qquad V_4=(-1,1,1)\,;
\label{T11:vectors}
\ee
which we show in Fig. \ref{T11:toric:pic}

The theory is $\CN=1$ supersymmetric with $\SU(N)\times\SU(N)$ gauge group and two pairs of bifundamental chiral multiplets $A_{1,2}$ and $B_{1,2}$ transforming in the
$\left({\bf N},\overline{ {\bf N} }  \right)$ and $\left(\overline{ {\bf N} },{\bf N}\right)$ representations of the gauge groups, correspondingly. The corresponding quiver is shown on the Fig.\ref{conifold:quiver}. The superpotential term is given by 
\be
W\sim \epsilon_{ij}\epsilon_{kl} \Tr\left(A^i B^k A^j B^l  \right)\,.
\label{conifold:superpotential}
\ee

The global symmetries of the theory include $\U(1)_R$ $R$-symmetry factor, two $\SU(2)$ flavor symmetries rotating the $A$ and $B$ doublets, correspondingly, and a $\U(1)_B$ baryonic symmetry. These symmetries can also be identified with the isometries of $T^{1,1}$. It is further convenient to use instead of $\SU(2)$ flavor symmetries Cartans of their combinations, which we denote by $\U(1)_1$ and $\U(1)_2$. The matter fields' charges under these global symmetries are summarized in the table below.
\begin{center}
	\begin{tabular}{|c||c|c|c|c|}
		\hline
		Field & $\U\left(1\right)_{R}$ & $\U\left(1\right)_{B}$ & $\U\left(1\right)_{1}$ & $\U\left(1\right)_{2}$\\
		\hline 
		$A_{1}$ & $1/2$ & $1$ & $1$ & $1$\\
		
		$A_{2}$ & $1/2$ & $1$ & $-1$ & $-1$\\
		\hline
		
		$B_{1}$ & $1/2$ & $-1$ & $1$ & $-1$\\
		
		$B_{2}$ & $1/2$ & $-1$ & $-1$ & $1$\\
		\hline
	\end{tabular}
\end{center}
Also notice that conifold theory is a particular case of the infinite $Y_{pq}$ class of theories, corresponding to $p=1$ and $q=0$. We will be considering this class of theories in section \ref{sec:ypq} with less detail.

We now have all the ingredients at our disposal to write down the integral representation (\ref{index:general}) of the superconformal index for the conifold theory.

\be
\mathcal{I}&=&\kappa_{G}\oint\prod_{i=1}^{N}\frac{dz_{i}^{\left(1\right)}}{2\pi iz_{i}^{\left(1\right)}}\frac{dz_{i}^{\left(2\right)}}{2\pi iz_{i}^{\left(2\right)}}
\prod_{a=1}^{2}\prod_{i\neq j}^{N}\Gamma^{-1}\left(\frac{z_{i}^{\left(a\right)}}{z_{j}^{\left(a\right)}};p,q\right) 
\nn\\
&\times&\prod\limits_{i,j=1}^N\Gamma\left(\left(pq\right)^{1/4}\frac{z_{i}^{\left(1\right)}}{z_{j}^{\left(2\right)}} v_{\scalebox{0.6}{$F_1$}}
v_{\scalebox{0.6}{$F_2$}}v_{\scalebox{0.6}{$B$}};p,q\right)
\Gamma\left(\left(pq\right)^{1/4}\frac{z_{i}^{\left(1\right)}}{z_{j}^{\left(2\right)}}v^{-1}_{\scalebox{0.6}{$F_1$}}
v_{\scalebox{0.6}{$F_2$}}^{-1}v_{\scalebox{0.6}{$B$}};p,q\right) 
\nn\\
&&\times\Gamma\left(\left(pq\right)^{1/4}\frac{z_{j}^{\left(2\right)}}{z_{i}^{\left(1\right)}}v_{\scalebox{0.6}{$F_1$}}
v_{\scalebox{0.6}{$F_2$}}^{-1}v^{-1}_{\scalebox{0.6}{$B$}};p,q\right)
\Gamma\left(\left(pq\right)^{1/4}\frac{z_{j}^{\left(2\right)}}{z_{i}^{\left(1\right)}}v^{-1}_{\scalebox{0.6}{$F_1$}}
v_{\scalebox{0.6}{$F_2$}}v^{-1}_{\scalebox{0.6}{$B$}};p,q\right)\,,
\label{index:T11:def}
\ee
where we turned on fugacities $v_{\scalebox{0.6}{$F_{1,2}$}}$ for the flavor symmetries $\U(1)_{1,2}$, and
$v_{\scalebox{0.6}{$B$}}$ for the baryonic symmetry $\U(1)_{B}$. As usual, we introduce the corresponding chemical potentials according to 
(\ref{chem:exp}),
\be
v_{\scalebox{0.6}{$F_{i}$}}\equiv e^{2\pi i \xi_{F_i}}\,,\qquad v_{\scalebox{0.6}{$B$}}\equiv e^{2\pi i \xi_{B}}\,;
\label{chem:pot:T11}
\ee
The expression for $\kappa_G$ is given by (\ref{kappa:expression}) 
\begin{equation}
\kappa_{G}=\frac{\left(p;p\right)_{\infty}^{2(N-1)}\left(q;q\right)_{\infty}^{2(N-1)}}{\left( N! \right)^2}\,.
\end{equation}
Throughout this paper we focus on the case where $p=q$ and use the chemical potential $\tau$ defined in (\ref{chem:exp}).
From expression (\ref{index:T11:def}), it is obvious that the following parameterization of the chemical potentials would be convenient to work with: 
\be
\Delta_{1}&=&\xi_{F_{1}}-\xi_{F_{2}}-\xi_{B}+\frac{1}{2}\tau \,,\nn\\
\Delta_{2}&=&-\xi_{F_{1}}+\xi_{F_{2}}-\xi_{B}+\frac{1}{2}\tau \,,\nn\\
\Delta_{3}&=&\xi_{F_{1}}+\xi_{F_{2}}+\xi_{B}+\frac{1}{2}\tau \,,\nn\\
\Delta_{4}&=&-\xi_{F_{1}}-\xi_{F_{2}}+\xi_{B}+\frac{1}{2}\tau\,.
\label{chem:pot:param:T11}
\ee 
In particular, with this parametrization the BAEs (\ref{BAE:general}),(\ref{BAE:operator:toric}) for the conifold case may then be written in the following compact form
\be
Q_{l}^{\left(1\right)}=\Lambda^{(1)}
\prod_{i=1}^{N}\frac{\theta_{0}\left(u_{j}^{\left(2\right)}-u_{l}^{\left(1\right)}+\Delta_{1};\tau\right)\theta_{0}
\left(u_{j}^{\left(2\right)}-u_{l}^{\left(1\right)}+\Delta_{2};\tau\right)}{\theta_{0}\left(u_{l}^{\left(1\right)}-
u_{j}^{\left(2\right)}+\Delta_{3};\tau\right)\theta_{0}\left(u_{l}^{\left(1\right)}-u_{j}^{\left(2\right)}+\Delta_{4};\tau\right)}
=1
\nn\\
Q_{l}^{\left(2\right)}=\Lambda^{(2)}
\prod_{j=1}^{N}\frac{\theta_{0}\left(u_{j}^{\left(1\right)}-u_{l}^{\left(2\right)}+\Delta_{3};\tau\right)\theta_{0}\left(u_{j}^{\left(1\right)}
-u_{l}^{\left(2\right)}+\Delta_{4};\tau\right)}{\theta_{0}\left(u_{l}^{\left(2\right)}-u_{j}^{\left(1\right)}+\Delta_{1};\tau\right)\theta_{0}
\left(u_{l}^{\left(2\right)}-u_{j}^{\left(1\right)}+\Delta_{2};\tau\right)}
=1\,,
\label{conifoldBAEs}
\ee
Notice that upon substituting the functions $B(u)$ given in (\ref{P:function}) into the BAE operators (\ref{BAE:operator:toric}) we will get certain terms depending on $\Delta$'s but independent of the holonomies $u_i^{(1,2)}$ in the exponents. However, we ignored these terms in the expressions written above since they can always be adsorbed into a redefinition of the Lagrange multipliers $\Lambda^{(i)}\equiv e^{2\pi i\lambda^{(i)}}$.

It is noteworthy that the parametrization (\ref{chem:pot:param:T11}) can be obtained in two natural steps. First, one should perform a mixing of the flavor and baryonic symmetries. This leads to the new basis of $\U(1)_{1,2,3}$ global symmetries with the corresponding chemical potentials $\tilde{\xi}_{1,2,3}$ turned on for each of them. The charges of the conifold theory matter fields in this new basis are summarized in the table below.
\begin{center}
	\begin{tabular}{|c||c|c|c|c|}
		\hline
		Field & $\U(1)_{R}$ & $\U\left(1\right)_1$ & $\U\left(1\right)_{2}$ & $\U\left(1\right)_{3}$\\
		\hline 
		$A_{1}$ & $1/2$ & $0$ & $0$ & $1$\\
		
		$A_{2}$ & $1/2$ & $-1$ & $-1$ & $-1$\\
		\hline
		
		$B_{1}$ & $1/2$ & $0$ & $1$ & $0$\\
		
		$B_{2}$ & $1/2$ & $1$ & $0$ & $0$\\
		\hline
	\end{tabular}
\end{center}
Actually, the proper choice of mixed $\U(1)$-symmetries can be seen directly from the geometric data of the toric manifold. To this end, one should apply algorithms defining the matter content charges from the toric diagrams as proposed in \cite{Butti:2005vn}. The same parametrization of the global symmetries has been used in the calculation of the superconformal indices of toric theories at the high-temperature limit in \cite{Amariti:2019mgp}. 
In Section \ref{Oth} we will be considering many examples of toric theories, and we will always be using a choice of global symmetries that is determined by the corresponding toric data. The relevant algorithm will be briefly reviewed in that section. 

Finally, after the flavor symmetry mixing we should perform the shift (\ref{delta:def})
\be
\Delta_a=\tilde{\xi}_a+r_a \tau\,,
\ee
leading exactly to expressions (\ref{chem:pot:param:T11}). It will often prove important to remember in future derivations that the chemical potentials $\Delta_a$ are not independent, rather they satisfy the constraint
\be
\sum\limits_{a=1}^4\Delta_a=2\tau\,.
\label{Delta:constraint:T11}
\ee

Let us now show how the basic solution (\ref{BAE:sol:u:gen}) satisfies the BAEs (\ref{conifoldBAEs}). In particular, let us substitute the basic solution into the first BAE and apply a chain of relations similar to the one used in (\ref{theta:simplification}). We first wish to concentrate on the transformation 
of $\theta$-functions:
\be
\prod\limits_{j=1}^N\frac{\thf{\frac{\tau}{N}(l-j)+\Delta_{1}}{\tau}\thf{\frac{\tau}{N}(l-j)+\Delta_{2}}{\tau}}
{\thf{\frac{\tau}{N}(j-l)+\Delta_{3}}{\tau}\thf{\frac{\tau}{N}(j-l)+\Delta_{4}}{\tau}}=
\nn\\
G\left( \{\Delta\} \right)
\frac{\prod\limits_{k=-1}^{l-N}\left(q^\frac{2k}{N}e^{2\pi i\left( \Delta_1+\Delta_2 \right)}\right)}
{\prod\limits_{k=-1}^{1-l}\left(q^\frac{2k}{N}e^{2\pi i\left( \Delta_3+\Delta_4 \right)}\right)}=G\left( \{\Delta\} \right) e^{2\pi i(N-1)(\tau-\Delta_1-\Delta_2)}\,,
\label{theta:transform:T11}
\ee
where for the notational brevity we introduced 
\be
G\left( \{\Delta\} \right)\equiv\prod\limits_{k=0}^{N-1}\frac{\thf{\frac{\tau}{N}k+\Delta_{1}}{\tau}\thf{\frac{\tau}{N}k+\Delta_{2}}{\tau}}
{\thf{\frac{\tau}{N}k+\Delta_{3}}{\tau}\thf{\frac{\tau}{N}k+\Delta_{4}}{\tau}}=\frac{\thf{\Delta_1}{\frac{\tau}{N}}\thf{\Delta_1}{\frac{\tau}{N}}}
{\thf{\Delta_3}{\frac{\tau}{N}}\thf{\Delta_4}{\frac{\tau}{N}}}
\label{G:function}
\ee
Using (\ref{theta:transform:T11}) in the first BAE (\ref{conifoldBAEs}), and similar relations in the second, it turns out that all $l$-dependences drop out and we are left simply with
\be
Q_l^{(1)}&=&\Lambda^{(1)}G(\{\Delta\})e^{2\pi i (N-1)\left( \tau-\Delta_1-\Delta_2 \right)}=1\,,\nn\\ 
Q_l^{(2)}&=&\Lambda^{(2)}G^{-1}(\{\Delta\})e^{-2\pi i (N-1)\left( \tau-\Delta_1-\Delta_2 \right)}=1\,.
\label{T11:lagrange:solution}
\ee
The equations above merely define the values of the Lagrange multipliers $\Lambda^{(1,2)}$; these are irrelevant for the large-$N$ contribution of the basic solution, as discussed in Section \ref{Gen}. However, it is an interesting observation that the values of the Lagrange multipliers obtained from (\ref{T11:lagrange:solution}) precisely reproduce the results of \cite{Hosseini:2016cyf} at the high-temperature limit $\tau\to 0$.\footnote{To see how the results match, one should carefully 
take into account constant shifts in the BAE (\ref{conifoldBAEs}) coming from the $B(u)$ functions (\ref{P:function}).} 

After showing how the basic solution (\ref{BAE:sol:u:gen}) solves the BAEs (\ref{conifoldBAEs}) of the conifold theory, we are now ready to directly evaluate the index (\ref{index:T11:def}) using our BAE formula (\ref{index:BAE:gen}). In principle, calculations of all possible contributions were performed in Section \ref{Gen}. Nevertheless, to illustrate our estimates for the contribution of the Jacobian (\ref{jacobian:general}) to the index, let us take a look at the particular explicit form the Jacobian assumes for the conifold theory case, that is 
\begin{equation}
	H=\det\left(\begin{array}{cc}
	J_{1} & \mathcal{O}\left(1\right)_{N\times N}\\
	\mathcal{O}\left(1\right)_{N\times N} & J_{2}
	\end{array}\right)\,.
\end{equation}
Here, each of the diagonal blocks $J_i$ has the form 
	\begin{equation}
		J_{i}=\left(\begin{array}{ccccc}
		A_1^{(i)} & \mathcal{O}\left(1\right) & \cdots & \mathcal{O}\left(1\right) & 1\\
		\mathcal{O}\left(1\right) & A_2^{(i)} &  & \vdots & 1\\
		\vdots &  & \ddots & \vdots & \vdots\\
		\mathcal{O}\left(1\right) & \cdots & \cdots & A_{N-1}^{(i)} & 1\\
		 -A_{N}^{(i)} & -A_N^{(i)} & \cdots &  -A_N^{(i)} & 1
		\end{array}\right)\,.
	\end{equation}
where the elements $A_j^{(i)}$ are all of order $N$ and are given in (\ref{jacobian:matrix:elements}). Notice that $J_i$ coincides with the Jacobian appearing in the calculations for the case of $\CN=4$ SYM. It can be clearly seen from the expression above that indeed, as expected, the logarithm of the Jacobian $\log H$ contributes at order $N\log N$ at most. 

Now using the results of Section \ref{Gen}, we can directly obtain expression (\ref{index:BAE:gen:final}) with the function $\Theta$ given by
\be
\Theta\left(\{\Delta\};\tau\right)&=&\frac{2}{3\tau}\left(\tau-\frac{1}{2}\right)\left(\tau-1\right)+
   \nn\\
   &&\hspace{1cm}\sum\limits_{i=1}^4\frac{1}{3\tau^{2}}\left(\left[\Delta_{i}\right]_{\tau}-\tau\right)\left(\left[\Delta_{i}\right]_{\tau}-\tau+
   \frac{1}{2}\right)\left(\left[\Delta_{i}\right]_{\tau}-\tau+1\right)\,,
   \label{Theta:T11:def}
\ee
where the function $\Dbr{}{\tau}$ is defined in (\ref{square:brackets}). However, this form may be drastically simplified if one  remembers that our chemical potentials $\Delta_a$ satisfy the constraint (\ref{Delta:constraint:T11}), leading to
\be
\Dbr{4}{\tau}=2\tau-1-\left[ \Delta_1+\Delta_2+\Delta_3 \right]_\tau\,.
\label{T11:chem:pot:constraint:2}
\ee
To simplify the function $\Theta$ written above, we then need to understand which of the stripes shown in Fig. \ref{stripe:pic} does the sum $\Dbr{1}{\tau}+\Dbr{2}{\tau}+\Dbr{3}{\tau}$ belong to. In particular, there are three possibilities
\be
	\textrm{Case I}&:\,\,\,\,\,\,\,\,\,\,\,\textrm{Im}\left(-\frac{1}{\tau}\right)>\textrm{Im}\left(\frac{\left[\Delta_{1}\right]_{\tau}+\left[\Delta_{2}\right]_{\tau}+\left[\Delta_{3}\right]_{\tau}}{\tau}\right)>0\,,
	\nn\\
	\textrm{Case II}&:\,\,\,\,\,\,\,\,\,\,\,\textrm{Im}\left(-\frac{2}{\tau}\right)>\textrm{Im}\left(\frac{\left[\Delta_{1}\right]_{\tau}+\left[\Delta_{2}\right]_{\tau}+\left[\Delta_{3}\right]_{\tau}}{\tau}\right)>\textrm{Im}\left(-\frac{1}{\tau}\right)\,,\\
	\textrm{Case III}&:\,\,\,\,\,\,\,\,\,\,\,\textrm{Im}\left(-\frac{3}{\tau}\right)>\textrm{Im}\left(\frac{\left[\Delta_{1}\right]_{\tau}+\left[\Delta_{2}\right]_{\tau}+\left[\Delta_{3}\right]_{\tau}}{\tau}\right)>\textrm{Im}\left(-\frac{2}{\tau}\right)\,.
\label{T11:cases}
\ee
These three cases assign different expressions relating $\left[\Delta_{1}+\Delta_{2}+\Delta_{3}\right]_{\tau}$ to the sum $\left[\Delta_{1}\right]_{\tau}+\left[\Delta_{2}\right]_{\tau}
+\left[\Delta_{3}\right]_{\tau}$. The latter expression should acquire an extra integer real shift bringing it to the fundamental domain in the complex $\Delta$-plane, {\it i.e.} to  $\textrm{Im}\left(-\frac{1}{\tau}\right)>\textrm{Im}\left(\frac{\Delta}{\tau}\right)>0$ represented by the green strip in Fig. \ref{stripe:pic}. The three different  relations are   
\begin{equation*}
\left[\Delta_{1}+\Delta_{2}+\Delta_{3}\right]_{\tau}=\left\{ \begin{array}{ll}
\left[\Delta_{1}\right]_{\tau}+\left[\Delta_{2}\right]_{\tau}+\left[\Delta_{3}\right]_{\tau}\,\,\,\,\,\,\,\,\,\,\,\,\,\,&\textrm{Case I}\\
\left[\Delta_{1}\right]_{\tau}+\left[\Delta_{2}\right]_{\tau}+\left[\Delta_{3}\right]_{\tau}+1\,\,\,\,\,\,\,\,\,\,\,\,\,\,&\textrm{Case II}\\
\left[\Delta_{1}\right]_{\tau}+\left[\Delta_{2}\right]_{\tau}+\left[\Delta_{3}\right]_{\tau}+2\,\,\,\,\,\,\,\,\,\,\,\,\,\,&\textrm{Case III}
\end{array}\right.\,,
\end{equation*}
Upon substituting these expressions back into the chemical potentials constraint (\ref{T11:chem:pot:constraint:2}) and into (\ref{Theta:T11:def}), we obtain the final expression for the $\Theta$-function:
\be
\label{T11Theta}
\Theta\left(\Delta_{i};\tau\right)=\left\{ \begin{array}{ll}
	F\left(\Dbr{i}{\tau};\tau\right)\,\,\,\,\,\,\,\,\,\,\,\,\,\,&\textrm{Case I}\\
	F\left(\Dbr{i}{\tau}+\frac{1}{2};\tau\right)-1+\frac{1}{2\tau}\,\,\,\,\,\,\,\,\,\,\,\,\,\,&\textrm{Case II}\\
        F\left(\Dbr{i}{\tau}+1;\tau\right)-2\,\,\,\,\,\,\,\,\,\,\,\,\,\,&\textrm{Case III}
\end{array}\right.
\ee
where $\Dbr{4}{\tau}$ is defined in (\ref{T11:chem:pot:constraint:2}) and the function $F\left( \Dbr{ }{\tau};\tau\right)$ is defined as follows:
\begin{equation}
	F\left(\Dbr{i}{\tau};\tau\right)=\tau^{-2}\left(\left[\Delta_{1}\right]_{\tau}\left[\Delta_{2}\right]_{\tau}\left[\Delta_{3}\right]_{\tau}+
	\left[\Delta_{1}\right]_{\tau}\left[\Delta_{2}\right]_{\tau}\left[\Delta_{4}\right]_{\tau}+\left[\Delta_{1}\right]_{\tau}\left[\Delta_{3}\right]_{\tau}
\left[\Delta_{4}\right]_{\tau}+\left[\Delta_{2}\right]_{\tau}\left[\Delta_{3}\right]_{\tau}\left[\Delta_{4}\right]_{\tau}\right) \,.
\label{T11:f:function}
\end{equation}
Expressions (\ref{T11Theta}) together with (\ref{T11:f:function}) and (\ref{index:BAE:gen:final}) completely define the leading term in the superconformal index of the conifold theory. To find the final expression for the index, it is still necessary to perform an $r$-extremization as prescribed by (\ref{index:BAE:gen:final}). However, it is quite a technically involved task due to the Stokes phenomenon and at the moment we postpone this question to the future. 
In Section \ref{Extrem} we will consider the extremization procedure for the particular case of equal chemical potentials $\Delta_a$. 
Notice that in case I our large-$N$ index behavior (\ref{T11Theta}) is in line with the Cardy-like limit computations (\ref{antonio:result})  performed in \cite{Amariti:2019mgp}. 

Finally, as we have mentioned in the beginning of this section the conifold theory is a particular example of an infinite $Y^{pq}$ class of toric theories with the specification $p=1,\, q=0$. As we will see further in Section \ref{sec:ypq}, the general result (\ref{Theta:Ypq:final}) for the large-$N$ index of $Y^{pq}$ theories indeed reduces to the $T^{1,1}$ result (\ref{T11Theta}) upon this identification.

\

\section{Other toric models}
\label{Oth}

In this section we extend our analysis to more toric theories and find the large-$N$ asymptotic forms of their indices using the general prescription of section \ref{Gen}, thereby obtaining the entropy functions of the dual black holes. We consider the models discussed in \cite{Amariti:2019mgp}, and for most of them (where it is not cumbersome) present the expressions for the entropy function in all the different domains, as in the previous section. 

Before going into particular examples of toric quiver theories, let us consider some of their general properties which will be of particular use for us in this section. Toric quiver gauge theories arise as the world-volume theories of D3 branes probing the tip of a toric cone over various five-dimensional Sasaki-Einstein manifolds. In the near-horizon limit, these constructions provide us with the holographic dual description of the corresponding gauge theories. A nice and useful property of the toric models is that all their data can be connected to the geometry of the cones \cite{Hanany:2005ve,Franco:2005rj}. In particular, all this data can be encoded in the toric diagrams of the corresponding cones, which are convex polytopes $P$ built from vectors of the form $V_i=(\cdot,\cdot,1)$. We summarize below the algorithm of \cite{Butti:2005vn} used to extract the assignment of $U(1)$ charges for the chiral fields, as well as other gauge theory data. To illustrate this algorithm, we show all the steps in the example of the conifold theory whose toric diagram is shown in Fig. \ref{T11:toric:pic} and the corresponding vectors $V_I$ are given in \eqref{T11:vectors}. 

As a first step, and without any extra constructions, we can extract the total number of gauge group nodes $F$ from the area of the polytope $P$:
\be
F=2\mathrm{Area}\left( P \right)\,.
\label{toric:gauge:number}
\ee
Another important fact is that the total number of global symmetries (including $R$-symmetry) $d$ is equal to the number of external points in the diagram. In the case of the $T^{1,1}$ diagram, we can see that the total number of gauge group nodes is $2\times 1$ and the total number of global symmetries is $4$, as we expect from the gauge theory content. 

Next, we continue with defining two-component vectors that connect nodes of the toric diagram: $v_i\equiv V_{i+1}-V_i$, and assign numbers for each of them as follows: $a_i=\{a_i^{R},a_i^I \}$ where $I=1,\dots,(d-1)$. In the case of the conifold theory, these vectors are given by 
\be
v_1=(1,0)\,;\quad v_2=(-1,1)\,;\quad v_3=(-1,0)\,;\quad v_4=(1,-1)\,;
\label{T11:v:vectors}
\ee
To each pair of vectors $(ij)=\left(v_i,v_j\right)$ such that $v_j$ can be rotated into $v_i$ clockwise with less than a $180^{\circ}$ turn, we associate a chiral field with a multiplicity of $\left|\det \left(v_i^{T} v_j^{T}\right)\right|$ and charges $a_{i+1}+a_{i+2}+\dots+a_j$. In the conifold example, this gives four pairs in total:
\begin{center}
	\begin{tabular}{|c||c|c|c|c|}
		\hline
		pair             & $(12)$ & $(23)$ & $(34)$ & $(41)$ \\
		\hline 
		
		multiplicity & 1      & 1      & 1      & 1 \\
		\hline		
		charge       & $a_2$  & $a_3$  & $a_4$  & $a_1$ \\
		\hline
		field        & $B_1$  & $A_1$ &  $A_2$  & $B_2$  \\
		\hline
	\end{tabular}
\end{center}
In the last line of the table we identify a specific bifundamental multiplet of the conifold theory for each of the
pairs. The algorithm for this identification requires an introduction for dimer constructions \cite{Hanany:2005ve}
and we omit it here for simplicity\footnote{Notice that for an index computation the precise identification of pairs with particular fields of the quiver theory is not even required.}  

Now, in order to fix the charges of the fields we are free to make any choice of the numbers $a_i$ that satisfies the constraints:
\be
\sum_{i=1}^d a_i^R=2\,,\qquad \sum_{i=1}^d a_i^I=0.
\ee
For example, in the case of the conifold theory we can choose the following set of numbers:
\be
a_i^R&=&\frac{1}{2}\,,\qquad i=1,\dots,4\,;\\
a_i^I&=& \left\{\begin{array}{lll}
	1,~ ~  ~ ~\mathrm{if}~ ~      & i=I \quad~ & i,I=1,2,3\,;\\
	0, ~ ~ ~ ~\mathrm{if}~ ~     & i\neq I   & i,I=1,2,3\,;\\
	-1, ~ ~ \mathrm{if}~ ~    & i=4       & I=1,2,3\,;
\end{array}\right.
\ee
Then, $a_i^R$ will set the assignment of $R$-charges according to the table of pairs and $a_i^I$ will assign the charges under the $(d-1)$ global symmetries $\U(1)_I$. The choice presented above precisely reproduces the charges used in Section \ref{Coni}. 

The final ingredients we will need to extract from the toric data are the triangular anomalies of the $U(1)_I$ global symmetries $C_{IJK}$. They are simply given by the area of the triangles built from the vertices $\left( V_I,V_J,V_K \right)$ of the toric diagram \cite{Benvenuti:2006xg}. 

Now let us discuss the form of the large-$N$ index we expect to obtain. As we have mentioned in the introduction, the Cardy-like limit computations of \cite{Amariti:2019mgp} has led to the general result (\ref{antonio:result}) for the indices of the toric theories. This result, in turn, supported the conjectured form of the entropy function of multi-charged $\mathrm{AdS}_5$ black holes proposed in \cite{Hosseini:2017mds}. However, our computations are slightly different since we do not limit ourselves to any chamber of the chemical potentials. Still, we expect to obtain the same form for this expression, with $\Delta_a$ substituted by $\Dbr{a}{\tau}$, at least in some regions of the $\Delta$-space, similarly to the $\CN=4$ SYM case in \cite{Benini:2018ywd}. 
In terms of the function $\Theta$, defining the index according to (\ref{index:general:BAE:basic}) we expect the following contribution of the basic solution in one of the chambers:
\be
\Theta\left( \{\Delta_i\};\tau \right)=\frac{C_{IJK}\Dbr{I}{\tau} \Dbr{J}{\tau} \Dbr{K}{\tau} }{6\tau^2}\,.
\label{expected:answer}
\ee
It can be easily checked that in the case of the conifold theory, (\ref{T11Theta}) precisely reproduces this expression in case I. As we will see in this section, the same situation takes place for every toric theory we consider. 

\

\subsection{The family $Y^{pq}$}
\label{sec:ypq}
\begin{figure}[h]
        \centering
        \begin{minipage}{0.43\textwidth}
        \centering
        \includegraphics[width=.6\linewidth]{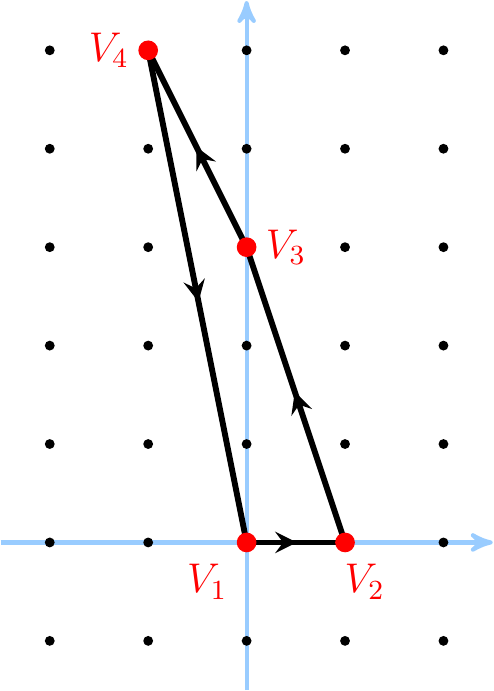} 
        \caption{Toric digram of $Y^{pq}$. We show $p=3,q=2$ on this picture. }
        \label{ypq:toric:pic}
        \end{minipage}\hspace{20mm}
        \begin{minipage}{0.43\textwidth}
        \centering
        \includegraphics[width=0.95\linewidth]{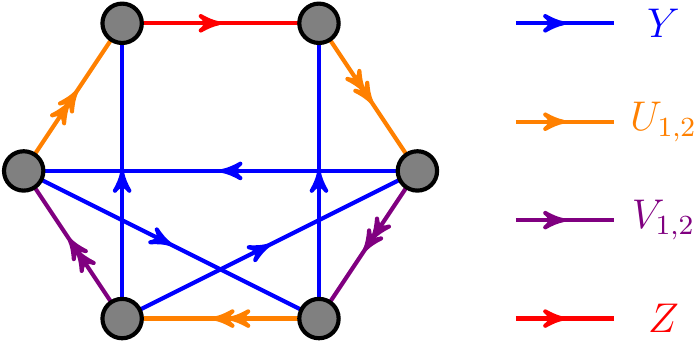} 
        \caption{Quiver diagram of the $Y^{32}$ theory.}
        \label{ypq:quiver}
        \end{minipage}
\end{figure}

Let us start our discussion with the infinite family of models $Y^{pq}$, given by quiver gauge
theories with $2p$ gauge groups and bifundamental chiral fields \cite{Benvenuti:2004dy}. In Fig. \ref{ypq:quiver} we show an example of the  $Y^{32}$ quiver gauge theory. The notation $Y_{pq}$ denotes a toric manifold with the corresponding diagram shown in Fig. \ref{ypq:toric:pic}, and is parametrized by the vectors:
\be
V_1=(0,0,1)\,,\quad V_2=(1,0,1)\,,\quad V_3=(0,p,1)\,,\quad V_4=(-1,p+q,1)\,. 
\ee

Using the algorithm summarized above, the charges of the fields under the three global $\U(1)$ symmetries 
and the $\U(1)_R$ $R$-symmetry can be identified from the form of the toric diagram shown on Fig. \ref{ypq:toric:pic} as follows,  
\begin{center}
	\begin{tabular}{|c||c|c|c|c|c|}
		\hline
		Field & Multiplicity & $U\left(1\right)_{1}$ & $U\left(1\right)_{2}$ & $U\left(1\right)_{3}$ & $U\left(1\right)_{R}$\\
		\hline 
		
		$Y$ & $p+q$ & 1 & 0 & 0 & $\frac{1}{2}$\\
		
		$U_{1}$ & $p$ & 0 & 1 & 0 & $\frac{1}{2}$\\
		
		$Z$ & $p-q$ & 0 & 0 & 1 & $\frac{1}{2}$\\
		
		$U_{2}$ & $p$ & -1 & -1 & -1 & $\frac{1}{2}$\\
		
		$V_{1}$ & $q$ & 0 & 1 & 1 & 1\\
		
		$V_{2}$ & $q$ & -1 & -1 & 0 & 1\\
		\hline
	\end{tabular}
\end{center}
As before, we turn on the chemical potentials $\xi_i$ for each of the three $\U(1)_i$ global symmetries and $\tau$ for the $R$ symmetry, and perform the shift (\ref{delta:def}): 
\be
\Delta_{i}&=&\xi_{i}+\frac{1}{2}\tau\,\,\,\,\,\,\left(i=1,2,3\right),
\nn\\
\Delta_{4}&=&2\tau-\sum_{i=1}^{3}\Delta_{i}=-\sum_{i=1}^{3}\xi_i+\frac{1}{2}\tau\,,
\label{DshiftY}
\ee
where we have also defined the usual auxiliary chemical potential $\Delta_4$.

Now we have everything we need in order to evaluate the superconformal index using the expression (\ref{index:BAE:gen:final}), with the function $\Theta\left( \{\Delta_i\};\tau \right)$ (\ref{index:general:BAE:basic}) given for the case of the $Y_{pq}$ matter content by the following lengthy expression:
\be
&\Theta\left(\{\Delta_{i}\};\tau\right)=2p\frac{1}{3\tau}\left(\tau-\frac{1}{2}\right)\left(\tau-1\right)+
\left(p+q\right)\frac{1}{3\tau^{2}}\left(\left[\Delta_{1}\right]_{\tau}-\tau\right)\left(\left[\Delta_{1}\right]_{\tau}-\tau+\frac{1}{2}\right)\times
\nn\\
&\left(\left[\Delta_{1}\right]_{\tau}-\tau+1\right)+
p\frac{1}{3\tau^{2}}\left(\left[\Delta_{2}\right]_{\tau}-\tau\right)\left(\left[\Delta_{2}\right]_{\tau}-\tau+\frac{1}{2}\right)\left(\left[\Delta_{2}\right]_{\tau}-\tau+1\right)+
\nn\\
&\left(p-q\right)\frac{1}{3\tau^{2}}\left(\left[\Delta_{3}\right]_{\tau}-\tau\right)\left(\left[\Delta_{3}\right]_{\tau}-\tau+\frac{1}{2}\right)\left(\left[\Delta_{3}\right]_{\tau}-\tau+1\right)+p\frac{1}{3\tau^{2}}\left(\left[\Delta_{4}\right]_{\tau}-\tau\right)\times
\nn\\
&\left(\left[\Delta_{4}\right]_{\tau}-\tau+\frac{1}{2}\right)\left(\left[\Delta_{4}\right]_{\tau}-\tau+1\right)+q\frac{1}{3\tau^{2}}\left(\left[\Delta_{2}+\Delta_{3}\right]_{\tau}-\tau\right)\left(\left[\Delta_{2}+\Delta_{3}\right]_{\tau}-\tau+\frac{1}{2}\right)\times
\nn\\
&\left(\left[\Delta_{2}+\Delta_{3}\right]_{\tau}-\tau+1\right)+q\frac{1}{3\tau^{2}}\left(\left[2\tau-\Delta_{1}-\Delta_{2}\right]_{\tau}-\tau\right)\left(\left[2\tau-\Delta_{1}-\Delta_{2}\right]_{\tau}-\tau+\frac{1}{2}\right)\times
\nn\\
&\left(\left[2\tau-\Delta_{1}-\Delta_{2}\right]_{\tau}-\tau+1\right)\,.
\label{Theta:Ypq:start}
\ee
Just as in the case of the conifold theory, this expression can be simplified and written in a compact form. In the expression above we have the bracket functions of various sums of chemical potentials, like $\left[\Delta_2+\Delta_3 \right]_\tau$ for example. In order to simplify them we need to understand how these bracket functions are expressed in terms of the bracket functions of the summands, like $\Dbr{2}{\tau}+\Dbr{3}{\tau}$ in the example of the previous sentence. To do it, we first distinguish between different cases similarly to the previous section, and summarize the results in the table below.
\begin{center}
	\begin{tabular}{|c||c|c|c|}
		\hline
		&  $\Dbr{1}{\tau}+\Dbr{2}{\tau}+\Dbr{3}{\tau}$   & $\Dbr{1}{\tau}+\Dbr{2}{\tau}$ & $\Dbr{2}{\tau}+\Dbr{3}{\tau}$\\
		\hline 
		
		case I & 1  & 1 & 1 \\
		
		case IIa & 2 & 1 & 1 \\
		
		case IIb & 2 & 2 & 1 \\
		
		case IIc & 2 & 1 & 2  \\
		
		case IId & 2 & 2 & 2 \\
		
		case III  & 3 & 2 & 2 \\
		\hline
	\end{tabular}
\end{center}
In this table we fill in the number of the strip in the complex plane to which the particular sum of $\Dbr{i}{\tau}$ belongs, with the $n$-th strip being defined as follows:
\be
\Im\left(- \frac{n}{\tau} \right)>\Im\left(\frac{\Delta}{\tau} \right)>\Im\left(- \frac{n-1}{\tau} \right)\,.
\label{nth:stripe}
\ee
More explicitly, the green stripe shown in Fig. \ref{stripe:pic} corresponds to $n=1$, while the strips to its left (right) correspond to $n=2,3,\dots$ ($n=0,-1,-2,\dots$). As a simple example that illustrates the notations used in this type of table, which we will also employ below when analyzing other theories, let us consider the first line of the table corresponding to case I. As we see, all the relevant sums of $\Dbr{i}{\tau}$ are in the first strip, hence we get that case I is given by:
\be
&\textrm{Case I}:\,\,\,\,\,\,\,\,\,\,\,\textrm{Im}\left(-\frac{1}{\tau}\right)>\textrm{Im}\left(\frac{\left[\Delta_{1}\right]_{\tau}+\left[\Delta_{2}\right]_{\tau}+\left[\Delta_{3}\right]_{\tau}}{\tau}\right)>0\,,
\nn\\
&\textrm{Im}\left(-\frac{1}{\tau}\right)>\textrm{Im}\left(\frac{\left[\Delta_{1}\right]_{\tau}+\left[\Delta_{2}\right]_{\tau}}{\tau}\right)>0\,,\,\,\,\,\textrm{Im}\left(-\frac{1}{\tau}\right)>\textrm{Im}\left(\frac{\left[\Delta_{2}\right]_{\tau}+\left[\Delta_{3}\right]_{\tau}}{\tau}\right)>0\,.
\ee
Next, after we distinguished between the various cases, we find the following relations for the bracket functions:
\begin{equation}
\left[\Delta_{1}+\Delta_{2}+\Delta_{3}\right]_{\tau}=\left\{ \begin{array}{ll}
\left[\Delta_{1}\right]_{\tau}+\left[\Delta_{2}\right]_{\tau}+\left[\Delta_{3}\right]_{\tau}\,\,\,\,\,\,\,\,\,\,\,\,\,\,&\textrm{Case I}\\
\left[\Delta_{1}\right]_{\tau}+\left[\Delta_{2}\right]_{\tau}+\left[\Delta_{3}\right]_{\tau}+1\,\,\,\,\,\,\,\,\,\,\,\,\,\,&\textrm{Cases IIa, IIb, IIc, IId}\\
\left[\Delta_{1}\right]_{\tau}+\left[\Delta_{2}\right]_{\tau}+\left[\Delta_{3}\right]_{\tau}+2\,\,\,\,\,\,\,\,\,\,\,\,\,\,&\textrm{Case III}
\end{array}\right.\,,
\label{Yrel1}
\end{equation}
\begin{equation}
\left[\Delta_{1}+\Delta_{2}\right]_{\tau}=\left\{ \begin{array}{ll}
\left[\Delta_{1}\right]_{\tau}+\left[\Delta_{2}\right]_{\tau}\,\,\,\,\,\,\,\,\,\,\,\,\,\,&\textrm{Cases I, IIa, IIc}\\
\left[\Delta_{1}\right]_{\tau}+\left[\Delta_{2}\right]_{\tau}+1\,\,\,\,\,\,\,\,\,\,\,\,\,\,&\textrm{Cases IIb, IId, III}
\end{array}\right.
\label{Yrel2}
\end{equation}
and 
\begin{equation}
\left[\Delta_{2}+\Delta_{3}\right]_{\tau}=\left\{ \begin{array}{ll}
\left[\Delta_{2}\right]_{\tau}+\left[\Delta_{3}\right]_{\tau}\,\,\,\,\,\,\,\,\,\,\,\,\,\,&\textrm{Cases I, IIa, IIb}\\
\left[\Delta_{2}\right]_{\tau}+\left[\Delta_{3}\right]_{\tau}+1\,\,\,\,\,\,\,\,\,\,\,\,\,\,&\textrm{Cases IIc, IId, III}
\end{array}\right.\:.
\label{Yrel3}
\end{equation}
Substituting these relations into (\ref{Theta:Ypq:start}), we get the final form of the function $\Theta\left(\Delta_{i};\tau\right)$ in the different cases:
\begin{equation}
\Theta\left(\Delta_{i};\tau\right)=\left\{ \begin{array}{c}
\Theta_{\textrm{I}}\left(\Delta_{i};\tau\right)\,\,\,\,\,\,\,\,\,\,\,\,\,\,\textrm{Case I}\\
\Theta_{\textrm{IIa}}\left(\Delta_{i};\tau\right)\,\,\,\,\,\,\,\,\,\,\,\,\,\,\textrm{Case IIa}\\
\Theta_{\textrm{IIb}}\left(\Delta_{i};\tau\right)\,\,\,\,\,\,\,\,\,\,\,\,\,\,\textrm{Case IIb}\\
\Theta_{\textrm{IIc}}\left(\Delta_{i};\tau\right)\,\,\,\,\,\,\,\,\,\,\,\,\,\,\textrm{Case IIc}\\
\Theta_{\textrm{IId}}\left(\Delta_{i};\tau\right)\,\,\,\,\,\,\,\,\,\,\,\,\,\,\textrm{Case IId}\\
\Theta_{\textrm{III}}\left(\Delta_{i};\tau\right)\,\,\,\,\,\,\,\,\,\,\,\,\,\,\textrm{Case III}
\end{array}\right.
\label{Theta:Ypq:final}
\end{equation}
where, defining for convenience the functions ($a=\left\{ 1,2,3,4\right\} $): 
\be
F_{p}\left(\left\{ x_{a}\right\} ,c,\tau\right)\equiv p\tau^{-2}\left(x_{1}x_{2}x_{3}+x_{1}x_{2}x_{4}+x_{1}x_{3}x_{4}+x_{2}x_{3}x_{4}+c\right),
\\
F_{q}\left(\left\{ x_{a}\right\} ,c_{1},c_{2},c_{3},c_{4},\tau\right)\equiv q\tau^{-2}\left[\left(x_{1}-x_{3}+c_{1}\right)\left(x_{2}+c_{2}\right)\left(x_{4}+c_{3}\right)+c_{4}\right],
\ee
$\Theta_{K}\left(\Delta_{i};\tau\right)$ are given by: 
\be
&\Theta_{\textrm{I}}\left(\Delta_{i};\tau\right)=F_{p}\left(\left\{ \left[\Delta_{a}\right]_{\tau}\right\} ,0,\tau\right)+F_{q}\left(\left\{ \left[\Delta_{a}\right]_{\tau}\right\} ,0,0,0,0,\tau\right),
\nn\\
&\Theta_{\textrm{IIa}}\left(\Delta_{i};\tau\right)=F_{p}\left(\left\{ \left[\Delta_{a}\right]_{\tau}+\frac{1}{2}\right\} ,\tau\left(\frac{1}{2}-\tau\right),\tau\right)+F_{q}\left(\left\{ \left[\Delta_{a}\right]_{\tau}\right\} ,0,0,1,0,\tau\right),
\nn\\
&\Theta_{\textrm{IIb}}\left(\Delta_{i};\tau\right)=F_{p}\left(\left\{ \left[\Delta_{a}\right]_{\tau}+\frac{1}{2}\right\} ,\tau\left(\frac{1}{2}-\tau\right),\tau\right)+F_{q}\left(\left\{ \left[\Delta_{a}\right]_{\tau}\right\} ,1,1,1,\left[\Delta_{3}\right]_{\tau}\left(2\tau-\left[\Delta_{3}\right]_{\tau}\right)-\tau^{2},\tau\right),
\nn\\
&\Theta_{\textrm{IIc}}\left(\Delta_{i};\tau\right)=F_{p}\left(\left\{ \left[\Delta_{a}\right]_{\tau}+\frac{1}{2}\right\} ,\tau\left(\frac{1}{2}-\tau\right),\tau\right)+F_{q}\left(\left\{ \left[\Delta_{a}\right]_{\tau}\right\} ,-1,1,1,-\left[\Delta_{1}\right]_{\tau}\left(2\tau-\left[\Delta_{1}\right]_{\tau}\right)+\tau^{2},\tau\right),
\nn\\
&\Theta_{\textrm{IId}}\left(\Delta_{i};\tau\right)=F_{p}\left(\left\{ \left[\Delta_{a}\right]_{\tau}+\frac{1}{2}\right\} ,\tau\left(\frac{1}{2}-\tau\right),\tau\right)+F_{q}\left(\left\{ \left[\Delta_{a}\right]_{\tau}\right\} ,0,1,0,0,\tau\right),
\nn\\
&\Theta_{\textrm{III}}\left(\Delta_{i};\tau\right)=F_{p}\left(\left\{ \left[\Delta_{a}\right]_{\tau}+1\right\} ,-2\tau^{2},\tau\right)+F_{q}\left(\left\{ \left[\Delta_{a}\right]_{\tau}\right\} ,0,1,1,0,\tau\right).
\ee
Notice that $\Theta_{\textrm{I}}\left(\Delta_{i};\tau\right)$, which can be written more explicitly as 
\be
\Theta_{\textrm{I}}\left(\Delta_{i};\tau\right)=\tau^{-2}\left(p\left[\Delta_{1}\right]_{\tau}\left[\Delta_{2}\right]_{\tau}\left[\Delta_{3}\right]_{\tau}+\left(p+q\right)\left[\Delta_{1}\right]_{\tau}\left[\Delta_{2}\right]_{\tau}\left[\Delta_{4}\right]_{\tau}+\right.
\nn\\
\left.+p\left[\Delta_{1}\right]_{\tau}\left[\Delta_{3}\right]_{\tau}\left[\Delta_{4}\right]_{\tau}+\left(p-q\right)\left[\Delta_{2}\right]_{\tau}\left[\Delta_{3}\right]_{\tau}\left[\Delta_{4}\right]_{\tau}\right),
\ee
has exactly the same structure as the expected result \eqref{expected:answer}. In addition to that,
it can be shown that the corresponding coefficients are exactly the triangular anomalies which can
be obtained from the toric diagram in Fig. \ref{ypq:toric:pic}. Note also that exactly the same result has been obtained in \cite{Amariti:2019mgp} using the Cardy limit of the superconformal index, and that by specifying to $p=1$, $q=0$ the expressions in (\ref{Theta:Ypq:final}) reduce to (\ref{Theta:T11:def}) obtained in the conifold case.

\

\subsection{The family $L^{pqr}$ $\left(p\neq r\right)$}

We next turn to the family of models $L^{pqr}$, built from quiver gauge theories with $p+q$ gauge groups and bifundamental chiral fields \cite{Benvenuti:2005ja,Butti:2005sw,Franco:2005sm}. The notation $L^{pqr}$ denotes a toric manifold with the corresponding diagram parametrized by the vectors:
\be
V_{1}=(0,0,1)\,,\quad V_{2}=(1,0,1)\,,\quad V_{3}=(P,s,1)\,,\quad V_{4}=(-k,q,1)\,,
\ee
where we have $Pq=r-ks$, $s=p+q-r$ and we assume that $p\neq r$. Using our previous discussion, the charges of the fields under the global symmetries can be read from the toric diagram, and are given in the table below. 
\begin{center}
	\begin{tabular}{|c||c|c|c|c|c|}
		\hline
		Field & Multiplicity & $U\left(1\right)_{1}$ & $U\left(1\right)_{2}$ & $U\left(1\right)_{3}$ & $U\left(1\right)_{R}$\\
		\hline 
		
		$Y$ & $q$ & 1 & 0 & 0 & $\frac{1}{2}$\\
		
		$W_{2}$ & $s$ & 0 & 1 & 0 & $\frac{1}{2}$\\
		
		$Z$ & $p$ & 0 & 0 & 1 & $\frac{1}{2}$\\
		
		$X_{2}$ & $r$ & -1 & -1 & -1 & $\frac{1}{2}$\\
		
		$W_{1}$ & $q-s$ & 0 & 1 & 1 & 1\\
		
		$X_{1}$ & $q-r$ & -1 & -1 & 0 & 1\\
		\hline
	\end{tabular}
\end{center}
As in \eqref{DshiftY}, we turn on the chemical potentials $\xi_i$ and $\tau$ for the $\U(1)_i$ and $\U(1)_R$ symmetries and perform the shift (\ref{delta:def}): 
\be
\Delta_{i}&=&\xi_{i}+\frac{1}{2}\tau\,\,\,\,\,\,\left(i=1,2,3\right),
\nn\\
\Delta_{4}&=&2\tau-\sum_{i=1}^{3}\Delta_{i}=-\sum_{i=1}^{3}\xi_i+\frac{1}{2}\tau\,,
\ee
where we have defined $\Delta_4$ as before. Then, the index is given at large $N$ by \eqref{index:BAE:gen:final}, where 
\be
&\Theta\left(\Delta_{i};\tau\right)=\left(p+q\right)\frac{1}{3\tau}\left(\tau-\frac{1}{2}\right)\left(\tau-1\right)+q\frac{1}{3\tau^{2}}\left(\left[\Delta_{1}\right]_{\tau}-\tau\right)\left(\left[\Delta_{1}\right]_{\tau}-\tau+\frac{1}{2}\right)\left(\left[\Delta_{1}\right]_{\tau}-\tau+1\right)+
\nn\\
&s\frac{1}{3\tau^{2}}\left(\left[\Delta_{2}\right]_{\tau}-\tau\right)\left(\left[\Delta_{2}\right]_{\tau}-\tau+\frac{1}{2}\right)\left(\left[\Delta_{2}\right]_{\tau}-\tau+1\right)+p\frac{1}{3\tau^{2}}\left(\left[\Delta_{3}\right]_{\tau}-\tau\right)\left(\left[\Delta_{3}\right]_{\tau}-\tau+\frac{1}{2}\right)\left(\left[\Delta_{3}\right]_{\tau}-\tau+1\right)+
\nn\\
&r\frac{1}{3\tau^{2}}\left(\left[\Delta_{4}\right]_{\tau}-\tau\right)\left(\left[\Delta_{4}\right]_{\tau}-\tau+\frac{1}{2}\right)\left(\left[\Delta_{4}\right]_{\tau}-\tau+1\right)+
\nn\\
&\left(q-s\right)\frac{1}{3\tau^{2}}\left(\left[\Delta_{2}+\Delta_{3}\right]_{\tau}-\tau\right)\left(\left[\Delta_{2}+\Delta_{3}\right]_{\tau}-\tau+\frac{1}{2}\right)\left(\left[\Delta_{2}+\Delta_{3}\right]_{\tau}-\tau+1\right)+
\nn\\
&\left(q-r\right)\frac{1}{3\tau^{2}}\left(\left[2\tau-\Delta_{1}-\Delta_{2}\right]_{\tau}-\tau\right)\left(\left[2\tau-\Delta_{1}-\Delta_{2}\right]_{\tau}-\tau+\frac{1}{2}\right)\left(\left[2\tau-\Delta_{1}-\Delta_{2}\right]_{\tau}-\tau+1\right).
\label{thL}
\ee
To simplify \eqref{thL} as we did above, we notice that the different cases here are exactly the same as for the $Y^{pq}$ family. Therefore, we can use the same classification along with the relations \eqref{Yrel1}, \eqref{Yrel2} and \eqref{Yrel3}, and get
\begin{equation*}
\Theta\left(\Delta_{i};\tau\right)=\left\{ \begin{array}{c}
\Theta_{\textrm{I}}\left(\Delta_{i};\tau\right)\,\,\,\,\,\,\,\,\,\,\,\,\,\,\textrm{Case I}\\
\Theta_{\textrm{IIa}}\left(\Delta_{i};\tau\right)\,\,\,\,\,\,\,\,\,\,\,\,\,\,\textrm{Case IIa}\\
\Theta_{\textrm{IIb}}\left(\Delta_{i};\tau\right)\,\,\,\,\,\,\,\,\,\,\,\,\,\,\textrm{Case IIb}\\
\Theta_{\textrm{IIc}}\left(\Delta_{i};\tau\right)\,\,\,\,\,\,\,\,\,\,\,\,\,\,\textrm{Case IIc}\\
\Theta_{\textrm{IId}}\left(\Delta_{i};\tau\right)\,\,\,\,\,\,\,\,\,\,\,\,\,\,\textrm{Case IId}\\
\Theta_{\textrm{III}}\left(\Delta_{i};\tau\right)\,\,\,\,\,\,\,\,\,\,\,\,\,\,\textrm{Case III}
\end{array}\right.
\end{equation*}
where
\be
&\Theta_{\textrm{I}}\left(\Delta_{i};\tau\right)=\tau^{-2}\left(s\left[\Delta_{1}\right]_{\tau}\left[\Delta_{2}\right]_{\tau}\left[\Delta_{3}\right]_{\tau}+q\left[\Delta_{1}\right]_{\tau}\left[\Delta_{2}\right]_{\tau}\left[\Delta_{4}\right]_{\tau}+r\left[\Delta_{1}\right]_{\tau}\left[\Delta_{3}\right]_{\tau}\left[\Delta_{4}\right]_{\tau}+p\left[\Delta_{2}\right]_{\tau}\left[\Delta_{3}\right]_{\tau}\left[\Delta_{4}\right]_{\tau}\right)\,,
\nn\\
&\Theta_{\textrm{IIa}}\left(\Delta_{i};\tau\right)=\tau^{-2}\left\{ s\left[\Delta_{1}\right]_{\tau}\left[\Delta_{2}\right]_{\tau}\left[\Delta_{3}\right]_{\tau}+q\left[\Delta_{1}\right]_{\tau}\left[\Delta_{2}\right]_{\tau}\left(\left[\Delta_{4}\right]_{\tau}+1\right)+r\left[\left(\left[\Delta_{1}\right]_{\tau}+1\right)\left(\left[\Delta_{3}\right]_{\tau}+1\right)+\left[\Delta_{2}\right]_{\tau}\right]\times\right.
\nn\\
&\left.\left(\left[\Delta_{4}\right]_{\tau}+1\right)+p\left[\Delta_{2}\right]_{\tau}\left[\Delta_{3}\right]_{\tau}\left(\left[\Delta_{4}\right]_{\tau}+1\right)\right\} -r\,,
\nn\\
&\Theta_{\textrm{IIb}}\left(\Delta_{i};\tau\right)=\tau^{-2}\left\{ \left(s-q\right)\left[\Delta_{1}\right]_{\tau}\left[\Delta_{2}\right]_{\tau}\left[\Delta_{3}\right]_{\tau}+q\left(\left[\Delta_{1}\right]_{\tau}+1\right)\left(\left[\Delta_{2}\right]_{\tau}+1\right)\left(\left[\Delta_{4}\right]_{\tau}+\left[\Delta_{3}\right]_{\tau}+1\right)+\right.
\nn\\
&\left.r\left[\Delta_{3}\right]_{\tau}\left[\left(\left[\Delta_{1}\right]_{\tau}+1\right)\left[\Delta_{4}\right]_{\tau}-\left[\Delta_{2}\right]_{\tau}\right]+p\left[\Delta_{2}\right]_{\tau}\left[\Delta_{3}\right]_{\tau}\left(\left[\Delta_{4}\right]_{\tau}+1\right)\right\} -q\,,
\nn\\
&\Theta_{\textrm{IIc}}\left(\Delta_{i};\tau\right)=\tau^{-2}\left\{ -r\left[\Delta_{1}\right]_{\tau}\left[\Delta_{2}\right]_{\tau}\left[\Delta_{3}\right]_{\tau}+q\left[\Delta_{1}\right]_{\tau}\left[\Delta_{2}\right]_{\tau}\left(\left[\Delta_{4}\right]_{\tau}+\left[\Delta_{3}\right]_{\tau}+1\right)+r\left[\Delta_{1}\right]_{\tau}\left[\left(\left[\Delta_{3}\right]_{\tau}+1\right)\times\right.\right.
\nn\\
&\left.\left.\left[\Delta_{4}\right]_{\tau}-\left[\Delta_{2}\right]_{\tau}\right]+p\left(\left[\Delta_{2}\right]_{\tau}+1\right)\left(\left[\Delta_{3}\right]_{\tau}+1\right)\left(\left[\Delta_{1}\right]_{\tau}+\left[\Delta_{4}\right]_{\tau}+1\right)\right\} -p\,,
\nn\\
&\Theta_{\textrm{IId}}\left(\Delta_{i};\tau\right)=\tau^{-2}\left\{ -r\left[\Delta_{1}\right]_{\tau}\left[\Delta_{2}\right]_{\tau}\left[\Delta_{3}\right]_{\tau}+\left[\left(q-r\right)\left(\left[\Delta_{1}\right]_{\tau}+1\right)\left(\left[\Delta_{2}\right]_{\tau}+1\right)+r\left[\Delta_{1}\right]_{\tau}\left[\Delta_{2}\right]_{\tau}\right]\left(\left[\Delta_{4}\right]_{\tau}+\right.\right.
\nn\\
&\left.\left.\left[\Delta_{3}\right]_{\tau}+1\right)+r\left[\Delta_{1}\right]_{\tau}\left[\left(\left[\Delta_{3}\right]_{\tau}+1\right)\left[\Delta_{4}\right]_{\tau}-\left[\Delta_{2}\right]_{\tau}\right]+p\left(\left[\Delta_{2}\right]_{\tau}+1\right)\left(\left[\Delta_{3}\right]_{\tau}+1\right)\left(\left[\Delta_{1}\right]_{\tau}+\left[\Delta_{4}\right]_{\tau}+1\right)\right\} -s\,,
\nn\\
&\Theta_{\textrm{III}}\left(\Delta_{i};\tau\right)=\tau^{-2}\left\{ -r\left[\Delta_{1}\right]_{\tau}\left[\Delta_{2}\right]_{\tau}\left[\Delta_{3}\right]_{\tau}+\left[\left(q-r\right)\left(\left[\Delta_{1}\right]_{\tau}+1\right)\left(\left[\Delta_{2}\right]_{\tau}+1\right)+r\left[\Delta_{1}\right]_{\tau}\left[\Delta_{2}\right]_{\tau}\right]\left(\left[\Delta_{4}\right]_{\tau}+\right.\right.
\nn\\
&\left.\left[\Delta_{3}\right]_{\tau}+2\right)+r\left[\Delta_{1}\right]_{\tau}\left[\left(\left[\Delta_{3}\right]_{\tau}+1\right)\left(\left[\Delta_{4}\right]_{\tau}+1\right)-\left[\Delta_{2}\right]_{\tau}\right]+r\left(\left[\Delta_{1}\right]_{\tau}+\left[\Delta_{2}\right]_{\tau}+\left[\Delta_{3}\right]_{\tau}\right)\left(\left[\Delta_{4}\right]_{\tau}-1\right)+
\nn\\
&\left.p\left(\left[\Delta_{2}\right]_{\tau}+1\right)\left(\left[\Delta_{3}\right]_{\tau}+1\right)\left(\left[\Delta_{1}\right]_{\tau}+\left[\Delta_{4}\right]_{\tau}+2\right)\right\} -s-r\left(1-2\tau^{-1}\right)^{2}\,.
\ee
We notice that here as well, $\Theta_{\textrm{I}}\left(\Delta_{i};\tau\right)$ (which corresponds to Case I) has the expected form \eqref{expected:answer} with the coefficients given by the triangular anomalies, and matches the result obtained in \cite{Amariti:2019mgp} using the Cardy limit of the index.

\

\subsection{The family $X^{pq}$}

This family is built from quiver gauge theories with $2p+1$ gauge groups and bifundamental chiral fields \cite{Hanany:2005hq}. The toric diagram of the manifold $X^{pq}$ is parametrized by the vectors:
\begin{equation*}
V_{1}=(1-q,1,1)\,,\,\,V_{2}=(-1,0,1)\,,\,\,V_{3}=(q-p,-1,1)\,,\,\,V_{4}=(0,-1,1)\,,\,\,V_{5}=(p,1,1)\,,
\end{equation*}
and as discussed above, the corresponding charges of the fields under the global symmetries are as follows,  
\begin{center}
	\begin{tabular}{|c||c|c|c|c|c|c|}
		\hline
		Field & Multiplicity & $U\left(1\right)_{1}$ & $U\left(1\right)_{2}$ & $U\left(1\right)_{3}$ & $U\left(1\right)_{4}$ & $U\left(1\right)_{R}$\\
		\hline 
		
		$A_{1}$ & $p+q-1$ & 1 & 0 & 0 & 0 & $\frac{2}{5}$\\
		
		$A_{2}$ & 1 & 0 & 1 & 0 & 0 & $\frac{2}{5}$\\
		
		$A_{3}$ & 1 & 0 & 0 & 1 & 0 & $\frac{2}{5}$\\
		
		$A_{4}$ & $p-q$ & 0 & 0 & 0 & 1 & $\frac{2}{5}$\\
		
		$A_{5}$ & $p$ & -1 & -1 & -1 & -1 & $\frac{2}{5}$\\
		
		$A_{12}$ & 1 & 1 & 1 & 0 & 0 & $\frac{4}{5}$\\
		
		$A_{23}$ & $p-1$ & 0 & 1 & 1 & 0 & $\frac{4}{5}$\\
		
		$A_{34}$ & 1 & 0 & 0 & 1 & 1 & $\frac{4}{5}$\\
		
		$A_{45}$ & $q$ & -1 & -1 & -1 & 0 & $\frac{4}{5}$\\
		
		$A_{234}$ & $q-1$ & 0 & 1 & 1 & 1 & $\frac{6}{5}$\\
		\hline
	\end{tabular}
\end{center}
We turn on the chemical potentials $\xi_i$ and $\tau$ for the $\U(1)_i$ and $\U(1)_R$ symmetries and perform the shift \eqref{delta:def}: 
\be
\Delta_{i}&=&\xi_{i}+\frac{2}{5}\tau\,\,\,\,\,\,\left(i=1,\dots,4\right),
\nn\\
\Delta_{5}&=&2\tau-\sum_{i=1}^{4}\Delta_{i}=-\sum_{i=1}^{4}\xi_i+\frac{2}{5}\tau\,,
\ee
where we have defined the auxiliary $\Delta_5$ in the usual way. Then, the index is given at large $N$ by \eqref{index:BAE:gen:final}, where 
\be
&\Theta\left(\Delta_{i};\tau\right)=\left(2p+1\right)\frac{1}{3\tau}\left(\tau-\frac{1}{2}\right)\left(\tau-1\right)+\left(p+q-1\right)\frac{1}{3\tau^{2}}\left(\left[\Delta_{1}\right]_{\tau}-\tau\right)\left(\left[\Delta_{1}\right]_{\tau}-\tau+\frac{1}{2}\right)\times
\nn\\
&\left(\left[\Delta_{1}\right]_{\tau}-\tau+1\right)+\frac{1}{3\tau^{2}}\left(\left[\Delta_{2}\right]_{\tau}-\tau\right)\left(\left[\Delta_{2}\right]_{\tau}-\tau+\frac{1}{2}\right)\left(\left[\Delta_{2}\right]_{\tau}-\tau+1\right)+\frac{1}{3\tau^{2}}\left(\left[\Delta_{3}\right]_{\tau}-\tau\right)\times
\nn\\
&\left(\left[\Delta_{3}\right]_{\tau}-\tau+\frac{1}{2}\right)\left(\left[\Delta_{3}\right]_{\tau}-\tau+1\right)+\left(p-q\right)\frac{1}{3\tau^{2}}\left(\left[\Delta_{4}\right]_{\tau}-\tau\right)\left(\left[\Delta_{4}\right]_{\tau}-\tau+\frac{1}{2}\right)\left(\left[\Delta_{4}\right]_{\tau}-\tau+1\right)+
\nn\\
&p\frac{1}{3\tau^{2}}\left(\left[\Delta_{5}\right]_{\tau}-\tau\right)\left(\left[\Delta_{5}\right]_{\tau}-\tau+\frac{1}{2}\right)\left(\left[\Delta_{5}\right]_{\tau}-\tau+1\right)+\frac{1}{3\tau^{2}}\left(\left[\Delta_{1}+\Delta_{2}\right]_{\tau}-\tau\right)\left(\left[\Delta_{1}+\Delta_{2}\right]_{\tau}-\tau+\frac{1}{2}\right)\times
\nn\\
&\left(\left[\Delta_{1}+\Delta_{2}\right]_{\tau}-\tau+1\right)+\left(p-1\right)\frac{1}{3\tau^{2}}\left(\left[\Delta_{2}+\Delta_{3}\right]_{\tau}-\tau\right)\left(\left[\Delta_{2}+\Delta_{3}\right]_{\tau}-\tau+\frac{1}{2}\right)\left(\left[\Delta_{2}+\Delta_{3}\right]_{\tau}-\tau+1\right)+
\nn\\
&\frac{1}{3\tau^{2}}\left(\left[\Delta_{3}+\Delta_{4}\right]_{\tau}-\tau\right)\left(\left[\Delta_{3}+\Delta_{4}\right]_{\tau}-\tau+\frac{1}{2}\right)\left(\left[\Delta_{3}+\Delta_{4}\right]_{\tau}-\tau+1\right)+q\frac{1}{3\tau^{2}}\left(\left[\Delta_{4}+\Delta_{5}\right]_{\tau}-\tau\right)\times
\nn\\
&\left(\left[\Delta_{4}+\Delta_{5}\right]_{\tau}-\tau+\frac{1}{2}\right)\left(\left[\Delta_{4}+\Delta_{5}\right]_{\tau}-\tau+1\right)+\left(q-1\right)\frac{1}{3\tau^{2}}\left(\left[\Delta_{2}+\Delta_{3}+\Delta_{4}\right]_{\tau}-\tau\right)\times
\nn\\
&\left(\left[\Delta_{2}+\Delta_{3}+\Delta_{4}\right]_{\tau}-\tau+\frac{1}{2}\right)\left(\left[\Delta_{2}+\Delta_{3}+\Delta_{4}\right]_{\tau}-\tau+1\right)\,.
\ee
As before, to simplify this expression we need to consider the different domains of the bracket functions. However, for compactness we only present here the result for Case Ia, in which 
\begin{equation*}
\textrm{Case Ia}:\,\,\,\,\,\,\,\,\,\,\,\textrm{Im}\left(-\frac{1}{\tau}\right)>\textrm{Im}\left(\frac{1}{\tau}\sum_{i=1}^{4}\left[\Delta_{i}\right]_{\tau}\right)>0
\end{equation*}
(note that all the sub-sums are also in the same strip). For this case, we have
\be
&\Theta_{\textrm{Ia}}\left(\Delta_{i};\tau\right)=\tau^{-2}\left(\left[\Delta_{1}\right]_{\tau}\left[\Delta_{2}\right]_{\tau}\left[\Delta_{3}\right]_{\tau}+p\left[\Delta_{1}\right]_{\tau}\left[\Delta_{2}\right]_{\tau}\left[\Delta_{4}\right]_{\tau}+p\left[\Delta_{1}\right]_{\tau}\left[\Delta_{3}\right]_{\tau}\left[\Delta_{4}\right]_{\tau}+\left[\Delta_{2}\right]_{\tau}\left[\Delta_{3}\right]_{\tau}\left[\Delta_{4}\right]_{\tau}+\right.
\nn\\
&\left(p+q-1\right)\left[\Delta_{1}\right]_{\tau}\left[\Delta_{2}\right]_{\tau}\left[\Delta_{5}\right]_{\tau}+\left(p+q\right)\left[\Delta_{1}\right]_{\tau}\left[\Delta_{3}\right]_{\tau}\left[\Delta_{5}\right]_{\tau}+2\left[\Delta_{2}\right]_{\tau}\left[\Delta_{3}\right]_{\tau}\left[\Delta_{5}\right]_{\tau}+
\nn\\
&p\left[\Delta_{1}\right]_{\tau}\left[\Delta_{4}\right]_{\tau}\left[\Delta_{5}\right]_{\tau}+\left.\left(p-q+1\right)\left[\Delta_{2}\right]_{\tau}\left[\Delta_{4}\right]_{\tau}\left[\Delta_{5}\right]_{\tau}+\left(p-q\right)\left[\Delta_{3}\right]_{\tau}\left[\Delta_{4}\right]_{\tau}\left[\Delta_{5}\right]_{\tau}\right),
\ee
which is of the same form as \eqref{expected:answer} with the coefficients given by the triangular anomalies, and matches the result obtained in \cite{Amariti:2019mgp} using the Cardy limit of the index.

\

\subsection{The model SPP}

We consider the SPP (short for "suspended pinch point") model, the theory living on a stack of $N$ D3-branes probing the SPP conical singularity $x^2y=wz$. This gauge theory is described by the quiver given in Fig. \ref{SPPfig} below. 
\begin{figure}[h]
	\centering
	\includegraphics[scale=0.8]{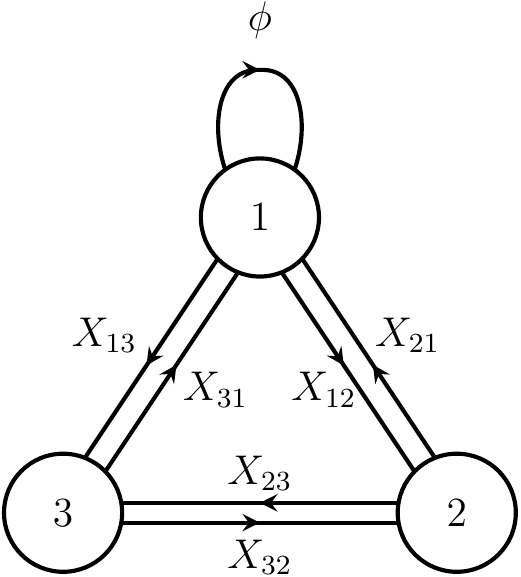}
	\caption{Quiver diagram corresponding to the SPP model. Each node represents an $SU(N)$ gauge group and each $X_{ij}$ field transforms in the bifundamental representation, {\it i.e.} in the $\boldsymbol{N}$ of the $i$-th node and in the $\boldsymbol{\overline{N}}$ of the $j$-th node. The field $\phi$ transforms in the adjoint ($\boldsymbol{N}^2-1$) representation.
	}
	\label{SPPfig}
\end{figure}

The toric manifold in this case is parametrized by the vectors:
\begin{equation*}
V_{1}=(1,1,1)\,,\,\,V_{2}=(1,0,1)\,,\,\,V_{3}=(0,0,1)\,,\,\,V_{4}=(2,0,1)\,,\,\,V_{5}=(1,0,1)\,,
\end{equation*}
and the corresponding charges of the fields under the global symmetries are given in the following table:  
\begin{center}
	\begin{tabular}{|c||c|c|c|c|c|}
		\hline
		Field & $U\left(1\right)_{1}$ & $U\left(1\right)_{2}$ & $U\left(1\right)_{3}$ & $U\left(1\right)_{4}$ & $U\left(1\right)_{R}$\\
		\hline  
		$X_{32}$ & 1 & 0 & 0 & 0 & $\frac{2}{5}$\\
		
		$X_{23}$ & 0 & 1 & 0 & 0 & $\frac{2}{5}$\\
		
		$X_{13}$ & 0 & 0 & 1 & 0 & $\frac{2}{5}$\\
		
		$X_{12}$ & 0 & 0 & 0 & 1 & $\frac{2}{5}$\\
		
		$X_{31}$ & -1 & -1 & -1 & 0 & $\frac{4}{5}$\\
		
		$X_{21}$ & -1 & -1 & 0 & -1 & $\frac{4}{5}$\\
		
		$\phi$ & 1 & 1 & 0 & 0 & $\frac{4}{5}$\\
		\hline
	\end{tabular}
\end{center}
We turn on the chemical potentials $\xi_i$ and $\tau$ for the $\U(1)_i$ and $\U(1)_R$ symmetries and perform the shift \eqref{delta:def}: 
\be
\Delta_{i}&=&\xi_{i}+\frac{2}{5}\tau\,\,\,\,\,\,\left(i=1,\dots,4\right),
\nn\\
\Delta_{5}&=&2\tau-\sum_{i=1}^{4}\Delta_{i}=-\sum_{i=1}^{4}\xi_i+\frac{2}{5}\tau\,,
\ee
where we have defined the auxiliary $\Delta_5$ in the usual way. Then, the index is given at large $N$ by \eqref{index:BAE:gen:final}, where 
\be
&\Theta\left(\Delta_{i};\tau\right)=3\frac{1}{3\tau}\left(\tau-\frac{1}{2}\right)\left(\tau-1\right)+\frac{1}{3\tau^{2}}\left(\left[\Delta_{1}\right]_{\tau}-\tau\right)\left(\left[\Delta_{1}\right]_{\tau}-\tau+\frac{1}{2}\right)\left(\left[\Delta_{1}\right]_{\tau}-\tau+1\right)+
\nn\\
&\frac{1}{3\tau^{2}}\left(\left[\Delta_{2}\right]_{\tau}-\tau\right)\left(\left[\Delta_{2}\right]_{\tau}-\tau+\frac{1}{2}\right)\left(\left[\Delta_{2}\right]_{\tau}-\tau+1\right)+\frac{1}{3\tau^{2}}\left(\left[\Delta_{3}\right]_{\tau}-\tau\right)\left(\left[\Delta_{3}\right]_{\tau}-\tau+\frac{1}{2}\right)\times
\nn\\
&\left(\left[\Delta_{3}\right]_{\tau}-\tau+1\right)+\frac{1}{3\tau^{2}}\left(\left[\Delta_{4}\right]_{\tau}-\tau\right)\left(\left[\Delta_{4}\right]_{\tau}-\tau+\frac{1}{2}\right)\left(\left[\Delta_{4}\right]_{\tau}-\tau+1\right)+
\nn\\
&\frac{1}{3\tau^{2}}\left(\left[\Delta_{5}+\Delta_{4}\right]_{\tau}-\tau\right)\left(\left[\Delta_{5}+\Delta_{4}\right]_{\tau}-\tau+\frac{1}{2}\right)\left(\left[\Delta_{5}+\Delta_{4}\right]_{\tau}-\tau+1\right)+
\nn\\
&\frac{1}{3\tau^{2}}\left(\left[\Delta_{5}+\Delta_{3}\right]_{\tau}-\tau\right)\left(\left[\Delta_{5}+\Delta_{3}\right]_{\tau}-\tau+\frac{1}{2}\right)\left(\left[\Delta_{5}+\Delta_{3}\right]_{\tau}-\tau+1\right)+
\nn\\
&\frac{1}{3\tau^{2}}\left(\left[\Delta_{1}+\Delta_{2}\right]_{\tau}-\tau\right)\left(\left[\Delta_{1}+\Delta_{2}\right]_{\tau}-\tau+\frac{1}{2}\right)\left(\left[\Delta_{1}+\Delta_{2}\right]_{\tau}-\tau+1\right).
\label{thSPP}
\ee
As usual, to simplify \eqref{thSPP} we need to consider the different domains of the bracket functions. Using the notations presented in subsection \ref{sec:ypq}, we define the different cases as follows:
\begin{center}
	\begin{tabular}{|c||c|c|c|}
		\hline
		& $\Dbr{1}{\tau}+\Dbr{2}{\tau}+\Dbr{3}{\tau}$ & $\Dbr{1}{\tau}+\Dbr{2}{\tau}$ & $\Dbr{1}{\tau}+\Dbr{2}{\tau}+\Dbr{4}{\tau}$\\
		\hline 
		
		case Ia & 1  & 1 & 1 \\
		
		case Ib & 1  & 1 & 2 \\
		
		case IIa & 2 & 1 & 1 \\
		
		case IIb & 2 & 1 & 2 \\
		
		case IIc & 2 & 2 & 2  \\
		
		case IId & 2 & 2 & 3 \\
		
		case IIIa  & 3 & 2 & 2 \\
		
		case IIIb  & 3 & 2 & 3 \\
		\hline
	\end{tabular}
\end{center}
Similarly to subsection \ref{sec:ypq}, we find the following relations for the bracket functions:
\begin{equation*}
S_{123}\equiv\left[\Delta_{1}+\Delta_{2}+\Delta_{3}\right]_{\tau}=\left\{ \begin{array}{c}
\left[\Delta_{1}\right]_{\tau}+\left[\Delta_{2}\right]_{\tau}+\left[\Delta_{3}\right]_{\tau}\,\,\,\,\,\,\,\,\,\,\,\,\,\,\textrm{Cases Ia, Ib}\\
\left[\Delta_{1}\right]_{\tau}+\left[\Delta_{2}\right]_{\tau}+\left[\Delta_{3}\right]_{\tau}+1\,\,\,\,\,\,\,\,\,\,\,\,\,\,\textrm{Cases IIa, IIb, IIc, IId}\\
\left[\Delta_{1}\right]_{\tau}+\left[\Delta_{2}\right]_{\tau}+\left[\Delta_{3}\right]_{\tau}+2\,\,\,\,\,\,\,\,\,\,\,\,\,\,\textrm{Cases IIIa, IIIb}
\end{array}\right.\,,
\end{equation*}
\begin{equation*}
S_{12}\equiv\left[\Delta_{1}+\Delta_{2}\right]_{\tau}=\left\{ \begin{array}{c}
\left[\Delta_{1}\right]_{\tau}+\left[\Delta_{2}\right]_{\tau}\,\,\,\,\,\,\,\,\,\,\,\,\,\,\textrm{Cases Ia, Ib, IIa, IIb}\\
\left[\Delta_{1}\right]_{\tau}+\left[\Delta_{2}\right]_{\tau}+1\,\,\,\,\,\,\,\,\,\,\,\,\,\,\textrm{Cases IIc, IId, IIIa, IIIb}
\end{array}\right.
\end{equation*}
and 
\begin{equation*}
S_{124}\equiv\left[\Delta_{1}+\Delta_{2}+\Delta_{4}\right]_{\tau}=\left\{ \begin{array}{c}
\left[\Delta_{1}\right]_{\tau}+\left[\Delta_{2}\right]_{\tau}+\left[\Delta_{4}\right]_{\tau}\,\,\,\,\,\,\,\,\,\,\,\,\,\,\textrm{Cases Ia, IIa}\\
\left[\Delta_{1}\right]_{\tau}+\left[\Delta_{2}\right]_{\tau}+\left[\Delta_{4}\right]_{\tau}+1\,\,\,\,\,\,\,\,\,\,\,\,\,\,\textrm{Cases Ib, IIb, IIc, IIIa}\\
\left[\Delta_{1}\right]_{\tau}+\left[\Delta_{2}\right]_{\tau}+\left[\Delta_{4}\right]_{\tau}+2\,\,\,\,\,\,\,\,\,\,\,\,\,\,\textrm{Cases IId, IIIb}
\end{array}\right.\,.
\end{equation*}
Substituting these relations into \eqref{thSPP}, we get the expression for $\Theta\left(\Delta_{i};\tau\right)$ in the different cases:
\begin{equation*}
\Theta\left(\Delta_{i};\tau\right)=\left\{ \begin{array}{c}
\Theta_{\textrm{Ia}}\left(\Delta_{i};\tau\right)\,\,\,\,\,\,\,\,\,\,\,\,\,\,\textrm{Case Ia}\\
\Theta_{\textrm{Ib}}\left(\Delta_{i};\tau\right)\,\,\,\,\,\,\,\,\,\,\,\,\,\,\textrm{Case Ib}\\
\Theta_{\textrm{IIa}}\left(\Delta_{i};\tau\right)\,\,\,\,\,\,\,\,\,\,\,\,\,\,\textrm{Case IIa}\\
\Theta_{\textrm{IIb}}\left(\Delta_{i};\tau\right)\,\,\,\,\,\,\,\,\,\,\,\,\,\,\textrm{Case IIb}\\
\Theta_{\textrm{IIc}}\left(\Delta_{i};\tau\right)\,\,\,\,\,\,\,\,\,\,\,\,\,\,\textrm{Case IIc}\\
\Theta_{\textrm{IId}}\left(\Delta_{i};\tau\right)\,\,\,\,\,\,\,\,\,\,\,\,\,\,\textrm{Case IId}\\
\Theta_{\textrm{IIIa}}\left(\Delta_{i};\tau\right)\,\,\,\,\,\,\,\,\,\,\,\,\,\,\textrm{Case IIIa}\\
\Theta_{\textrm{IIIb}}\left(\Delta_{i};\tau\right)\,\,\,\,\,\,\,\,\,\,\,\,\,\,\textrm{Case IIIb}
\end{array}\right.
\end{equation*}
where
\be
&\Theta_{\textrm{Ia}}\left(\Delta_{i};\tau\right)=\tau^{-2}\left\{ \left[\Delta_{1}\right]_{\tau}\left[\Delta_{2}\right]_{\tau}\left[\Delta_{4}\right]_{\tau}+\left[\Delta_{3}\right]_{\tau}S_{12}\left(2\tau-S_{123}-1\right)+\right.
\nn\\
&\left.\left(\left[\Delta_{1}\right]_{\tau}\left[\Delta_{2}\right]_{\tau}+\left[\Delta_{4}\right]_{\tau}S_{12}\right)\left(2\tau-S_{124}-1\right)\right\},
\nn\\
&\Theta_{\textrm{Ib}}\left(\Delta_{i};\tau\right)=\tau^{-2}\left\{ \left[\Delta_{1}\right]_{\tau}\left[\Delta_{2}\right]_{\tau}\left[\Delta_{4}\right]_{\tau}+\left[\Delta_{3}\right]_{\tau}S_{12}\left(2\tau-S_{123}-1\right)+\left(\left[\Delta_{1}\right]_{\tau}\left[\Delta_{2}\right]_{\tau}+\left[\Delta_{4}\right]_{\tau}S_{12}\right)\times\right.
\nn\\
&\left(2\tau-S_{124}-2\right)+\left[\Delta_{1}\right]_{\tau}\left(2\tau-\left[\Delta_{1}\right]_{\tau}-2\right)+\left[\Delta_{2}\right]_{\tau}\left(2\tau-\left[\Delta_{2}\right]_{\tau}-2\right)+
\nn\\
&\left.\left[\Delta_{4}\right]_{\tau}\left(2\tau-\left[\Delta_{4}\right]_{\tau}-2\right)\right\} -\left(1-\tau^{-1}\right)^{2}\,,
\nn\\
&\Theta_{\textrm{IIa}}\left(\Delta_{i};\tau\right)=\tau^{-2}\left\{ \left[\Delta_{1}\right]_{\tau}\left[\Delta_{2}\right]_{\tau}\left(\left[\Delta_{4}\right]_{\tau}-1\right)+\left[\Delta_{3}\right]_{\tau}\left(S_{12}+1\right)\left(2\tau-S_{123}-1\right)+\right.
\nn\\
&\left(\left[\Delta_{1}\right]_{\tau}\left[\Delta_{2}\right]_{\tau}+\left[\Delta_{4}\right]_{\tau}\left(S_{12}-1\right)\right)\left(2\tau-S_{124}-2\right)+
\nn\\
&\left.\left[\Delta_{1}\right]_{\tau}\left(2\tau-\left[\Delta_{1}\right]_{\tau}-2\right)+\left[\Delta_{2}\right]_{\tau}\left(2\tau-\left[\Delta_{2}\right]_{\tau}-2\right)+\left[\Delta_{4}\right]_{\tau}\left(2\tau-\left[\Delta_{4}\right]_{\tau}-2\right)\right\} -\left(1-\tau^{-1}\right)^{2}\,,
\nn\\
&\Theta_{\textrm{IIb}}\left(\Delta_{i};\tau\right)=\tau^{-2}\left\{ \left[\Delta_{1}\right]_{\tau}\left[\Delta_{2}\right]_{\tau}\left(\left[\Delta_{4}\right]_{\tau}-2\right)+\left[\Delta_{3}\right]_{\tau}\left(S_{12}+1\right)\left(2\tau-S_{123}-1\right)+\right.
\nn\\
&\left(\left[\Delta_{1}\right]_{\tau}\left[\Delta_{2}\right]_{\tau}+\left[\Delta_{4}\right]_{\tau}S_{12}\right)\left(2\tau-S_{124}-2\right)+
\nn\\
&\left.2\left[\Delta_{1}\right]_{\tau}\left(2\tau-\left[\Delta_{1}\right]_{\tau}-2\right)+2\left[\Delta_{2}\right]_{\tau}\left(2\tau-\left[\Delta_{2}\right]_{\tau}-2\right)+\left[\Delta_{4}\right]_{\tau}\left(2\tau-\left[\Delta_{4}\right]_{\tau}-2\right)\right\} -2\left(1-\tau^{-1}\right)^{2}\,,
\nn\\
&\Theta_{\textrm{IIc}}\left(\Delta_{i};\tau\right)=\tau^{-2}\left\{ \left[\Delta_{1}\right]_{\tau}\left[\Delta_{2}\right]_{\tau}\left[\Delta_{4}\right]_{\tau}+\left[\Delta_{3}\right]_{\tau}S_{12}\left(2\tau-S_{123}-1\right)+\right.
\nn\\
&\left(\left[\Delta_{1}\right]_{\tau}\left[\Delta_{2}\right]_{\tau}+\left[\Delta_{4}\right]_{\tau}\left(S_{12}-1\right)\right)\left(2\tau-S_{124}-2\right)+
\nn\\
&\left.\left[\Delta_{1}\right]_{\tau}\left(2\tau-\left[\Delta_{1}\right]_{\tau}-2\right)+\left[\Delta_{2}\right]_{\tau}\left(2\tau-\left[\Delta_{2}\right]_{\tau}-2\right)+\left[\Delta_{4}\right]_{\tau}\left(2\tau-\left[\Delta_{4}\right]_{\tau}-2\right)\right\} -\left(1-\tau^{-1}\right)^{2}\,,
\nn\\
&\Theta_{\textrm{IId}}\left(\Delta_{i};\tau\right)=\tau^{-2}\left\{ \left[\Delta_{1}\right]_{\tau}\left[\Delta_{2}\right]_{\tau}\left(\left[\Delta_{4}\right]_{\tau}+1\right)+\left[\Delta_{3}\right]_{\tau}S_{12}\left(2\tau-S_{123}-1\right)+\right.
\nn\\
&\left(\left[\Delta_{1}\right]_{\tau}\left[\Delta_{2}\right]_{\tau}+\left[\Delta_{4}\right]_{\tau}S_{12}\right)\left(2\tau-S_{124}-2\right)+\left[\Delta_{1}\right]_{\tau}\left(2\tau-\left[\Delta_{1}\right]_{\tau}-2\right)+
\nn\\
&\left.\left[\Delta_{2}\right]_{\tau}\left(2\tau-\left[\Delta_{2}\right]_{\tau}-2\right)+\left[\Delta_{4}\right]_{\tau}\left(2\tau-\left[\Delta_{4}\right]_{\tau}-2\right)+\left(S_{12}-1\right)\left(2\tau-S_{12}-3\right)\right\} -\left(1-\tau^{-1}\right)^{2}-\left(1-2\tau^{-1}\right)^{2}\,,
\nn\\
&\Theta_{\textrm{IIIa}}\left(\Delta_{i};\tau\right)=\tau^{-2}\left\{ \left[\Delta_{1}\right]_{\tau}\left[\Delta_{2}\right]_{\tau}\left[\Delta_{4}\right]_{\tau}+\left[\Delta_{3}\right]_{\tau}\left(S_{12}+1\right)\left(2\tau-S_{123}-1\right)+\right.
\nn\\
&\left(\left[\Delta_{1}\right]_{\tau}\left[\Delta_{2}\right]_{\tau}+\left[\Delta_{4}\right]_{\tau}\left(S_{12}-1\right)\right)\left(2\tau-S_{124}-2\right)+\left[\Delta_{1}\right]_{\tau}\left(2\tau-\left[\Delta_{1}\right]_{\tau}-2\right)+
\nn\\
&\left.\left[\Delta_{2}\right]_{\tau}\left(2\tau-\left[\Delta_{2}\right]_{\tau}-2\right)+\left[\Delta_{4}\right]_{\tau}\left(2\tau-\left[\Delta_{4}\right]_{\tau}-2\right)+\left(S_{12}-1\right)\left(2\tau-S_{12}-3\right)\right\} -\left(1-\tau^{-1}\right)^{2}-\left(1-2\tau^{-1}\right)^{2}\,,
\nn\\
&\Theta_{\textrm{IIIb}}\left(\Delta_{i};\tau\right)=\tau^{-2}\left\{ \left[\Delta_{1}\right]_{\tau}\left[\Delta_{2}\right]_{\tau}\left(\left[\Delta_{4}\right]_{\tau}-1\right)+\left[\Delta_{3}\right]_{\tau}\left(S_{12}+1\right)\left(2\tau-S_{123}-1\right)+\right.
\nn\\
&\left(\left[\Delta_{1}\right]_{\tau}\left[\Delta_{2}\right]_{\tau}+\left[\Delta_{4}\right]_{\tau}S_{12}\right)\left(2\tau-S_{124}-2\right)+2\left[\Delta_{1}\right]_{\tau}\left(2\tau-\left[\Delta_{1}\right]_{\tau}-2\right)+2\left[\Delta_{2}\right]_{\tau}\left(2\tau-\left[\Delta_{2}\right]_{\tau}-2\right)+
\nn\\
&\left.\left[\Delta_{4}\right]_{\tau}\left(2\tau-\left[\Delta_{4}\right]_{\tau}-2\right)+\left(S_{12}-1\right)\left(2\tau-S_{12}-5\right)\right\} -9\tau^{-2}+10\tau^{-1}-3\,.
\ee
Here, $\Theta_{\textrm{Ia}}\left(\Delta_{i};\tau\right)$ (which corresponds to Case Ia) is of the same form as \eqref{expected:answer} with the coefficients given by the triangular anomalies, and matches the result obtained in \cite{Amariti:2019mgp} using the Cardy limit of the index.

\

\subsection{The models $\mathbb{F}_{0}$, $\textrm{dP}_{1}$ and $\textrm{dP}_{2}$}

The large-$N$ indices of these models can be easily obtained from those corresponding to some of the theories we considered above, and will therefore not be analyzed separately. In particular, the large-$N$ index of the model $\mathbb{F}_{0}$ is just twice that of the conifold theory, while the large-$N$ indices of the models $\textrm{dP}_{1}$ and $\textrm{dP}_{2}$ are the same as those of the theories $Y^{21}$ and $X^{21}$, respectively.

\

\subsection{The model $\textrm{dP}_{3}$}

We turn to the $\textrm{dP}_{3}$ model, the theory living on a stack of $N$ D3-branes probing the tip of the Calabi-Yau cone whose base is $\textrm{dP}_{3}$ ({\it i.e.} the third del Pezzo surface). This gauge theory is described by the quiver given in Fig. \ref{dP3fig} below.  
\begin{figure}
	[htbp ]
	\center\includegraphics[scale=0.8]{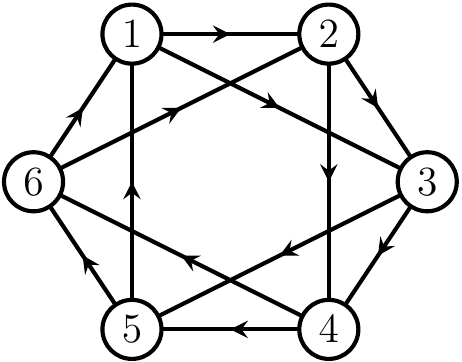}
	\caption{Quiver diagram corresponding to the $\textrm{dP}_{3}$ model. Each node represents an $SU(N)$ gauge group and each $X_{ij}$ field transforms in the bifundamental representation, {\it i.e.} in the $\boldsymbol{N}$ of the $i$-th node and in the $\boldsymbol{\overline{N}}$ of the $j$-th node. 
	}
	\label{dP3fig}
\end{figure}

The toric manifold in this case is parametrized by the vectors:
\begin{equation*}
V_{1}=(1,1,1)\,,\,\,V_{2}=(0,1,1)\,,\,\,V_{3}=(-1,0,1)\,,\,\,V_{4}=(-1,-1,1)\,,\,\,V_{5}=(0,-1,1)\,,\,\,V_{6}=(1,0,1)\,,
\end{equation*}
and the corresponding charges of the fields under the global symmetries are given in the following table: 
\begin{center}
	\begin{tabular}{|c||c|c|c|c|c|c|}
		\hline
		Field & $U\left(1\right)_{1}$ & $U\left(1\right)_{2}$ & $U\left(1\right)_{3}$ & $U\left(1\right)_{4}$ & $U\left(1\right)_{5}$ & $U\left(1\right)_{R}$\\
		\hline 
		
		$X_{61}$ & 1 & 0 & 0 & 0 & 0 & $\frac{1}{3}$\\
		
		$X_{56}$ & 0 & 1 & 0 & 0 & 0 & $\frac{1}{3}$\\
		
		$X_{45}$ & 0 & 0 & 1 & 0 & 0 & $\frac{1}{3}$\\
		
		$X_{34}$ & 0 & 0 & 0 & 1 & 0 & $\frac{1}{3}$\\
		
		$X_{23}$ & 0 & 0 & 0 & 0 & 1 & $\frac{1}{3}$\\
		
		$X_{12}$ & -1 & -1 & -1 & -1 & -1 & $\frac{1}{3}$\\
		
		$X_{24}$ & 1 & 1 & 0 & 0 & 0 & $\frac{2}{3}$\\
		
		$X_{13}$ & 0 & 1 & 1 & 0 & 0 & $\frac{2}{3}$\\
		
		$X_{62}$ & 0 & 0 & 1 & 1 & 0 & $\frac{2}{3}$\\
		
		$X_{51}$ & 0 & 0 & 0 & 1 & 1 & $\frac{2}{3}$\\
		
		$X_{46}$ & -1 & -1 & -1 & -1 & 0 & $\frac{2}{3}$\\
		
		$X_{35}$ & 0 & -1 & -1 & -1 & -1 & $\frac{2}{3}$\\
		\hline
	\end{tabular}
\end{center}
We turn on the chemical potentials $\xi_i$ and $\tau$ for the $\U(1)_i$ and $\U(1)_R$ symmetries and perform the shift \eqref{delta:def}: 
\be
\Delta_{i}&=&\xi_{i}+\frac{1}{3}\tau\,\,\,\,\,\,\left(i=1,\dots,5\right),
\nn\\
\Delta_{6}&=&2\tau-\sum_{i=1}^{5}\Delta_{i}=-\sum_{i=1}^{5}\xi_i+\frac{1}{3}\tau\,,
\ee
where we have defined the auxiliary $\Delta_6$ in the usual way. Then, the index is given at large $N$ by \eqref{index:BAE:gen:final}, where 
\be
&\Theta\left(\Delta_{i};\tau\right)=6\frac{1}{3\tau}\left(\tau-\frac{1}{2}\right)\left(\tau-1\right)+\frac{1}{3\tau^{2}}\left(\left[\Delta_{1}\right]_{\tau}-\tau\right)\left(\left[\Delta_{1}\right]_{\tau}-\tau+\frac{1}{2}\right)\left(\left[\Delta_{1}\right]_{\tau}-\tau+1\right)+
\nn\\
&\frac{1}{3\tau^{2}}\left(\left[\Delta_{2}\right]_{\tau}-\tau\right)\left(\left[\Delta_{2}\right]_{\tau}-\tau+\frac{1}{2}\right)\left(\left[\Delta_{2}\right]_{\tau}-\tau+1\right)+\frac{1}{3\tau^{2}}\left(\left[\Delta_{3}\right]_{\tau}-\tau\right)\left(\left[\Delta_{3}\right]_{\tau}-\tau+\frac{1}{2}\right)\left(\left[\Delta_{3}\right]_{\tau}-\tau+1\right)+
\nn\\
&\frac{1}{3\tau^{2}}\left(\left[\Delta_{4}\right]_{\tau}-\tau\right)\left(\left[\Delta_{4}\right]_{\tau}-\tau+\frac{1}{2}\right)\left(\left[\Delta_{4}\right]_{\tau}-\tau+1\right)+\frac{1}{3\tau^{2}}\left(\left[\Delta_{5}\right]_{\tau}-\tau\right)\left(\left[\Delta_{5}\right]_{\tau}-\tau+\frac{1}{2}\right)\left(\left[\Delta_{5}\right]_{\tau}-\tau+1\right)+
\nn\\
&\frac{1}{3\tau^{2}}\left(\left[\Delta_{6}\right]_{\tau}-\tau\right)\left(\left[\Delta_{6}\right]_{\tau}-\tau+\frac{1}{2}\right)\left(\left[\Delta_{6}\right]_{\tau}-\tau+1\right)+
\nn\\
&\frac{1}{3\tau^{2}}\left(\left[\Delta_{1}+\Delta_{2}\right]_{\tau}-\tau\right)\left(\left[\Delta_{1}+\Delta_{2}\right]_{\tau}-\tau+\frac{1}{2}\right)\left(\left[\Delta_{1}+\Delta_{2}\right]_{\tau}-\tau+1\right)+
\nn\\
&\frac{1}{3\tau^{2}}\left(\left[\Delta_{2}+\Delta_{3}\right]_{\tau}-\tau\right)\left(\left[\Delta_{2}+\Delta_{3}\right]_{\tau}-\tau+\frac{1}{2}\right)\left(\left[\Delta_{2}+\Delta_{3}\right]_{\tau}-\tau+1\right)+
\nn\\
&\frac{1}{3\tau^{2}}\left(\left[\Delta_{3}+\Delta_{4}\right]_{\tau}-\tau\right)\left(\left[\Delta_{3}+\Delta_{4}\right]_{\tau}-\tau+\frac{1}{2}\right)\left(\left[\Delta_{3}+\Delta_{4}\right]_{\tau}-\tau+1\right)+
\nn\\
&\frac{1}{3\tau^{2}}\left(\left[\Delta_{4}+\Delta_{5}\right]_{\tau}-\tau\right)\left(\left[\Delta_{4}+\Delta_{5}\right]_{\tau}-\tau+\frac{1}{2}\right)\left(\left[\Delta_{4}+\Delta_{5}\right]_{\tau}-\tau+1\right)+
\nn\\
&\frac{1}{3\tau^{2}}\left(\left[\Delta_{5}+\Delta_{6}\right]_{\tau}-\tau\right)\left(\left[\Delta_{5}+\Delta_{6}\right]_{\tau}-\tau+\frac{1}{2}\right)\left(\left[\Delta_{5}+\Delta_{6}\right]_{\tau}-\tau+1\right)+
\nn\\
&\frac{1}{3\tau^{2}}\left(\left[\Delta_{6}+\Delta_{1}\right]_{\tau}-\tau\right)\left(\left[\Delta_{6}+\Delta_{1}\right]_{\tau}-\tau+\frac{1}{2}\right)\left(\left[\Delta_{6}+\Delta_{1}\right]_{\tau}-\tau+1\right)\,.
\ee
To simplify this expression, we need to consider the different cases for the bracket functions. However, for compactness we only present here the result for Case Ia, in which 
\begin{equation*}
\textrm{Case Ia}:\,\,\,\,\,\,\,\,\,\,\,\textrm{Im}\left(-\frac{1}{\tau}\right)>\textrm{Im}\left(\frac{1}{\tau}\sum_{i=1}^{5}\left[\Delta_{i}\right]_{\tau}\right)>0
\end{equation*}
(note that all the sub-sums are also in the same strip). For this case, we have
\be
&\Theta\left(\Delta_{i};\tau\right)_{\textrm{Ia}}=\tau^{-2}\left(\left[\Delta_{1}\right]_{\tau}\left[\Delta_{2}\right]_{\tau}\left[\Delta_{3}\right]_{\tau}+2\left[\Delta_{1}\right]_{\tau}\left[\Delta_{2}\right]_{\tau}\left[\Delta_{4}\right]_{\tau}+2\left[\Delta_{1}\right]_{\tau}\left[\Delta_{3}\right]_{\tau}\left[\Delta_{4}\right]_{\tau}+\left[\Delta_{2}\right]_{\tau}\left[\Delta_{3}\right]_{\tau}\left[\Delta_{4}\right]_{\tau}+\right.
\nn\\
&2\left[\Delta_{1}\right]_{\tau}\left[\Delta_{2}\right]_{\tau}\left[\Delta_{5}\right]_{\tau}+3\left[\Delta_{1}\right]_{\tau}\left[\Delta_{3}\right]_{\tau}\left[\Delta_{5}\right]_{\tau}+2\left[\Delta_{2}\right]_{\tau}\left[\Delta_{3}\right]_{\tau}\left[\Delta_{5}\right]_{\tau}+2\left[\Delta_{1}\right]_{\tau}\left[\Delta_{4}\right]_{\tau}\left[\Delta_{5}\right]_{\tau}
\nn\\
&2\left[\Delta_{2}\right]_{\tau}\left[\Delta_{4}\right]_{\tau}\left[\Delta_{5}\right]_{\tau}+\left[\Delta_{3}\right]_{\tau}\left[\Delta_{4}\right]_{\tau}\left[\Delta_{5}\right]_{\tau}+\left[\Delta_{1}\right]_{\tau}\left[\Delta_{2}\right]_{\tau}\left[\Delta_{6}\right]_{\tau}+2\left[\Delta_{1}\right]_{\tau}\left[\Delta_{3}\right]_{\tau}\left[\Delta_{6}\right]_{\tau}+
\nn\\
&2\left[\Delta_{2}\right]_{\tau}\left[\Delta_{3}\right]_{\tau}\left[\Delta_{6}\right]_{\tau}+2\left[\Delta_{1}\right]_{\tau}\left[\Delta_{4}\right]_{\tau}\left[\Delta_{6}\right]_{\tau}+3\left[\Delta_{2}\right]_{\tau}\left[\Delta_{4}\right]_{\tau}\left[\Delta_{6}\right]_{\tau}+2\left[\Delta_{3}\right]_{\tau}\left[\Delta_{4}\right]_{\tau}\left[\Delta_{6}\right]_{\tau}+
\nn\\
&\left.\left[\Delta_{1}\right]_{\tau}\left[\Delta_{5}\right]_{\tau}\left[\Delta_{6}\right]_{\tau}+2\left[\Delta_{2}\right]_{\tau}\left[\Delta_{5}\right]_{\tau}\left[\Delta_{6}\right]_{\tau}+2\left[\Delta_{3}\right]_{\tau}\left[\Delta_{5}\right]_{\tau}\left[\Delta_{6}\right]_{\tau}+\left[\Delta_{4}\right]_{\tau}\left[\Delta_{5}\right]_{\tau}\left[\Delta_{6}\right]_{\tau}\right),
\ee
which is of the same form as \eqref{expected:answer} with the coefficients given by the triangular anomalies, and matches the result obtained in \cite{Amariti:2019mgp} using the Cardy limit of the index.

\

\subsection{The model $\textrm{(P)dP}_{4}$}

We consider the $\textrm{(P)dP}_{4}$ model, defined using the fourth del Pezzo surface. The quiver corresponding to this theory is given in Fig. \ref{dP4fig} below.
\begin{figure}
	[htbp ]
	\center\includegraphics[scale=0.8]{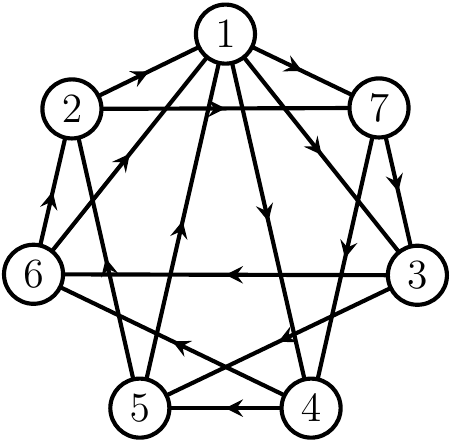}
	\caption{Quiver diagram corresponding to the $\textrm{(P)dP}_{4}$ model. Each node represents an $SU(N)$ gauge group and each $X_{ij}$ field transforms in the bifundamental representation, {\it i.e.} in the $\boldsymbol{N}$ of the $i$-th node and in the $\boldsymbol{\overline{N}}$ of the $j$-th node. 
	}
	\label{dP4fig}
\end{figure}

The toric manifold in this case is parametrized by the vectors:
\begin{equation*}
V_{1}=(0,0,1)\,,\,\,V_{2}=(1,0,1)\,,\,\,V_{3}=(2,0,1)\,,\,\,V_{4}=(2,1,1)\,,\,\,V_{5}=(1,2,1)\,,\,\,V_{6}=(0,2,1)\,,\,\,V_{7}=(0,1,1)\,,
\end{equation*}
and the corresponding charges of the fields under the global symmetries are given in the following table: 
\begin{center}
	\begin{tabular}{|c||c|c|c|c|c|c|c|}
		\hline
		Field & $U\left(1\right)_{1}$ & $U\left(1\right)_{2}$ & $U\left(1\right)_{3}$ & $U\left(1\right)_{4}$ & $U\left(1\right)_{5}$ & $U\left(1\right)_{6}$ & $U\left(1\right)_{R}$\\
		\hline 
		
		$X_{73}$ & 0 & 1 & 0 & 0 & 0 & 0 & $\frac{2}{7}$\\
		
		$X_{35}$ & 0 & 0 & 1 & 0 & 0 & 0 & $\frac{2}{7}$\\
		
		$X_{46}$ & 0 & 0 & 0 & 1 & 0 & 0 & $\frac{2}{7}$\\
		
		$X_{74}$ & 0 & 0 & 0 & 0 & 1 & 0 & $\frac{2}{7}$\\
		
		$X_{21}$ & -1 & -1 & -1 & -1 & -1 & -1 & $\frac{2}{7}$\\
		
		$X_{52}$ & 1 & 1 & 0 & 0 & 0 & 0 & $\frac{4}{7}$\\
		
		$X_{61}$ & 0 & 1 & 1 & 0 & 0 & 0 & $\frac{4}{7}$\\
		
		$X_{27}$ & 0 & 0 & 1 & 1 & 0 & 0 & $\frac{4}{7}$\\
		
		$X_{51}$ & 0 & 0 & 0 & 1 & 1 & 0 & $\frac{4}{7}$\\
		
		$X_{62}$ & 0 & 0 & 0 & 0 & 1 & 1 & $\frac{4}{7}$\\
		
		$X_{45}$ & -1 & -1 & -1 & -1 & -1 & 0 & $\frac{4}{7}$\\
		
		$X_{36}$ & 0 & -1 & -1 & -1 & -1 & -1 & $\frac{4}{7}$\\
		
		$X_{14}$ & 1 & 1 & 1 & 0 & 0 & 0 & $\frac{6}{7}$\\
		
		$X_{13}$ & 0 & 0 & 0 & 1 & 1 & 1 & $\frac{6}{7}$\\
		
		$X_{17}$ & 0 & -1 & -1 & -1 & -1 & 0 & $\frac{6}{7}$\\
		\hline
	\end{tabular}
\end{center}
We turn on the chemical potentials $\xi_i$ and $\tau$ for the $\U(1)_i$ and $\U(1)_R$ symmetries and perform the shift \eqref{delta:def}: 
\be
\Delta_{i}&=&\xi_{i}+\frac{2}{7}\tau\,\,\,\,\,\,\left(i=1,\dots,6\right),
\nn\\
\Delta_{7}&=&2\tau-\sum_{i=1}^{6}\Delta_{i}=-\sum_{i=1}^{6}\xi_i+\frac{2}{7}\tau\,,
\ee
where we have defined the auxiliary $\Delta_7$ in the usual way. Then, the index is given at large $N$ by \eqref{index:BAE:gen:final}, where 
\be
&\Theta\left(\Delta_{i};\tau\right)=7\frac{1}{3\tau}\left(\tau-\frac{1}{2}\right)\left(\tau-1\right)+\frac{1}{3\tau^{2}}\left(\left[\Delta_{2}\right]_{\tau}-\tau\right)\left(\left[\Delta_{2}\right]_{\tau}-\tau+\frac{1}{2}\right)\left(\left[\Delta_{2}\right]_{\tau}-\tau+1\right)+
\nn\\
&\frac{1}{3\tau^{2}}\left(\left[\Delta_{3}\right]_{\tau}-\tau\right)\left(\left[\Delta_{3}\right]_{\tau}-\tau+\frac{1}{2}\right)\left(\left[\Delta_{3}\right]_{\tau}-\tau+1\right)+\frac{1}{3\tau^{2}}\left(\left[\Delta_{4}\right]_{\tau}-\tau\right)\left(\left[\Delta_{4}\right]_{\tau}-\tau+\frac{1}{2}\right)\left(\left[\Delta_{4}\right]_{\tau}-\tau+1\right)+
\nn\\
&\frac{1}{3\tau^{2}}\left(\left[\Delta_{5}\right]_{\tau}-\tau\right)\left(\left[\Delta_{5}\right]_{\tau}-\tau+\frac{1}{2}\right)\left(\left[\Delta_{5}\right]_{\tau}-\tau+1\right)+\frac{1}{3\tau^{2}}\left(\left[\Delta_{7}\right]_{\tau}-\tau\right)\left(\left[\Delta_{7}\right]_{\tau}-\tau+\frac{1}{2}\right)\left(\left[\Delta_{7}\right]_{\tau}-\tau+1\right)+
\nn\\
&\frac{1}{3\tau^{2}}\left(\left[\Delta_{1}+\Delta_{2}\right]_{\tau}-\tau\right)\left(\left[\Delta_{1}+\Delta_{2}\right]_{\tau}-\tau+\frac{1}{2}\right)\left(\left[\Delta_{1}+\Delta_{2}\right]_{\tau}-\tau+1\right)+
\nn\\
&\frac{1}{3\tau^{2}}\left(\left[\Delta_{2}+\Delta_{3}\right]_{\tau}-\tau\right)\left(\left[\Delta_{2}+\Delta_{3}\right]_{\tau}-\tau+\frac{1}{2}\right)\left(\left[\Delta_{2}+\Delta_{3}\right]_{\tau}-\tau+1\right)+
\nn\\
&\frac{1}{3\tau^{2}}\left(\left[\Delta_{3}+\Delta_{4}\right]_{\tau}-\tau\right)\left(\left[\Delta_{3}+\Delta_{4}\right]_{\tau}-\tau+\frac{1}{2}\right)\left(\left[\Delta_{3}+\Delta_{4}\right]_{\tau}-\tau+1\right)+
\nn\\
&\frac{1}{3\tau^{2}}\left(\left[\Delta_{4}+\Delta_{5}\right]_{\tau}-\tau\right)\left(\left[\Delta_{4}+\Delta_{5}\right]_{\tau}-\tau+\frac{1}{2}\right)\left(\left[\Delta_{4}+\Delta_{5}\right]_{\tau}-\tau+1\right)+
\nn\\
&\frac{1}{3\tau^{2}}\left(\left[\Delta_{5}+\Delta_{6}\right]_{\tau}-\tau\right)\left(\left[\Delta_{5}+\Delta_{6}\right]_{\tau}-\tau+\frac{1}{2}\right)\left(\left[\Delta_{5}+\Delta_{6}\right]_{\tau}-\tau+1\right)+
\nn\\
&\frac{1}{3\tau^{2}}\left(\left[\Delta_{6}+\Delta_{7}\right]_{\tau}-\tau\right)\left(\left[\Delta_{5}+\Delta_{6}\right]_{\tau}-\tau+\frac{1}{2}\right)\left(\left[\Delta_{5}+\Delta_{6}\right]_{\tau}-\tau+1\right)+
\nn\\
&\frac{1}{3\tau^{2}}\left(\left[\Delta_{7}+\Delta_{1}\right]_{\tau}-\tau\right)\left(\left[\Delta_{6}+\Delta_{1}\right]_{\tau}-\tau+\frac{1}{2}\right)\left(\left[\Delta_{6}+\Delta_{1}\right]_{\tau}-\tau+1\right)+
\nn\\
&\frac{1}{3\tau^{2}}\left(\left[\Delta_{1}+\Delta_{2}+\Delta_{3}\right]_{\tau}-\tau\right)\left(\left[\Delta_{1}+\Delta_{2}+\Delta_{3}\right]_{\tau}-\tau+\frac{1}{2}\right)\left(\left[\Delta_{1}+\Delta_{2}+\Delta_{3}\right]_{\tau}-\tau+1\right)+
\nn\\
&\frac{1}{3\tau^{2}}\left(\left[\Delta_{4}+\Delta_{5}+\Delta_{6}\right]_{\tau}-\tau\right)\left(\left[\Delta_{4}+\Delta_{5}+\Delta_{6}\right]_{\tau}-\tau+\frac{1}{2}\right)\left(\left[\Delta_{4}+\Delta_{5}+\Delta_{6}\right]_{\tau}-\tau+1\right)+
\nn\\
&\frac{1}{3\tau^{2}}\left(\left[\Delta_{7}+\Delta_{1}+\Delta_{6}\right]_{\tau}-\tau\right)\left(\left[\Delta_{7}+\Delta_{1}+\Delta_{6}\right]_{\tau}-\tau+\frac{1}{2}\right)\left(\left[\Delta_{7}+\Delta_{1}+\Delta_{6}\right]_{\tau}-\tau+1\right)\,.
\ee
To simplify this expression, we need to consider the different cases for the bracket functions. However, for compactness we only present here the result for Case Ia, in which 
\begin{equation*}
\textrm{Case Ia}:\,\,\,\,\,\,\,\,\,\,\,\textrm{Im}\left(-\frac{1}{\tau}\right)>\textrm{Im}\left(\frac{1}{\tau}\sum_{i=1}^{6}\left[\Delta_{i}\right]_{\tau}\right)>0
\end{equation*}
(note that all the sub-sums are also in the same strip). For this case, we have
\be
&\Theta\left(\Delta_{i};\tau\right)_{\textrm{Ia}}=\tau^{-2}\left(\left[\Delta_{1}\right]_{\tau}\left[\Delta_{2}\right]_{\tau}\left[\Delta_{3}\right]_{\tau}+2\left[\Delta_{1}\right]_{\tau}\left[\Delta_{2}\right]_{\tau}\left[\Delta_{4}\right]_{\tau}+2\left[\Delta_{1}\right]_{\tau}\left[\Delta_{3}\right]_{\tau}\left[\Delta_{4}\right]_{\tau}+\left[\Delta_{2}\right]_{\tau}\left[\Delta_{3}\right]_{\tau}\left[\Delta_{4}\right]_{\tau}+\right.
\nn\\
&2\left[\Delta_{1}\right]_{\tau}\left[\Delta_{2}\right]_{\tau}\left[\Delta_{5}\right]_{\tau}+3\left[\Delta_{1}\right]_{\tau}\left[\Delta_{3}\right]_{\tau}\left[\Delta_{5}\right]_{\tau}+2\left[\Delta_{2}\right]_{\tau}\left[\Delta_{3}\right]_{\tau}\left[\Delta_{5}\right]_{\tau}+2\left[\Delta_{1}\right]_{\tau}\left[\Delta_{4}\right]_{\tau}\left[\Delta_{5}\right]_{\tau}+
\nn\\
&2\left[\Delta_{2}\right]_{\tau}\left[\Delta_{4}\right]_{\tau}\left[\Delta_{5}\right]_{\tau}+\left[\Delta_{3}\right]_{\tau}\left[\Delta_{4}\right]_{\tau}\left[\Delta_{5}\right]_{\tau}+\left[\Delta_{1}\right]_{\tau}\left[\Delta_{2}\right]_{\tau}\left[\Delta_{6}\right]_{\tau}+2\left[\Delta_{1}\right]_{\tau}\left[\Delta_{3}\right]_{\tau}\left[\Delta_{6}\right]_{\tau}+
\nn\\
&2\left[\Delta_{2}\right]_{\tau}\left[\Delta_{3}\right]_{\tau}\left[\Delta_{6}\right]_{\tau}+2\left[\Delta_{1}\right]_{\tau}\left[\Delta_{4}\right]_{\tau}\left[\Delta_{6}\right]_{\tau}+3\left[\Delta_{2}\right]_{\tau}\left[\Delta_{4}\right]_{\tau}\left[\Delta_{6}\right]_{\tau}+2\left[\Delta_{3}\right]_{\tau}\left[\Delta_{4}\right]_{\tau}\left[\Delta_{6}\right]_{\tau}+
\nn\\
&\left[\Delta_{1}\right]_{\tau}\left[\Delta_{5}\right]_{\tau}\left[\Delta_{6}\right]_{\tau}+2\left[\Delta_{2}\right]_{\tau}\left[\Delta_{5}\right]_{\tau}\left[\Delta_{6}\right]_{\tau}+2\left[\Delta_{3}\right]_{\tau}\left[\Delta_{5}\right]_{\tau}\left[\Delta_{6}\right]_{\tau}+\left[\Delta_{4}\right]_{\tau}\left[\Delta_{5}\right]_{\tau}\left[\Delta_{6}\right]_{\tau}+
\nn\\
&\left[\Delta_{1}\right]_{\tau}\left[\Delta_{3}\right]_{\tau}\left[\Delta_{7}\right]_{\tau}+2\left[\Delta_{2}\right]_{\tau}\left[\Delta_{3}\right]_{\tau}\left[\Delta_{7}\right]_{\tau}+2\left[\Delta_{1}\right]_{\tau}\left[\Delta_{4}\right]_{\tau}\left[\Delta_{7}\right]_{\tau}+4\left[\Delta_{2}\right]_{\tau}\left[\Delta_{4}\right]_{\tau}\left[\Delta_{7}\right]_{\tau}+3\left[\Delta_{3}\right]_{\tau}\left[\Delta_{4}\right]_{\tau}\left[\Delta_{7}\right]_{\tau}+
\nn\\
&2\left[\Delta_{1}\right]_{\tau}\left[\Delta_{5}\right]_{\tau}\left[\Delta_{7}\right]_{\tau}+4\left[\Delta_{2}\right]_{\tau}\left[\Delta_{5}\right]_{\tau}\left[\Delta_{7}\right]_{\tau}+4\left[\Delta_{3}\right]_{\tau}\left[\Delta_{5}\right]_{\tau}\left[\Delta_{7}\right]_{\tau}+2\left[\Delta_{4}\right]_{\tau}\left[\Delta_{5}\right]_{\tau}\left[\Delta_{7}\right]_{\tau}+
\nn\\
&\left.\left[\Delta_{1}\right]_{\tau}\left[\Delta_{6}\right]_{\tau}\left[\Delta_{7}\right]_{\tau}+2\left[\Delta_{2}\right]_{\tau}\left[\Delta_{6}\right]_{\tau}\left[\Delta_{7}\right]_{\tau}+2\left[\Delta_{3}\right]_{\tau}\left[\Delta_{6}\right]_{\tau}\left[\Delta_{7}\right]_{\tau}+\left[\Delta_{4}\right]_{\tau}\left[\Delta_{6}\right]_{\tau}\left[\Delta_{7}\right]_{\tau}\right),
\ee
which is of the same form as \eqref{expected:answer} with the coefficients given by the triangular anomalies, and matches the result obtained in \cite{Amariti:2019mgp} using the Cardy limit of the index.

\

\section{Extremization and Black Hole Entropy}
\label{Extrem}

As discussed in the introduction, black hole entropy can be obtained from the Legendre transform 
of the entropy function\cite{Hosseini:2017mds,Cabo-Bizet:2018ehj}. The latter may be formulated in terms of the dual
CFT parameters, and has been shown to coincide with the Cardy-like limit of the index for the case of $\CN=4$ SYM dual 
to the $\mathrm{AdS}_5\times S^5$ black hole with three electric charges $Q_{1,2,3}$ and two angular momenta $J_{1,2}$ \cite{Gutowski:2004yv,Gutowski:2004ez}.
Its entropy function in particular was shown to be equal to
\be
\mathcal{S}=-\frac{\pi iN^{2}}{\omega_{1}\omega_{2}}X_{1}X_{2}X_{3}\,,
\label{sym:entropy:function}
\ee
where $\omega_{1,2}$ are chemical potentials conjugate to two angular momenta and $X_{1,2,3}$ are the ones conjugate to three charges. In \cite{Benini:2018ywd} the large-$N$ behavior of the index was derived using a BAE analysis. In particular, it has been shown that the basic contribution to the index is dominant for the case of the dual single-centered black hole \footnote{By saying here that the basic solution is dominant for a single-centered black hole we precisely mean that in (\ref{index:BAE:gen:final})  $r=0$ gives the dominant contribution.}. Moreover, the authors have shown that (\ref{sym:entropy:function}) is reproduced by their large-$N$ index in this case upon the identification
\be
\omega_1=\omega_2=\tau\,,\qquad \Dbr{a}{\tau}=X_a\quad \mathrm{for}\quad a=1,\dots,d\,,
\label{entropy:function:id}
\ee
where $\Delta_a$ are the same chemical potentials (\ref{delta:def}) we use in the present work and $d=3$ for the SYM case. Their observations hold 
under an assumption that the $\Delta$s belong to the region analogous to the one we call ``case I'' through Sections \ref{Coni} and \ref{Oth}. 
In this case, $\Dbr{a}{\tau}$ satisfy the constraint
\be
\sum_{a=1}^d\Dbr{a}{\tau}=2\tau-1\,,
\label{Dbr:constraint}
\ee
which is precisely the constraint that should be satisfied by $X_a$. Another argument in favor of the identification (\ref{entropy:function:id}), as shown in \cite{Benini:2018ywd} for the case of SYM, is that $X_a$ should satisfy the same strip inequality as $\Dbr{a}{\tau}$ do, namely
\be
\Im\left(- \frac{1}{\omega} \right)>\Im\left( \frac{X_a}{\omega} \right)>0\,,\quad \mathrm{for} \quad \omega_1=\omega_2=\omega\,,~ a=1,2,3\,.
\ee
Similar arguments work also for black hole with $d>3$ charges, which on the dual side correspond to theories with $d$ global $\U(1)$ symmetries. We thus propose, based on the results of Sections \ref{Coni} and \ref{Oth}, the following entropy functions for black holes dual to toric theories:
\be
\mathcal{S}=-\frac{\pi iN^{2}}{6\omega_{1}\omega_{2}}C_{IJK}X_{I}X_{J}X_{K}\,,\quad I,J,K=1,\dots,d\,.
\label{entropy:function:toric}
\ee
Here, the coefficients $C_{IJK}$ are defined by the triangular anomalies of the corresponding $\U(1)$ global symmetries, and the chemical potentials $X_a$ satisfy the constraint (\ref{Dbr:constraint}),
\be
\sum\limits_{I=1}^d X_I=\omega_1+\omega_2-1\,.
\label{constraint:X}
\ee
Notice that this is exactly the proposal of \cite{Hosseini:2018dob} confirmed at the Cardy-like limit in \cite{Amariti:2019mgp}. 

Let us now briefly revise the procedure of extracting the entropy from the entropy functions $\CS$, giving a concrete illustration of the simplest case 
of $\CN=4$ SYM. The black hole entropy is computed as the Legendre transform of the entropy function $\CS$, given by the following function:
\be
\label{Legendre:transform:def}
\hat{\mathcal{S}}=\mathcal{S}-2\pi i\left(\sum_{I=1}^d Q_{I}X_{I}+\sum_{i=1}^2 J_{i}\omega_{i}\right)-2\pi i\Lambda\left(\sum_{I=1}^d X_{I}-\sum_{i=1}^2\omega_{i}+1\right)\,,
\ee
computed at the critical point
\be
\label{entropy:function:crit:gen}
Q_{I}+\Lambda=\frac{1}{2\pi i}\frac{\partial\mathcal{S}}{\partial X_{I}}\,,\qquad
J_{i}-\Lambda=\frac{1}{2\pi i}\frac{\partial\mathcal{S}}{\partial\omega_{i}}\,.
\ee
Here, as usual $Q_I$ are the black hole $d$ electric charges and $J_i$ are its two angular momenta. We additionally introduce a Lagrange multiplier $\Lambda$ to 
impose the constraint (\ref{constraint:X}). At this point one should, in principle, perform the technically challenging task of solving (\ref{entropy:function:crit:gen}) and substituting the solution into (\ref{Legendre:transform:def}). However, Ref. \cite{Hosseini:2017mds} proposed a nice trick to circumvent this complication and obtain the black hole entropy rather easily. Specifically, it was noticed that the entropy function (\ref{sym:entropy:function}) is a monomial of $X$ so it can be rewritten in the following form
 \be
 \mathcal{S}=\sum_{I}X_{I}\frac{\partial\mathcal{S}}{\partial X_{I}}+\sum_{i}\omega_{i}
 \frac{\partial\mathcal{S}}{\partial\omega_{i}}\,,
\label{entropy:func:homogenity}
\ee
In more general cases of toric theories, the entropy function (\ref{entropy:function:toric}) is a homogenous polynomial of degree three so the same argument applies. 
This relation together with the equations of motion (\ref{entropy:function:crit:gen}) allows us to express the black hole entropy (\ref{Legendre:transform:def}) 
in terms the Lagrange multiplier $\Lambda$ alone, as follows:
\be
\label{EntrCritLambd}
       S_{\textrm{BH}}=\left.\hat{\mathcal{S}}\right|_{\textrm{critical}}=-2\pi i\Lambda\,.
\ee
Now, the only thing left to do is to express $\Lambda$ in terms of the charges $Q_I$ and angular momenta $J_i$. For example, in the $\CN=4$ SYM case the equations of motion (\ref{entropy:function:crit:gen}) take the following form:
\be
Q_1+\Lambda&=&-\nu\frac{X_2 X_3}{\omega_1\omega_2},\,\quad Q_2+\Lambda=-\nu\frac{X_1 X_3}{\omega_1\omega_2},\,\quad 
Q_3+\Lambda=-\nu\frac{X_1 X_2}{\omega_1\omega_2},\,
\nn\\
J_1-\Lambda&=&\nu \frac{X_1X_2X_3}{\omega_1^2\omega_2}\,,\quad J_2-\Lambda=\nu \frac{X_1X_2X_3}{\omega_1\omega_2^2}\,,\quad \nu\equiv \frac{N^2}{2}\,.
\label{entropy:eom:sym}
\ee
Then it can be seen that the charges and angular momenta enjoy the following cubic relation:
\be
0=\left(Q_{1}+\Lambda\right)\left(Q_{2}+\Lambda\right)\left(Q_{3}+\Lambda\right)+\frac{N^{2}}{2}\left(J_{1}-\Lambda\right)\left(J_{2}-\Lambda\right)=\Lambda^{3}+p_{2}\Lambda^{2}+p_{1}\Lambda+p_{0}\,,
\label{cubic:relation}
\ee
where we introduced the coefficients $p_i$ defined as
\be
p_2&=&Q_1+Q_2+Q_3+\nu\,,\nn\\
p_1&=& Q_1Q_2+Q_1Q_3+Q_2Q_3-\nu\left( J_1+J_2 \right)\,,\nn\\
p_0&=& Q_1Q_2Q_3+\nu J_1J_2\,.
\label{P:coeff:sym}
\ee
From the requirement that the entropy be real and positive, one can easily deduce that the charges should satisfy the constraint 
$p_0=p_1p_2$ and that the entropy of the dual black hole is given by
\be
S_{\textrm{BH}}=-2\pi i\Lambda=2\pi\sqrt{p_{1}}\,,  
\label{sym:bh:entropy}
\ee
which precisely reproduces the supergravity result \cite{Gutowski:2004yv}. A similar analysis has also been performed for the black hole duals of 
$Y^{pp}$ and $L^{aba}$ theories in \cite{Amariti:2019mgp}. However, the algorithm described above is technically complicated for other examples 
of toric theories. As we will be seeing momentarily for the conifold theory case, one of the problems is that the cubic relation (\ref{cubic:relation}) 
does not always exist. At the same time, solving the equations of motion (\ref{entropy:function:crit:gen}) directly is technically too complicated. 
It would be interesting to understand if these problems can somehow be worked around in the future. 

Taking another route, however, we will try to analyze the structure of the large-$N$ index of the conifold theory in the space of chemical potentials $\Delta_I$, for the specific case dual to a black hole with equal charges. A similar analysis was performed in \cite{Benini:2018ywd}. In particular, the authors identified 
Stokes lines and regions, and found that for certain values of charges and angular momenta single-centered black hole solutions become unstable,
as the index corresponding to these values ends up sitting on a Stokes line.

\subsection{The case of equal charges}

Let us now consider the large-$N$ index of the conifold theory given in (\ref{T11Theta}) and (\ref{T11:f:function}) and concentrate on the following simple choice of chemical potentials: 
\be
 \Delta_{1}=\Delta_{2}=\Delta_{3}=\Delta_{4}=\Delta=\frac{2\tau-1}{4}\,,
\label{chem:pot:eq:charge}
\ee
where the last inequality comes from the constraint (\ref{Dbr:constraint}). Notice that we are now left with the single chemical potential $\tau$. Also notice that the index $\mathcal{I}\left(\frac{2\tau-1}{4};\tau\right)$
is periodic under the shift $\tau\rightarrow\tau+2$, from which it follows that we may restrict to $0\leq\textrm{Re}\tau<2$. As we will later see, the choice (\ref{chem:pot:eq:charge}) has a nice interpretation in terms of a dual black hole with equal charges $Q_{I}$. 

The contribution of the basic solution and its $T$-transforms to the logarithm of the superconformal index reduces in this case to  
        \be
        -\pi iN^{2}\Theta\left(\Delta;\tau+r\right)=-\frac{\pi iN^{2}}{\tau^{2}}\times
	\left\{\begin{array} {ll} 
	4\left[\Delta\right]_{\tau+r}^{3}\,,  \qquad &  \mathrm{Case~ I}\\
	4\left(\left[\Delta\right]_{\tau+r}^{3}+\frac{1}{2}\right)+\frac{\tau}{2}-\tau^2\,, \qquad &  \mathrm{Case~ II}\\
	4\left(\left[\Delta\right]_{\tau+r}^{3}+1\right)-2\tau^2\,, \qquad &  \mathrm{Case~ III}
	\end{array}\right.
	\label{T11Theta:eq:charg}
	\ee
The definition of the different cases may be found in (\ref{T11:cases}). To work with this expression, we must understand what are the values of $\Dbr{ }{\tau+r}$ for various $r\in \mathbb{Z}$ and which cases in (\ref{T11Theta:eq:charg}) they correspond to. In particular, it can be checked that we should distinguish between cases of even and odd $r$ as follows:
\be
        \left[\Delta\right]_{\tau+r}=\left[\frac{2\tau-1}{4}\right]_{\tau+r}=\left\{ \begin{array}{c}
        \Delta+\frac{r}{2}\qquad\textrm{\,\,\ensuremath{r} even}\\
        \Delta+\frac{r-1}{2}\qquad\textrm{\ensuremath{r} odd}
        \end{array}\right.
	\label{eq:charges:delta}
\ee
It can also be checked that for even $r$, one should use ``case I'' in (\ref{T11Theta:eq:charg}), while for odd $r$ ``case III'' is the suitable one.

Now we finally would like to understand the structure of the large-$N$ index in the complex plane of the single parameter we are left with, namely $\tau$. For this we need to understand which $\hat{r}$ are maximizing $\textrm{Im}\Theta\left(\Delta;\tau+r\right)$ in the different regions. If in some region there is no $r$ maximizing this function, one then needs to sum over infinitely many exponents contributing at the same order. This would require 
extra information and probably the inclusion of subleading terms to evaluate index. That situation precisely corresponds to being on the Stokes line. In our case, the imaginary part of the $\Theta$ function (\ref{T11Theta:eq:charg}) is given by

\be
\label{T11ImTheta}
\textrm{Im}\Theta\left(\Delta;\tau+r\right)=\left\{ \begin{array}{c}
\frac{1}{8}\textrm{Im}\tau\left(4+\frac{\textrm{Re}\tau+r}{\left|\tau+r\right|^{4}}-\frac{3}{\left|\tau+r\right|^{2}}\right)\,\,\,\,\,\textrm{\ensuremath{r} even}\\
\frac{1}{8}\textrm{Im}\tau\left(4-\frac{\textrm{Re}\tau+r}{\left|\tau+r\right|^{4}}-\frac{3}{\left|\tau+r\right|^{2}}\right)\,\,\,\,\,\textrm{\textrm{\ensuremath{r} odd}}
\end{array}\right.
\ee
This is remarkably similar to the SYM result \cite{Benini:2018ywd}, with only two differences between the two cases. 
One difference is that in $\CN=4$ SYM there are three possibilities with $r=0,1,2~ \mathrm{mod}~ 3$. For $r=1~\mathrm{mod}~3$, $\Dbr{ }{\tau}$ is not defined since this value is precisely on the Stokes line. The second is that the expressions for $r=0$ and $r=2~\mathrm{mod}~3$ exactly reproduce our expressions for even and odd $r$, correspondingly, up to an overall factor. However, this is irrelevant since this coefficient only defines the value of the index in the particular region but not the region itself. 

\begin{figure}[h]
        \centering
        \includegraphics[scale=1.25]{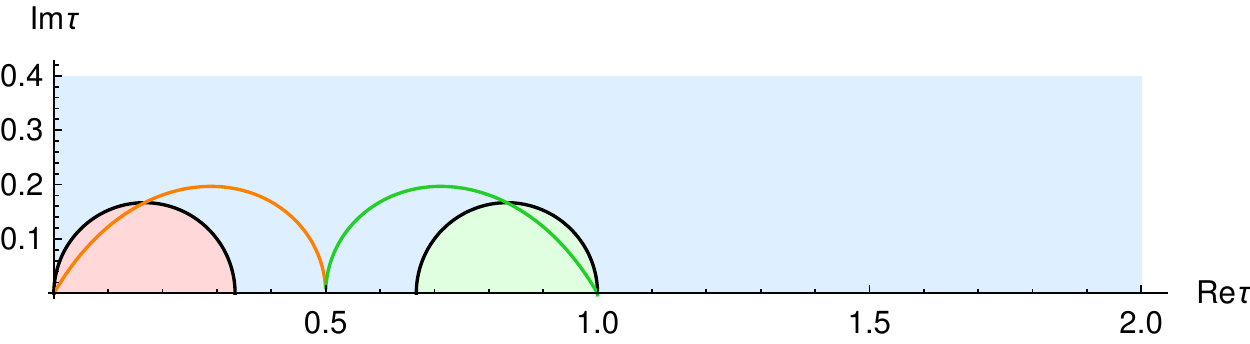}
        \caption{Stokes lines in the fundamental region of the $\tau$-plane. Left and right semicirlces correspond to the 
	regions where $r=0$ and $r=1$ contributions dominate the index of the conifold theory. These semicircles are defined in 
	Eqs. \ref{semicircle1} and \ref{semicircle2}, correspondingly. The region outside the semicircles has no dominant contribution. Colored 
	lines correspond to the critical points of the entropy function of the single-centered BPS black hole in the dual description.}
        \label{T11Stokes}
\end{figure}

As illustrated in Fig. \ref{T11Stokes}, there exists a maximum of $\textrm{Im}\Theta\left(\Delta;\tau+r\right)$ given in (\ref{T11ImTheta}) with an {\it even} $\widehat{r}$ if $\tau$ resides in the interior of the semicircle defined by
\be
        \label{semicircle1}
                \textrm{Re}\tau+\hat{r}_{\textrm{even}}>3\left|\tau+\hat{r}_{\textrm{even}}\right|^{2}\,.
\ee
There further exists a maximum with an {\it odd} $\widehat{r}$ if $\tau$ resides in the interior of the semicircle defined by
\be
 \label{semicircle2}
   -\textrm{Re}\tau-\hat{r}_{\textrm{odd}}>3\left|\tau+\hat{r}_{\textrm{odd}}\right|^{2}\,.
\ee
This pair of semicircles appearing in the fundamental range $0\leq\textrm{Re}\tau<2$ periodically features
in all the images of the fundamental range under $\tau\rightarrow\tau+2$.
\footnote{The same picture emerges for SYM, with the difference that there the periodicity is $\tau\rightarrow\tau+3$
and, correspondingly, the fundamental range is $0\leq\textrm{Re}\tau<3$.} The region external to the semicircles admits
no dominant contribution; there, the supremal value for $\textrm{Im}\Theta$ as a function of $r$ is approached asymptotically
in the limit where $r\rightarrow\pm\infty$ and is equal to $\frac{1}{2}\textrm{Im}\tau$, as can be readily seen from \eqref{T11ImTheta}.

Now, when we know the structure of the large-$N$ index for the conifold theory in the complex $\tau$-plane, we can also try to explore the position of the critical points of the entropy function in the same plane. In particular, according to our proposal (\ref{entropy:function:toric}) the entropy function of the black hole dual to the conifold theory should be given by
\be
    \label{T11entropyfnc}
    \mathcal{S}=-\frac{\pi iN^{2}}{\omega_{1}\omega_{2}}\left(X_{1}X_{2}X_{3}+X_{1}X_{2}X_{4}+X_{1}X_{3}X_{4}+X_{2}X_{3}X_{4}\right)\,,
\ee
with the chemical potentials $X_a$ satisfying the constraint (\ref{constraint:X}). 
The equations of motion (\ref{entropy:function:crit:gen}) in this case take the following form:
\be
Q_1+\Lambda&=&-\nu\frac{X_2X_3+X_2X_4+X_3X_4}{\omega_1\omega_2}\,,\quad 
Q_2+\Lambda=-\nu\frac{X_1X_3+X_1X_4+X_3X_4}{\omega_1\omega_2}\,,\quad\nn\\
Q_3+\Lambda&=&-\nu\frac{X_1X_2+X_1X_4+X_2X_4}{\omega_1\omega_2}\,,\quad
Q_4+\Lambda=-\nu\frac{X_1X_2+X_1X_3+X_2X_3}{\omega_1\omega_2}\,,\quad\nn\\ 
J_1-\Lambda&=&-\frac{1}{2\pi i}\frac{\CS}{\omega_1}\,,
\qquad J_2-\Lambda=-\frac{1}{2\pi i}\frac{\CS}{\omega_2}\,,\qquad \nu=\frac{N^2}{2}\,.
\label{entropyfn:eom:t11}
\ee
In general, it is very hard to solve these equations and a relation of the form (\ref{cubic:relation}) is also impossible to find. However, if we require equal chemical potentials $\Delta$ (\ref{chem:pot:eq:charge}) in the index, then the identification (\ref{entropy:function:id}) immediately implies:
\be
X_1=X_2=X_3=X_4=X=\frac{2\tau-1}{4}\,.
\label{X:equal:charge}
\ee
Substituting the above expression into the equations of motion (\ref{entropyfn:eom:t11}), and assuming $\omega_1=\omega_2=\tau$ we obtain that all electric charges are equal as well as two angular momenta are equal to each other:
 \be
&Q_1=Q_2=Q_3=Q_4=Q\,,\quad &Q+\Lambda=-3\nu\frac{X^2}{\tau^2}\,,\\
&J_1=J_2=J\,,\quad         &J-\Lambda=4\nu\frac{X^3}{\tau^3}\,.
\label{eom:T11:eq:charge}
 \ee
We are now able to use the usual trick and construct a simple cubic relation between the charges and the angular momenta:
\be
\left( Q+\Lambda \right)^3+c\left( J-\Lambda \right)^2=\Lambda^3+p_2\Lambda^2+p_1\Lambda+p_0\,,
\label{cubic:rel:eq:charge}
\ee
where 
\be
p_0=Q^3+cJ^2\,,\quad p_1=3Q^2-cJ\,,\quad p_2=3Q+c\,,\quad c\equiv\frac{27}{16}\nu\,.
\label{cubic:rel:eq:charge:coeff}
\ee
It is convenient to measure the charges in units of $c\nu$, so we introduce $Q\equiv c\nu\tQ\,,\,\,J\equiv c\nu\tJ\,,\,\, \Lambda\equiv c\nu\tL $. 
In this case all our expressions, except for the equations of motion (\ref{eom:T11:eq:charge}) appear to be the same as in the case of $\CN=4$ SYM in \cite{Benini:2018ywd}. In particular, we obtain the following constraint for the charges and entropy 
\be
0&=&p_1p_2-p_0=c^3\nu^3\left( 8\tQ^3+3\tQ^2-2\left( 3\tQ+1 \right)\tJ-\tJ^2 \right)\,,\nn\\
S&=&2\pi\sqrt{p_1}=2\pi c\nu\sqrt{3\tQ^2-\tJ}\,.
\label{charges:entropy:t11}
\ee
Due to the very similar form of the expressions, we will use the same parametrization of charges as in the case of $\CN=4$ SYM, that is
\be
\tQ=\mu+\frac{1}{2}\mu^2\,,\quad \tJ=-1-3\tQ+\left( 2\tQ+1 \right)^{3/2}=\frac{3}{2}\mu^2+\mu^3\,,\nn\\
p_1=c^2\left( 3\tQ^2-c\tJ \right)=c^2\left( \mu^3+\frac{3}{4}\mu^4 \right)\,,\quad \tL=i c^{-1}\sqrt{p_1}\,.
\label{entropy:parametrization}
\ee
Finally, we can also solve the equations of motion (\ref{eom:T11:eq:charge}) which gives us the parametrization of 
$\tau$ and $X$:
\be
 \tau=\frac{\tL+\tQ}{3\tJ+2\tQ-\tL}\,,\qquad X=\frac{3}{4}\frac{\tL-\tJ}{3\tJ+2\tQ-\tL}\,.
\label{eom:t11:sol}
\ee
It is interesting to note that the solution for $\tau$ is exactly the same as for the case of $\CN=4$ SYM \cite{Benini:2018ywd}. 
Moreover, even the parametrization we use in (\ref{entropy:parametrization}) is the same as in the SYM case, so the expression appears to be exactly the same. The form of the solution for the chemical potential $X$ is also identical to the one found in the case of $\CN=4$ SYM, up to an overall coefficient only. 

It is interesting to see how these results look like in Fig. \ref{T11Stokes}, where we show the structure of the large-$N$ index dual to the entropy of a black hole with equal charges. Using the parametrization (\ref{entropy:parametrization}) for the solution (\ref{eom:t11:sol}), the line of critical points is shown in orange in Fig. \ref{T11Stokes}. It is not surprising that the picture we obtain is identical to the one obtained for $\CN=4$ SYM in \cite{Benini:2018ywd}. In particular, we see that for small values of $\mu$ below a certain critical value $\mu<\mu_*$, the solution is in the region with no dominant contribution to the index. Hence, in this region we should take into 
account other solutions beyond the basic one to evaluate the index, and probably the leading order contribution will no longer be of order $N^2$. In \cite{Benini:2018ywd} it was proposed that such behavior signals the instability of small ({\it i.e.} with charge below some critical value, $Q<Q_*$) single-centered black hole towards other supergravity solutions. On the other hand, when we increase $\mu$ at some point it crosses the semicircle (\ref{semicircle1}) and the solution we found becomes dominant. Therefore, large single-centered black holes (with charge $Q$ above the threshold $Q_*$) appear to be stable solutions of supergravity. The critical value of $\mu_*$ can be found from the equation $\Re \tau=3|\tau|^2$ defining the boundary of the semicircle (\ref{semicircle1}). Using the parametrization (\ref{entropy:parametrization}) together with the solution (\ref{eom:t11:sol}) we can 
define the following values of critical charges and entropy:
\be
  \mu_{*}=\frac{2}{3},\quad\tau_{*}=\frac{1+i}{6},\quad Q_{*}=\frac{3}{2}\nu,\quad J_{*}=\frac{13}{8}\nu,\quad S_{*}=\frac{9\pi}{4}\nu\,.
\ee

An identical analysis can be performed for the interior of the semicircle (\ref{semicircle2}), where the large-$N$ index and the corresponding entropy function are dominated by the $r=1$ contribution and take the following form
        \be
        \label{indexcircle2}
        \log\mathcal{I}_{\infty}\left(\Delta;\tau\right)=-\pi iN^{2}\Theta\left(\Delta;\tau+1\right)=
	=-\pi iN^{2}\left[\frac{\left(2\tau+1\right)^{3}}{16(\tau+1)^{2}}-2\right]\,.
        \ee
In Fig.\ref{T11Stokes} we show the line of critical points in green. As we can see, the form of the curve just mirrors the line of critical points for the case of $r=0$. 

As we have mentioned repeatedly in this section, the picture we obtain for the case of equal black hole charges $Q_I$ is exactly identical to the one obtained in \cite{Benini:2018ywd}. It would be interesting to understand if the same pattern takes place for other toric theories. However, note that conifold theory is special because in general the requirement (\ref{X:equal:charge}) of equal chemical potentials does not translate into a condition of equal electric charges $Q_I$ of the dual black hole. This is due to more a complicated structure of the equations of motion (\ref{entropy:function:crit:gen}).

\

\section{Discussion and Outlook}
\label{Diss}

In this paper we used a BAE analysis to obtain the large-$N$ behavior of the superconformal index for a variety of toric gauge theories. We demonstrated that the basic solution (\ref{BAE:sol:sym}) to the BAEs of $\CN=4$ SYM can be directly generalized to (\ref{BAE:sol:more:general}), which solves the BAEs for any toric quiver theory. Using these solutions, we derived the relatively simple expressions (\ref{index:general:BAE:basic}) and (\ref{index:BAE:gen:final}) for their contribution to the large-$N$ limit of the superconformal index for arbitrary toric quivers. Studying these expressions for a number of theories, including the infinite families $Y^{pq},\, X^{pq}$ and $L^{pqr}$, we found a way to rewrite them in a form that coincides with (\ref{antonio:result}) for certain ranges of parameters. The latter is the result of the index computation in the Cardy-like limit \cite{Amariti:2019mgp}. Remarkably, this also reproduces a conjecture made in \cite{Hosseini:2018dob} for the entropy function of a multi-charged rotating $\mathrm{AdS}_5$ black hole. 

Finally, we also studied the large-$N$ index structure for the conifold theory in the specific case corresponding to a dual black hole with four equal charges and two equal angular momenta. In this case, we identified Stokes lines and regions in the space of the single complex parameter $\tau$, which is just the chemical potential dual to the angular momentum of the black hole. We further performed an extremization for this case. Curiously, our picture exactly reproduces the one found in the $\CN=4$ case presented in \cite{Benini:2018ywd}.

Many related open questions are still waiting to be answered. First, our results are based only on the basic solutions (\ref{BAE:sol:u:gen}) and their $\mathrm{SL}\left( 2,\mathbb{Z} \right)$ transforms. However, one should study closely the BAEs for particular theories and try to understand if there are other solutions, at which order of $N$ they contribute to the index, and what are the corresponding observables in the bulk of $\mathrm{AdS}$.

Another open question is the extremization procedure of the large-$N$ index. In this paper, we presented results for the very particular case of $T^{1,1}$ with all the chemical potentials being equal. The authors of \cite{Amariti:2019mgp} managed to perform an extremization in the cases of the $Y^{pp}$ and $L^{aba}$ theories. However, for all the other toric cases we have failed to perform a Legendre transform since the algorithms of \cite{Hosseini:2017mds} do not work for them and a straightforward solution is too complicated. It would be interesting to resolve these problems in one way or another. 

Finally, it would be interesting to understand if the BAE approach can be used in order to study superconformal indices of larger classes of theories. Of particular interest are theories with matter in the fundamental or anti-fundamental representations. It looks like the basic solution does not work in this case and that one needs to perform a more complicated analysis of the corresponding BAE equations.

\section*{Acknowledgments}

We would like to thank S.~S.~Razamat for collaborating with us during the early stages of this project, as well as for many fruitful discussions and valuable comments. 
We are also thankful to A.~Amariti, F.~Benini, D.~Cassani, P.~Milan and A.~Zaffaroni for interesting discussions during various stages of the project.
This work is supported by the Israel Science Foundation under grant No.~2289/18, and by the I-CORE Program of the Planning and Budgeting Committee. OS is also supported by the Daniel scholarship for PhD students. 

\bibliographystyle{./aug/ytphys}
\bibliography{./aug/refs}

\providecommand{\href}[2]{#2}\begingroup\raggedright\begin{thebibliography}{10}

\bibitem{Strominger:1996sh}
A.~Strominger and C.~Vafa, ``{Microscopic origin of the Bekenstein-Hawking
  entropy},'' \href{http://dx.doi.org/10.1016/0370-2693(96)00345-0}{{\em Phys.
  Lett.} {\bfseries B379} (1996) 99--104},
\href{http://arxiv.org/abs/hep-th/9601029}{{\ttfamily arXiv:hep-th/9601029
  [hep-th]}}.

\bibitem{Benini:2015noa}
F.~Benini and A.~Zaffaroni, ``{A topologically twisted index for
  three-dimensional supersymmetric theories},''
  \href{http://dx.doi.org/10.1007/JHEP07(2015)127}{{\em JHEP} {\bfseries 07}
  (2015) 127},
\href{http://arxiv.org/abs/1504.03698}{{\ttfamily arXiv:1504.03698 [hep-th]}}.

\bibitem{Benini:2016hjo}
F.~Benini and A.~Zaffaroni, ``{Supersymmetric partition functions on Riemann
  surfaces},'' {\em Proc. Symp. Pure Math.} {\bfseries 96} (2017) 13--46,
\href{http://arxiv.org/abs/1605.06120}{{\ttfamily arXiv:1605.06120 [hep-th]}}.

\bibitem{Closset:2016arn}
C.~Closset and H.~Kim, ``{Comments on twisted indices in 3d supersymmetric
  gauge theories},'' \href{http://dx.doi.org/10.1007/JHEP08(2016)059}{{\em
  JHEP} {\bfseries 08} (2016) 059},
\href{http://arxiv.org/abs/1605.06531}{{\ttfamily arXiv:1605.06531 [hep-th]}}.

\bibitem{Benini:2016rke}
F.~Benini, K.~Hristov, and A.~Zaffaroni, ``{Exact microstate counting for
  dyonic black holes in AdS4},''
  \href{http://dx.doi.org/10.1016/j.physletb.2017.05.076}{{\em Phys. Lett.}
  {\bfseries B771} (2017) 462--466},
\href{http://arxiv.org/abs/1608.07294}{{\ttfamily arXiv:1608.07294 [hep-th]}}.

\bibitem{Benini:2015eyy}
F.~Benini, K.~Hristov, and A.~Zaffaroni, ``{Black hole microstates in AdS$_{4}$
  from supersymmetric localization},''
  \href{http://dx.doi.org/10.1007/JHEP05(2016)054}{{\em JHEP} {\bfseries 05}
  (2016) 054},
\href{http://arxiv.org/abs/1511.04085}{{\ttfamily arXiv:1511.04085 [hep-th]}}.

\bibitem{Hosseini:2016tor}
S.~M. Hosseini and A.~Zaffaroni, ``{Large $N$ matrix models for 3d ${\cal N}=2$
  theories: twisted index, free energy and black holes},''
  \href{http://dx.doi.org/10.1007/JHEP08(2016)064}{{\em JHEP} {\bfseries 08}
  (2016) 064},
\href{http://arxiv.org/abs/1604.03122}{{\ttfamily arXiv:1604.03122 [hep-th]}}.

\bibitem{Hosseini:2017fjo}
S.~M. Hosseini, K.~Hristov, and A.~Passias, ``{Holographic microstate counting
  for AdS$_{4}$ black holes in massive IIA supergravity},''
  \href{http://dx.doi.org/10.1007/JHEP10(2017)190}{{\em JHEP} {\bfseries 10}
  (2017) 190},
\href{http://arxiv.org/abs/1707.06884}{{\ttfamily arXiv:1707.06884 [hep-th]}}.

\bibitem{Hosseini:2016cyf}
S.~M. Hosseini, A.~Nedelin, and A.~Zaffaroni, ``{The Cardy limit of the
  topologically twisted index and black strings in AdS$_{5}$},''
  \href{http://dx.doi.org/10.1007/JHEP04(2017)014}{{\em JHEP} {\bfseries 04}
  (2017) 014},
\href{http://arxiv.org/abs/1611.09374}{{\ttfamily arXiv:1611.09374 [hep-th]}}.

\bibitem{Hosseini:2018uzp}
S.~M. Hosseini, I.~Yaakov, and A.~Zaffaroni, ``{Topologically twisted indices
  in five dimensions and holography},''
  \href{http://dx.doi.org/10.1007/JHEP11(2018)119}{{\em JHEP} {\bfseries 11}
  (2018) 119},
\href{http://arxiv.org/abs/1808.06626}{{\ttfamily arXiv:1808.06626 [hep-th]}}.

\bibitem{Crichigno:2018adf}
P.~M. Crichigno, D.~Jain, and B.~Willett, ``{5d Partition Functions with A
  Twist},'' \href{http://dx.doi.org/10.1007/JHEP11(2018)058}{{\em JHEP}
  {\bfseries 11} (2018) 058},
\href{http://arxiv.org/abs/1808.06744}{{\ttfamily arXiv:1808.06744 [hep-th]}}.

\bibitem{Benini:2017oxt}
F.~Benini, H.~Khachatryan, and P.~Milan, ``{Black hole entropy in massive Type
  IIA},'' \href{http://dx.doi.org/10.1088/1361-6382/aa9f5b}{{\em Class. Quant.
  Grav.} {\bfseries 35} no.~3, (2018) 035004},
\href{http://arxiv.org/abs/1707.06886}{{\ttfamily arXiv:1707.06886 [hep-th]}}.

\bibitem{Azzurli:2017kxo}
F.~Azzurli, N.~Bobev, P.~M. Crichigno, V.~S. Min, and A.~Zaffaroni, ``{A
  universal counting of black hole microstates in AdS$_{4}$},''
  \href{http://dx.doi.org/10.1007/JHEP02(2018)054}{{\em JHEP} {\bfseries 02}
  (2018) 054},
\href{http://arxiv.org/abs/1707.04257}{{\ttfamily arXiv:1707.04257 [hep-th]}}.

\bibitem{Hosseini:2018usu}
S.~M. Hosseini, K.~Hristov, A.~Passias, and A.~Zaffaroni, ``{6D attractors and
  black hole microstates},'' \href{http://arxiv.org/abs/1809.10685}{{\ttfamily
  arXiv:1809.10685 [hep-th]}}.
[JHEP12,001(2018)].

\bibitem{Fluder:2019szh}
M.~Fluder, S.~M. Hosseini, and C.~F. Uhlemann, ``{Black hole microstate
  counting in Type IIB from 5d SCFTs},''
  \href{http://dx.doi.org/10.1007/JHEP05(2019)134}{{\em JHEP} {\bfseries 05}
  (2019) 134},
\href{http://arxiv.org/abs/1902.05074}{{\ttfamily arXiv:1902.05074 [hep-th]}}.

\bibitem{Hosseini:2016ume}
S.~M. Hosseini and N.~Mekareeya, ``{Large $N$ topologically twisted index:
  necklace quivers, dualities, and Sasaki-Einstein spaces},''
  \href{http://dx.doi.org/10.1007/JHEP08(2016)089}{{\em JHEP} {\bfseries 08}
  (2016) 089},
\href{http://arxiv.org/abs/1604.03397}{{\ttfamily arXiv:1604.03397 [hep-th]}}.

\bibitem{Liu:2018bac}
J.~T. Liu, L.~A. Pando~Zayas, and S.~Zhou, ``{Subleading Microstate Counting in
  the Dual to Massive Type IIA},''
\href{http://arxiv.org/abs/1808.10445}{{\ttfamily arXiv:1808.10445 [hep-th]}}.

\bibitem{Liu:2017vbl}
J.~T. Liu, L.~A. Pando~Zayas, V.~Rathee, and W.~Zhao, ``{One-Loop Test of
  Quantum Black Holes in anti–de Sitter Space},''
  \href{http://dx.doi.org/10.1103/PhysRevLett.120.221602}{{\em Phys. Rev.
  Lett.} {\bfseries 120} no.~22, (2018) 221602},
\href{http://arxiv.org/abs/1711.01076}{{\ttfamily arXiv:1711.01076 [hep-th]}}.

\bibitem{Liu:2017vll}
J.~T. Liu, L.~A. Pando~Zayas, V.~Rathee, and W.~Zhao, ``{Toward Microstate
  Counting Beyond Large N in Localization and the Dual One-loop Quantum
  Supergravity},'' \href{http://dx.doi.org/10.1007/JHEP01(2018)026}{{\em JHEP}
  {\bfseries 01} (2018) 026},
\href{http://arxiv.org/abs/1707.04197}{{\ttfamily arXiv:1707.04197 [hep-th]}}.

\bibitem{Zaffaroni:2019dhb}
A.~Zaffaroni, ``{Lectures on AdS Black Holes, Holography and Localization},''
\newblock 2019.
\newblock
\href{http://arxiv.org/abs/1902.07176}{{\ttfamily arXiv:1902.07176 [hep-th]}}.
\newblock

\bibitem{Gutowski:2004yv}
J.~B. Gutowski and H.~S. Reall, ``{General supersymmetric AdS(5) black
  holes},'' \href{http://dx.doi.org/10.1088/1126-6708/2004/04/048}{{\em JHEP}
  {\bfseries 04} (2004) 048},
\href{http://arxiv.org/abs/hep-th/0401129}{{\ttfamily arXiv:hep-th/0401129
  [hep-th]}}.

\bibitem{Gutowski:2004ez}
J.~B. Gutowski and H.~S. Reall, ``{Supersymmetric AdS(5) black holes},''
  \href{http://dx.doi.org/10.1088/1126-6708/2004/02/006}{{\em JHEP} {\bfseries
  02} (2004) 006},
\href{http://arxiv.org/abs/hep-th/0401042}{{\ttfamily arXiv:hep-th/0401042
  [hep-th]}}.

\bibitem{Chong:2005hr}
Z.~W. Chong, M.~Cvetic, H.~Lu, and C.~N. Pope, ``{General non-extremal rotating
  black holes in minimal five-dimensional gauged supergravity},''
  \href{http://dx.doi.org/10.1103/PhysRevLett.95.161301}{{\em Phys. Rev. Lett.}
  {\bfseries 95} (2005) 161301},
\href{http://arxiv.org/abs/hep-th/0506029}{{\ttfamily arXiv:hep-th/0506029
  [hep-th]}}.

\bibitem{Chong:2005da}
Z.~W. Chong, M.~Cvetic, H.~Lu, and C.~N. Pope, ``{Five-dimensional gauged
  supergravity black holes with independent rotation parameters},''
  \href{http://dx.doi.org/10.1103/PhysRevD.72.041901}{{\em Phys. Rev.}
  {\bfseries D72} (2005) 041901},
\href{http://arxiv.org/abs/hep-th/0505112}{{\ttfamily arXiv:hep-th/0505112
  [hep-th]}}.

\bibitem{Romelsberger:2005eg}
C.~Romelsberger, ``{Counting chiral primaries in N = 1, d=4 superconformal
  field theories},''
  \href{http://dx.doi.org/10.1016/j.nuclphysb.2006.03.037}{{\em Nucl. Phys.}
  {\bfseries B747} (2006) 329--353},
\href{http://arxiv.org/abs/hep-th/0510060}{{\ttfamily arXiv:hep-th/0510060
  [hep-th]}}.

\bibitem{Kinney:2005ej}
J.~Kinney, J.~M. Maldacena, S.~Minwalla, and S.~Raju, ``{An Index for 4
  dimensional super conformal theories},''
  \href{http://dx.doi.org/10.1007/s00220-007-0258-7}{{\em Commun. Math. Phys.}
  {\bfseries 275} (2007) 209--254},
\href{http://arxiv.org/abs/hep-th/0510251}{{\ttfamily arXiv:hep-th/0510251
  [hep-th]}}.

\bibitem{Rastelli:2016tbz}
L.~Rastelli and S.~S. Razamat, ``{The supersymmetric index in four
  dimensions},'' \href{http://dx.doi.org/10.1088/1751-8121/aa76a6}{{\em J.
  Phys.} {\bfseries A50} no.~44, (2017) 443013},
\href{http://arxiv.org/abs/1608.02965}{{\ttfamily arXiv:1608.02965 [hep-th]}}.

\bibitem{Nakayama:2005mf}
Y.~Nakayama, ``{Index for orbifold quiver gauge theories},''
  \href{http://dx.doi.org/10.1016/j.physletb.2006.03.045}{{\em Phys. Lett.}
  {\bfseries B636} (2006) 132--136},
\href{http://arxiv.org/abs/hep-th/0512280}{{\ttfamily arXiv:hep-th/0512280
  [hep-th]}}.

\bibitem{Gadde:2010en}
A.~Gadde, L.~Rastelli, S.~S. Razamat, and W.~Yan, ``{On the Superconformal
  Index of N=1 IR Fixed Points: A Holographic Check},''
  \href{http://dx.doi.org/10.1007/JHEP03(2011)041}{{\em JHEP} {\bfseries 03}
  (2011) 041},
\href{http://arxiv.org/abs/1011.5278}{{\ttfamily arXiv:1011.5278 [hep-th]}}.

\bibitem{Eager:2012hx}
R.~Eager, J.~Schmude, and Y.~Tachikawa, ``{Superconformal Indices,
  Sasaki-Einstein Manifolds, and Cyclic Homologies},''
  \href{http://dx.doi.org/10.4310/ATMP.2014.v18.n1.a3}{{\em Adv. Theor. Math.
  Phys.} {\bfseries 18} no.~1, (2014) 129--175},
\href{http://arxiv.org/abs/1207.0573}{{\ttfamily arXiv:1207.0573 [hep-th]}}.

\bibitem{Hosseini:2017mds}
S.~M. Hosseini, K.~Hristov, and A.~Zaffaroni, ``{An extremization principle for
  the entropy of rotating BPS black holes in AdS$_{5}$},''
  \href{http://dx.doi.org/10.1007/JHEP07(2017)106}{{\em JHEP} {\bfseries 07}
  (2017) 106},
\href{http://arxiv.org/abs/1705.05383}{{\ttfamily arXiv:1705.05383 [hep-th]}}.

\bibitem{Hosseini:2018dob}
S.~M. Hosseini, K.~Hristov, and A.~Zaffaroni, ``{A note on the entropy of
  rotating BPS AdS$_7\times S^4$ black holes},''
  \href{http://dx.doi.org/10.1007/JHEP05(2018)121}{{\em JHEP} {\bfseries 05}
  (2018) 121},
\href{http://arxiv.org/abs/1803.07568}{{\ttfamily arXiv:1803.07568 [hep-th]}}.

\bibitem{Choi:2018fdc}
S.~Choi, C.~Hwang, S.~Kim, and J.~Nahmgoong, ``{Entropy functions of BPS black
  holes in AdS$_4$ and AdS$_6$},''
\href{http://arxiv.org/abs/1811.02158}{{\ttfamily arXiv:1811.02158 [hep-th]}}.

\bibitem{Cabo-Bizet:2018ehj}
A.~Cabo-Bizet, D.~Cassani, D.~Martelli, and S.~Murthy, ``{Microscopic origin of
  the Bekenstein-Hawking entropy of supersymmetric AdS$_{\bf 5}$ black
  holes},''
\href{http://arxiv.org/abs/1810.11442}{{\ttfamily arXiv:1810.11442 [hep-th]}}.

\bibitem{Choi:2018hmj}
S.~Choi, J.~Kim, S.~Kim, and J.~Nahmgoong, ``{Large AdS black holes from
  QFT},''
\href{http://arxiv.org/abs/1810.12067}{{\ttfamily arXiv:1810.12067 [hep-th]}}.

\bibitem{DiPietro:2014bca}
L.~Di~Pietro and Z.~Komargodski, ``{Cardy formulae for SUSY theories in $d =$ 4
  and $d =$ 6},'' \href{http://dx.doi.org/10.1007/JHEP12(2014)031}{{\em JHEP}
  {\bfseries 12} (2014) 031},
\href{http://arxiv.org/abs/1407.6061}{{\ttfamily arXiv:1407.6061 [hep-th]}}.

\bibitem{ArabiArdehali:2019tdm}
A.~Arabi~Ardehali, ``{Cardy-like asymptotics of the 4d $ \mathcal{N}=4 $ index
  and AdS$_{5}$ blackholes},''
  \href{http://dx.doi.org/10.1007/JHEP06(2019)134}{{\em JHEP} {\bfseries 06}
  (2019) 134},
\href{http://arxiv.org/abs/1902.06619}{{\ttfamily arXiv:1902.06619 [hep-th]}}.

\bibitem{Honda:2019cio}
M.~Honda, ``{Quantum Black Hole Entropy from 4d Supersymmetric Cardy
  formula},'' \href{http://dx.doi.org/10.1103/PhysRevD.100.026008}{{\em Phys.
  Rev.} {\bfseries D100} no.~2, (2019) 026008},
\href{http://arxiv.org/abs/1901.08091}{{\ttfamily arXiv:1901.08091 [hep-th]}}.

\bibitem{Kim:2019yrz}
J.~Kim, S.~Kim, and J.~Song, ``{A 4d $N=1$ Cardy Formula},''
\href{http://arxiv.org/abs/1904.03455}{{\ttfamily arXiv:1904.03455 [hep-th]}}.

\bibitem{Cabo-Bizet:2019osg}
A.~Cabo-Bizet, D.~Cassani, D.~Martelli, and S.~Murthy, ``{The asymptotic growth
  of states of the 4d N=1 superconformal index},'' {\em Submitted to: J. High
  Energy Phys.} (2019) ,
\href{http://arxiv.org/abs/1904.05865}{{\ttfamily arXiv:1904.05865 [hep-th]}}.

\bibitem{Amariti:2019mgp}
A.~Amariti, I.~Garozzo, and G.~Lo~Monaco, ``{Entropy function from toric
  geometry},''
\href{http://arxiv.org/abs/1904.10009}{{\ttfamily arXiv:1904.10009 [hep-th]}}.

\bibitem{Benini:2018ywd}
F.~Benini and P.~Milan, ``{Black holes in 4d $\mathcal{N}=4$
  Super-Yang-Mills},''
\href{http://arxiv.org/abs/1812.09613}{{\ttfamily arXiv:1812.09613 [hep-th]}}.

\bibitem{Benini:2018mlo}
F.~Benini and P.~Milan, ``{A Bethe Ansatz type formula for the superconformal
  index},''
\href{http://arxiv.org/abs/1811.04107}{{\ttfamily arXiv:1811.04107 [hep-th]}}.

\bibitem{Closset:2017bse}
C.~Closset, H.~Kim, and B.~Willett, ``{$ \mathcal{N} $ = 1 supersymmetric
  indices and the four-dimensional A-model},''
  \href{http://dx.doi.org/10.1007/JHEP08(2017)090}{{\em JHEP} {\bfseries 08}
  (2017) 090},
\href{http://arxiv.org/abs/1707.05774}{{\ttfamily arXiv:1707.05774 [hep-th]}}.

\bibitem{Lezcano:2019pae}
A.~G. Lezcano and L.~A. Pando~Zayas, ``{Microstate Counting via Bethe Ans\"atze
  in the 4d ${\cal N}=1$ Superconformal Index},''
\href{http://arxiv.org/abs/1907.12841}{{\ttfamily arXiv:1907.12841 [hep-th]}}.

\bibitem{Dolan:2008qi}
F.~A. Dolan and H.~Osborn, ``{Applications of the Superconformal Index for
  Protected Operators and q-Hypergeometric Identities to N=1 Dual Theories},''
  \href{http://dx.doi.org/10.1016/j.nuclphysb.2009.01.028}{{\em Nucl. Phys.}
  {\bfseries B818} (2009) 137--178},
\href{http://arxiv.org/abs/0801.4947}{{\ttfamily arXiv:0801.4947 [hep-th]}}.

\bibitem{Felder}
G.~Felder and A.~N. Varchenko, ``{The elliptic $\Gamma$ function and $SL(3,Z)
  \times Z^{3}$},''. \url{http://cds.cern.ch/record/393292}.

\bibitem{Klebanov:1998hh}
I.~R. Klebanov and E.~Witten, ``{Superconformal field theory on three-branes at
  a Calabi-Yau singularity},''
  \href{http://dx.doi.org/10.1016/S0550-3213(98)00654-3}{{\em Nucl. Phys.}
  {\bfseries B536} (1998) 199--218},
\href{http://arxiv.org/abs/hep-th/9807080}{{\ttfamily arXiv:hep-th/9807080
  [hep-th]}}.

\bibitem{Hong:2018viz}
J.~Hong and J.~T. Liu, ``{The topologically twisted index of $ \mathcal{N} $ =
  4 super-Yang-Mills on T$^{2} \times S^{2}$ and the elliptic genus},''
  \href{http://dx.doi.org/10.1007/JHEP07(2018)018}{{\em JHEP} {\bfseries 07}
  (2018) 018},
\href{http://arxiv.org/abs/1804.04592}{{\ttfamily arXiv:1804.04592 [hep-th]}}.

\bibitem{klebanov1998superconformal}
I.~R. Klebanov and E.~Witten, ``Superconformal field theory on threebranes at a
  calabi-yau singularity,'' {\em Nuclear Physics B} {\bfseries 536} no.~1-2,
  (1998) 199--218.

\bibitem{Butti:2005vn}
A.~Butti and A.~Zaffaroni, ``{R-charges from toric diagrams and the equivalence
  of a-maximization and Z-minimization},''
  \href{http://dx.doi.org/10.1088/1126-6708/2005/11/019}{{\em JHEP} {\bfseries
  11} (2005) 019},
\href{http://arxiv.org/abs/hep-th/0506232}{{\ttfamily arXiv:hep-th/0506232
  [hep-th]}}.

\bibitem{Hanany:2005ve}
A.~Hanany and K.~D. Kennaway, ``{Dimer models and toric diagrams},''
\href{http://arxiv.org/abs/hep-th/0503149}{{\ttfamily arXiv:hep-th/0503149
  [hep-th]}}.

\bibitem{Franco:2005rj}
S.~Franco, A.~Hanany, K.~D. Kennaway, D.~Vegh, and B.~Wecht, ``{Brane dimers
  and quiver gauge theories},''
  \href{http://dx.doi.org/10.1088/1126-6708/2006/01/096}{{\em JHEP} {\bfseries
  01} (2006) 096},
\href{http://arxiv.org/abs/hep-th/0504110}{{\ttfamily arXiv:hep-th/0504110
  [hep-th]}}.

\bibitem{Benvenuti:2006xg}
S.~Benvenuti, L.~A. Pando~Zayas, and Y.~Tachikawa, ``{Triangle anomalies from
  Einstein manifolds},''
  \href{http://dx.doi.org/10.4310/ATMP.2006.v10.n3.a4}{{\em Adv. Theor. Math.
  Phys.} {\bfseries 10} no.~3, (2006) 395--432},
\href{http://arxiv.org/abs/hep-th/0601054}{{\ttfamily arXiv:hep-th/0601054
  [hep-th]}}.

\bibitem{Benvenuti:2004dy}
S.~Benvenuti, S.~Franco, A.~Hanany, D.~Martelli, and J.~Sparks, ``{An Infinite
  family of superconformal quiver gauge theories with Sasaki-Einstein duals},''
  \href{http://dx.doi.org/10.1088/1126-6708/2005/06/064}{{\em JHEP} {\bfseries
  06} (2005) 064},
\href{http://arxiv.org/abs/hep-th/0411264}{{\ttfamily arXiv:hep-th/0411264
  [hep-th]}}.

\bibitem{Benvenuti:2005ja}
S.~Benvenuti and M.~Kruczenski, ``{From Sasaki-Einstein spaces to quivers via
  BPS geodesics: L**p,q|r},''
  \href{http://dx.doi.org/10.1088/1126-6708/2006/04/033}{{\em JHEP} {\bfseries
  04} (2006) 033},
\href{http://arxiv.org/abs/hep-th/0505206}{{\ttfamily arXiv:hep-th/0505206
  [hep-th]}}.

\bibitem{Butti:2005sw}
A.~Butti, D.~Forcella, and A.~Zaffaroni, ``{The Dual superconformal theory for
  L**pqr manifolds},''
  \href{http://dx.doi.org/10.1088/1126-6708/2005/09/018}{{\em JHEP} {\bfseries
  09} (2005) 018},
\href{http://arxiv.org/abs/hep-th/0505220}{{\ttfamily arXiv:hep-th/0505220
  [hep-th]}}.

\bibitem{Franco:2005sm}
S.~Franco, A.~Hanany, D.~Martelli, J.~Sparks, D.~Vegh, and B.~Wecht, ``{Gauge
  theories from toric geometry and brane tilings},''
  \href{http://dx.doi.org/10.1088/1126-6708/2006/01/128}{{\em JHEP} {\bfseries
  01} (2006) 128},
\href{http://arxiv.org/abs/hep-th/0505211}{{\ttfamily arXiv:hep-th/0505211
  [hep-th]}}.

\bibitem{Hanany:2005hq}
A.~Hanany, P.~Kazakopoulos, and B.~Wecht, ``{A New infinite class of quiver
  gauge theories},''
  \href{http://dx.doi.org/10.1088/1126-6708/2005/08/054}{{\em JHEP} {\bfseries
  08} (2005) 054},
\href{http://arxiv.org/abs/hep-th/0503177}{{\ttfamily arXiv:hep-th/0503177
  [hep-th]}}.

\end{thebibliography}\endgroup

\end{document}